\documentclass[11pt, a4paper]{article}

\usepackage[utf8]{inputenc}
\usepackage[T1]{fontenc}
\usepackage{lmodern}

\usepackage[margin=1in]{geometry}
\usepackage{setspace}
\usepackage{titlesec}
\titleformat*{\section}{\large\bfseries}
\titleformat*{\subsection}{\normalsize\bfseries}
\usepackage{amsmath, amssymb, amsthm, amsfonts, mathtools, bm}
\allowdisplaybreaks

\usepackage[colorinlistoftodos, textsize=footnotesize]{todonotes}
\newcounter{DannyComment}

\usepackage[authoryear,round]{natbib}

\definecolor{darkblue}{rgb}{0.0, 0.0, 0.55}
\usepackage[pagebackref=true, colorlinks=true, allcolors=darkblue]{hyperref}
\renewcommand*{\backref}[1]{}
\renewcommand*{\backrefalt}[4]{%
    \ifcase #1%
    \or #2%
    \else #2%
    \fi%
}

\makeatletter
\newtheoremstyle{smallcaps}
  {}
  {}
  {\itshape}
  {}
  {\scshape}
  {.}
  { }
  {}

\renewenvironment{proof}[1][\proofname]{\par
  \pushQED{\qed}%
  \normalfont \topsep6\p@\@plus6\p@\relax
  \trivlist
  \item[\hskip\labelsep
        \scshape
    #1\@addpunct{.}]\ignorespaces
}{%
  \popQED\endtrivlist\@endpefalse
}
\makeatother
\theoremstyle{smallcaps}
\newtheorem{theorem}{Theorem}[section]
\newtheorem{lemma}[theorem]{Lemma}
\newtheorem{corollary}[theorem]{Corollary}
\newtheorem{claim}[theorem]{Claim}
\newtheorem{observation}[theorem]{Observation}

\newcommand{\pr}[1]{\operatorname{Pr}\left[#1\right]}
\newcommand{\ex}[1]{\mathbb{E}\left[#1\right]}

\newcommand{\expar}[1]{\mathbb{E}[#1]}
\newcommand{\exsub}[2]{\mathbb{E}_{#1}\left[#2\right]}
\newcommand{\prsub}[2]{\operatorname{Pr}_{#1}\left[#2\right]}
\newcommand{\prsubpar}[2]{\operatorname{Pr}_{#1}[#2]}
\newcommand{\exsubpar}[2]{\mathbb{E}_{#1}[#2]}
\newcommand{\exsubnop}[1]{\mathbb{E}_{#1}}
\newcommand{\MyAbove}[2]{\genfrac{}{}{0pt}{}{#1}{#2}}
\newcommand{\eps}{\epsilon}
\newcommand{\bbR}{\mathbb{R}}
\newcommand{\bbN}{\mathbb{N}}
\newcommand{\mydense}{\mathrm{dense}}
\newcommand{\mycomb}{\mathrm{combined}}
\newcommand{\mycyc}{\mathrm{cycle}}

\newcommand{\mypref}{\mathrm{prefix}}
\newcommand{\mysuff}{\mathrm{suffix}}
\newcommand{\mymid}{\mathrm{mid}}
\newcommand{\myeoq}{\mathrm{EOQ}}
\newcommand{\opt}{\mathrm{OPT}}

\title{\textbf{Economic Warehouse Lot Scheduling: \\
Breaking the {\boldmath $2$}-Approximation Barrier}}
\author{
    Danny Segev\thanks{School of Mathematical Sciences and Coller School of Management, Tel Aviv University, Tel Aviv 69978, Israel. Email: \texttt{segevdanny@tauex.tau.ac.il}. Supported by Israel Science Foundation grant 1407/20.}
}
\date{}

\begin{document}

\maketitle

\begin{abstract}
\noindent The economic warehouse lot scheduling problem is a foundational inventory-theory model, capturing computational challenges in dynamically coordinating replenishment decisions for multiple commodities subject to a shared capacity constraint. Even though this model has generated a vast body of literature over the last six decades, our algorithmic understanding has remained surprisingly limited. Indeed, for general problem instances, the best-known approximation guarantees have remained at a factor of $2$ since the mid-1990s. These guarantees were attained by the now-classic work of Anily [Operations Research,~1991] and Gallego, Queyranne, and Simchi-Levi [Operations Research,~1996] via the highly-structured class of ``stationary order sizes and stationary intervals'' (SOSI) policies, thereby avoiding direct competition against fully dynamic policies.

\medskip \noindent The main contribution of this paper resides in developing new analytical foundations and algorithmic techniques that enable such direct comparisons, leading to the first provable improvement over the $2$-approximation barrier. Leveraging these ideas, we design a constructive approach that allows us to balance cost and capacity at a finer granularity than previously possible via SOSI-based methods. Consequently, given any economic warehouse lot scheduling instance, we present a polynomial-time construction of a random capacity-feasible dynamic policy whose expected long-run average cost is within factor $2-\frac{17}{5000} + \eps$ of optimal.
\end{abstract}

\vspace{1em}
{\small \noindent \textbf{Keywords:} Inventory management, approximation algorithms, power-of-$2$ policies, dynamic replenishment, capacity-constrained scheduling.}

\thispagestyle{empty}

\newpage
\thispagestyle{empty}
\tableofcontents

\newpage
\setcounter{page}{1}
\section{Introduction}

The central objective of thiis paper is to advance our understanding of a true inventory management classic --- the economic warehouse lot scheduling problem --- by developing new algorithmic techniques and analytical ideas for efficiently constructing provably near-optimal replenishment policies. Since its introduction in the late 1950s, and through numerous subsequent formulations, this problem has generated a vast literature. Nevertheless, many of its surrounding computational questions remain poorly understood, with the inherently dynamic nature of optimal replenishment policies and their rich temporal structure rendering direct analysis elusive to date. Consequently, most existing work has either restricted attention to highly-structured policy classes or abandoned rigorous performance guarantees altogether, leaving the best-known approximation results untouchable for several decades. The present work addresses this long-standing gap by establishing new analytical foundations that enable head-to-head benchmarking against fully dynamic policies. Combined with further algorithmic progress and structural insights, these ideas yield a provable sub-$2$-approximation for general problem instances. This result constitutes the first improvement over the performance guarantees achieved by ``stationary order sizes and stationary intervals'' (SOSI) policies, which have been state-of-the-art since the mid-1990s~\citep{Anily91, GallegoQS96}.

Given the breadth of work that has accumulated around the economic warehouse lot scheduling problem, including heuristic methods, computational studies, and a wide range of applications, it is beyond the scope of this paper to provide a comprehensive literature review. For broader treatments, we refer readers to authoritative book chapters \citep{HadleyW1963, JohnsonM74, HaxC84, Zipkin00, SimchiLeviCB14, NahmiasL15}, as well as to the references therein; it may also be instructive to revisit early foundational contributions such as those of \citet{Holt58}, \citet{Homer1966}, \citet{PageP76}, and \citet{Zoller77}. At a high level, economic warehouse lot scheduling and most of its derivatives address lot-sizing and scheduling decisions for multiple commodities over a given planning horizon, with the objective of minimizing long-run average operating costs. Beyond marginal ordering and inventory holding costs, the defining challenge of this model lies in the interaction among commodities induced by a shared resource, referred to as ``warehouse space'', just to align ourselves with traditional terminology. In this regard, capacity-feasible replenishment policies must therefore ensure that, at any point in time, their aggregate inventory can be jointly packed within the warehouse capacity, with each commodity contributing toward this space requirement at an individual rate. This seemingly-mild coupling constraint fundamentally alters the structure of optimal policies, violating many computationally-useful properties that underlie classical single-commodity models. As a result, existing theory offers only limited guidance, leaving long-standing algorithmic questions regarding the efficient identification of provably-good capacity-feasible policies, as well as a substantial gap in our understanding of their structural properties.

To examine these questions in greater detail, Section~\ref{subsec:model_definition} presents a complete mathematical formulation of the economic warehouse lot scheduling problem in its most general form. Section~\ref{subsec:related_work} reviews central theoretical developments, including the classical SOSI-based approach, which provides the strongest approximation guarantees known to date, while simultaneously highlighting the key open questions that motivate our work. Finally, Section~\ref{subsec:contributions} outlines the main contributions of this paper, with detailed structural results, algorithmic constructions, and analytical arguments appearing in subsequent sections.

\subsection{Model formulation} \label{subsec:model_definition}

\paragraph{The economic order quantity model.} We begin by introducing the basic building block of our framework, the economic order quantity (EOQ) model, which captures single-commodity settings. In a nutshell, our objective is to identify an optimal dynamic replenishment policy for a single commodity over the continuous planning horizon $[0,\infty)$, so as to minimize its long-run average cost. Without loss of generality, this commodity is assumed to face a stationary demand rate of $1$, which must be met in full upon occurrence; lost sales and backorders are therefore not permitted. From a solution perspective, a ``dynamic'' policy ${\cal P}$ will be specified by two components:
\begin{itemize}
    \item A sequence of ordering points $0=\tau_0 < \tau_1 < \tau_2 < \cdots$ that covers the entire planning horizon $[0,\infty)$, in the sense that $\lim_{ k \to \infty } \tau_k = \infty$.

    \item The corresponding real-valued order quantities $q_0, q_1, q_2, \ldots$ placed at these points.
\end{itemize}
For a given replenishment policy ${\cal P}$, we make use of $I({\cal P},t)$ to designate its inventory level at time $t$. This function can be expressed as
$I({\cal P},t)=\sum_{k\ge 0:\tau_k\le t} q_k - t$, where the first term represents the total ordering quantity up to and
including time $t$, which is a finite sum due to having $\lim_{ k \to \infty } \tau_k = \infty$, and the second term corresponds to cumulative demand under the unit demand rate assumption. As such, we are indeed meeting our demand at all times if and only if $I( {\cal P}, t ) \geq 0$ for all $t \in [0,\infty)$. Policies satisfying this condition are said to be feasible, and we denote the collection of such policies by ${\cal F}$.

We next describe the cost structure associated with any feasible policy ${\cal P}\in{\cal F}$ and its resulting long-run average cost function. To this end, each of the ordering points $\tau_0, \tau_1, \tau_2, \ldots$ incurs a fixed setup cost of $K$, independent of the corresponding order quantity. Accordingly, letting
$N({\cal P},[0,t)) = |\{k\in\bbN_0:\tau_k<t\}|$ denote the number of orders placed over the interval $[0,t)$, the total ordering cost incurred on this interval is ${\cal K}({\cal P},[0,t)) = K\cdot N({\cal P},[0,t))$; clearly, the latter component alone favors infrequent ordering. In contrast, carried inventory incurs a linear holding cost, at rate $2H$ per unit of inventory per unit of time. Along this dimension, ${\cal H}({\cal P},[0,t))$ will denote our total holding cost over $[0,t)$, given by ${\cal H}( {\cal P}, [0,t)) = 2H \cdot \int_{[0,t)} I( {\cal P}, \tau ) \mathrm{d} \tau$, which penalizes high inventory levels and thus promotes more frequent replenishment. Taking these two components into account, the combined cost of a feasible policy ${\cal P}$ over $[0,t)$ will be designated by $C( {\cal P}, [0,t)) = {\cal K}( {\cal P}, [0,t)) + {\cal H}( {\cal P}, [0,t))$. The long-run average cost of this policy is then defined by
\begin{equation} \label{eqn:long_run_cost_single}
C( {\cal P} ) ~~=~~ \limsup_{t \to \infty} \frac{ C( {\cal P}, [0,t)) }{ t } \ .
\end{equation}
As an aside, it is easy to come up with feasible policies for which $\lim_{t\to\infty} \frac{ C( {\cal P}, [0,t)) }{ t }$ fails to exist, necessitating the use of $\limsup$ as part of definition~\eqref{eqn:long_run_cost_single}.

Under the economic order quantity model described above, our objective is to identify a feasible replenishment policy ${\cal P}$ that minimizes long-run average cost, that is, $C({\cal P})=\min_{\hat{\cal P}\in{\cal F}} C(\hat{\cal P})$. A classical and well-documented result, which can be found in relevant textbooks such as those of \citet[Sec.~3]{Zipkin00}, \citet[Sec.~2]{MuckstadtS10}, and \citet[Sec.~7.1]{SimchiLeviCB14}, states that this minimum is in fact attained by the extremely simple class of stationary order size and stationary interval (SOSI) policies. Such policies are parameterized by a single parameter, $T>0$, representing the constant time interval between successive orders. Once $T$ is fixed, orders are placed at the time points $0,T,2T,3T,\ldots$, each with quantity exactly $T$, so that inventory is depleted to zero at every ordering epoch. This structure allows the long-run average cost of a SOSI policy to be expressed in closed form as $C_{\myeoq}(T)=\frac{K}{T}+HT$, allowing the economic order quantity problem to admit an explicit closed-form solution. The next claim collects several well-known properties of this function, all following from elementary arguments.

\begin{claim} \label{clm:EOQ_properties}
The cost function $C_{\myeoq}(T) = \frac{ K }{ T } + HT$ satisfies the next few properties:
\begin{enumerate}
    \item $C_{\myeoq}$ is strictly convex.

    \item The unique minimizer of $C_{\myeoq}$ is $T^* = \sqrt{ K / H }$.

    \item For every $\alpha > 0$ and $T > 0$,
    \[ C_{\myeoq}( \alpha T ) ~\leq~ \max \left\{ \alpha, \frac{ 1 }{ \alpha } \right\} \cdot C_{\myeoq}(T) \ \ \text{and} \ \ \ C_{\myeoq}( \alpha T ) + C_{\myeoq} ( T / \alpha ) ~=~ \frac{ \alpha^2 + 1 }{ \alpha } \cdot C_{\myeoq}(T) \ . \]
\end{enumerate}
\end{claim}

\paragraph{The economic warehouse lot scheduling problem.} Given the optimality of SOSI policies for the economic order quantity model, it is natural to ask whether emphasizing fully dynamic policies is in fact necessary. Circling back to this issue in the sequel, we proceed by explaining how the EOQ framework provides a principled path to the economic warehouse lot scheduling problem, whose core challenge can be summarized by the following high-level question:
\begin{quote}
{\em How should multiple economic order quantity models be coordinated when different commodities are coupled through a shared resource?}
\end{quote}
Concretely, we wish to synchronize the joint lot sizing of $n$ distinct commodities, where each commodity $i\in[n]$ is governed by its own EOQ model, with ordering and holding cost parameters $K_i$ and $2H_i$, respectively. As in the single-commodity setting, choosing a dynamic replenishment policy ${\cal P}_i\in{\cal F}$ for this commodity induces a marginal long-run average cost of $C_i({\cal P}_i)$, defined according to equation~\eqref{eqn:long_run_cost_single}. The fundamental complication, however, is the presence of a shared ``warehouse space'' constraint: At any point in time, the inventory levels of all commodities must be jointly packable within a fixed warehouse capacity, with each commodity contributing toward this space requirement at an individual rate.

To formalize this constraint, we assume that each unit of commodity $i\in[n]$ requires $\gamma_i$ units of space when stored in a common warehouse of total capacity ${\cal V}$. Accordingly, for any joint dynamic replenishment policy ${\cal P}=({\cal P}_1,\ldots,{\cal P}_n)\in{\cal F}^n$, by recalling that $\{ I( {\cal P}_i, t) \}_{i \in [n]}$ designate its underlying inventory levels at time $t \in [0, \infty)$, it follows that the total warehouse space occupied at that time is given by $V({\cal P},t)=\sum_{i\in[n]}\gamma_i \cdot I({\cal P}_i,t)$. In turn, the maximum space ever required by this policy is $V_{\max}({\cal P})=\sup_{t\in[0,\infty)} V({\cal P},t)$, and we say that ${\cal P}$ is capacity-feasible when $V_{\max}({\cal P})\leq {\cal V}$. Clearly, this requirement is substantially stronger than simply imposing feasibility on each of the marginal policies ${\cal P}_1,\ldots,{\cal P}_n$ in isolation.

Putting the above components together, the economic warehouse lot scheduling problem is asking us to determine a capacity-feasible replenishment policy ${\cal P} = ( {\cal P}_1, \ldots, {\cal P}_n )$ whose long-run average
cost $C( {\cal P} ) = \sum_{i \in [n]} C_i( {\cal P}_i )$ is minimized. For notational convenience, we compactly formulate this model as
\begin{align} \label{eqn:model_warehouse} \tag{$\Pi$}
\begin{array}{ll}
{\displaystyle \min_{{\cal P} \in {\cal F}^n}} & C( {\cal P} ) \\
\text{s.t.} & V_{\max}( {\cal P} ) \leq {\cal V}
\end{array}
\end{align}
It is worth noting that the above-mentioned minimum is indeed attained by some capacity-feasible policy. Since the arguments involved are rather straightforward, the proof of this claim is omitted.

\subsection{Known results and open questions} \label{subsec:related_work}

We next review foundational results around the economic warehouse lot scheduling problem, focusing on those that are directly related to our basic research questions. The upcoming discussion proceeds along two complementary axes, one discussing algorithmic approaches aimed at efficiently identifying replenishment policies with provable performance guarantees, and the other addressing known intractability results; as noted earlier, such results have been very scarce. In regard to tangential research directions, given the breadth of this literature, we do not attempt to provide a comprehensive survey. Instead, readers who are interested in broader historical contexts, heuristic methods, and experimental studies are referred to numerous book chapters devoted to these topics \citep{HadleyW1963, JohnsonM74, HaxC84, Zipkin00, SimchiLeviCB14, NahmiasL15}.

\paragraph{Hardness results and representational issues.} Although our primary focus is algorithmic, it is important to briefly review known intractability results, in order to delineate the limits of achievable performance guarantees. In contrast to many deterministic inventory models whose computational complexity remains unresolved, \cite*{GallegoSS92} established definitive hardness results with respect to the efficient computation of optimal replenishment policies. Specifically, they presented a polynomial-time reduction from the $3$-partition problem~\citep{GareyJ75, GareyJ79}, showing that economic warehouse lot scheduling is strongly NP-hard in its decision problem setup. An immediate implication is that optimal replenishment policies cannot be computed in polynomial time unless $\mathrm{P}=\mathrm{NP}$. Moreover, under the same complexity assumption, this setting does not admit a fully polynomial-time approximation scheme~\citep[Sec.~8.3]{Vazirani01}.

Beyond complexity-related findings in terms of computation time, a concurrent obstacle to developing near-optimal policies arises from their representability. In particular, one may ask whether polynomially-bounded memory suffices to describe optimal or near-optimal dynamic policies. Put differently, even if computational efficiency was not a concern, it is unclear whether such policies admit succinct descriptions. One possible avenue toward resolving this question would be to establish the existence of a polynomially representable structure governing the sequences of ordering points $\tau_0^i, \tau_1^i, \tau_2^i, \ldots$ and quantities $q_0^i, q_1^i, q_2^i, \ldots$ associated with each commodity $i\in[n]$. Alternatively, we might hope to show that near-optimal behavior can be captured by a cyclic policy of bounded duration, involving only a polynomial number of ordering events. To the best of our knowledge, however, such structural results remain elusive, having been open for decades even in the most basic setting of just two commodities.

\paragraph{Constant-factor approximations via SOSI policies.} Prior to reviewing state-of-the-art approximation guarantees, it is worth clarifying a potential misconception suggested by parts of the surrounding literature. In particular, a cursory reading of many abstracts might give the impression that the economic warehouse lot scheduling problem is well understood, at least for stylized settings with two or three commodities. The key limitation, however, is that existing provably-good and efficient algorithmic results are focusing on highly-restricted classes of policies, rather than addressing the full space of arbitrarily-structured dynamic policies.

A turning point in our algorithmic understanding of such policies came with the influential work of \citet{Anily91}. Although her analysis was restricted to the class of SOSI policies, she established a rigorous connection between economic warehouse lot scheduling and the SOSI-constrained economic order quantity model, in which commodities interact through a warehouse space constraint on their peak inventory levels. The latter perspective enabled Anily to devise a polynomial-time construction of a SOSI policy whose long-run average cost is within factor $2$ of the minimum achievable within this class. Building on these ideas, the landmark contribution of \cite*{GallegoQS96} combined this framework with additional structural insights, most notably by introducing an elegant convex relaxation that yields a lower bound on the long-run average cost of any capacity-feasible dynamic policy. As a direct consequence, they were successful at approximating the economic warehouse lot scheduling problem in its utmost generality, proposing a polynomial-time algorithm for identifying a SOSI policy whose cost is within factor $2$ of optimal; this time, ``optimal'' refers to the best-possible dynamic policy!

\paragraph{Recent approximation scheme for \boldmath{$O(1)$} commodities.} Moving past constant-factor approximations, our very recent work~\citep{Segev26EWLSP_PTAS} developed novel dynamic programming ideas, proving that for constantly-many commodities, truly near-optimal policies are attainable. Specifically, for any $\eps > 0$, we proposed an $O( | {\cal I} |^{O(n)} \cdot 2^{ O( n^{6} / \eps^{5} ) } )$-time approach for computing an efficiently-representable capacity-feasible policy whose long-run average cost is within factor $1 + \eps$ of optimal. This long-awaited progress identifies with the notion of a polynomial-time approximation scheme (PTAS) for constantly-many commodities, i.e., when $n = O(1)$. As explained later on, even though the algorithmic ideas behind this result and their analysis are very different from those of the current paper, our approximation scheme will still play an important role as a black-box procedure in isolating several parametric regimes of the economic warehouse lot scheduling problem.

\paragraph{Main open questions.} Somewhat surprisingly, despite a steady stream of work over the past three decades --- largely focused on stylized settings, heuristic methods, and experimental investigations --- the algorithmic framework of \citet{Anily91} and \citet{GallegoQS96}, together with its lower-bounding mechanism, continues to represent the best-known approximation guarantees for computing dynamic replenishment policies. In light of this persistent gap, a set of fundamental questions has repeatedly emerged in related literature, textbooks, and course materials, lying at the core of our present work. These questions can be succinctly summarized as follows:
\begin{itemize}
    \item {\em Can these long-standing guarantees be surpassed for general problem instances?}

    \item {\em What algorithmic techniques and analytical ideas might enable such progress?}

    \item {\em Despite inherent representational challenges, can the flexibility of dynamic policies be exploited to obtain a sub-$2$-approximation?}
\end{itemize}

\subsection{Main contributions} \label{subsec:contributions}

This paper establishes the first provable improvement over the $2$-approximation barrier for the economic warehouse lot scheduling problem. Our main contribution resides in developing new analytical foundations and algorithmic techniques that enable direct comparisons against fully dynamic replenishment policies, rather than against restricted policy classes. Leveraging these ideas, we design a constructive approach that surpasses the long-standing approximation guarantees of \citet{Anily91} and \citet{GallegoQS96}, while relying on fundamentally different algorithmic principles. Specifically, as formalized in Theorem~\ref{thm:2_minus_delta}, we present a polynomial-time construction of a random capacity-feasible dynamic policy whose expected long-run average cost is within factor $2-\frac{17}{5000} + \eps$ of optimal, for general instances of the economic warehouse lot scheduling problem.

\begin{theorem} \label{thm:2_minus_delta}
For any $\eps \in (0,\frac{ 1 }{ 10 })$, we can compute in $O( | {\cal I} |^{\tilde{O}( 1/\eps^5)} \cdot 2^{ \tilde{O}( 1 / \eps^{35} ) } )$ time a random capacity-feasible replenishment policy ${\cal P}$, with an expected long-run average cost of
\[ \ex{ C({\cal P}) } ~~\leq~~ \left( 2-\frac{17}{5000} + \eps \right) \cdot \opt\eqref{eqn:model_warehouse} \ . \]
Here, $| {\cal I} |$ stands for the input length in its binary specification.
\end{theorem}

It is important to emphasize that, in terms of vicinity-to-optimum and running times, the above-mentioned guarantees are by no means attempting to optimize constants or to obtain the most efficient implementation. Rather, our presentation is deliberately designed to strike a balance between achieving long-awaited improvements over classical results and introducing new technical ideas in their simplest and most transparent form.

\paragraph{Technical highlights.} To place the overall discussion in context, Section~\ref{sec:anily_approx} reexamines the classical approximation framework of \citet{Anily91} and \citet{GallegoQS96} from a technical perspective that will serve our purposes later on. In particular, we identify the precise analytical bottlenecks responsible for the factor-$2$ barrier, tracing them to the reliance on global SOSI-based arguments and on quite coarse lower bounds. As such, this barrier appears to be methodological rather than inherent to the computational setting itself. Our three-pronged objective is to assess the limitations of existing techniques, to set the stage for their usefulness in specific parametric regimes, and to motivate the pursuit of fundamentally different ideas.

These new ideas are introduced in Section~\ref{sec:sub2-approx}, which develops a structural decomposition of economic warehouse lot scheduling instances, lying at the heart of our sub-$2$-approximation. Here, we partition the underlying commodities into ``volume classes'' according to their contribution to warehouse capacity under a near-optimal benchmark policy, and show that qualitatively different algorithmic strategies are required across distinct parametric regimes. This decomposition motivates us to eventually synthesize classical SOSI constructions, our recent approximation scheme for constantly-many commodities~\citep{Segev26EWLSP_PTAS}, and of primary importance, new probabilistic coordination mechanisms, moving beyond uniform scaling arguments and directly targeting the factor-$2$ barrier.

The most technically demanding regime of our decomposition is resolved in Section~\ref{sec:suffix_dense}, where we introduce a new matching-based construction, effectively simulating the behavior of a near-optimal benchmark policy without knowing its exact structure. This construction replaces global coordination constraints with carefully-engineered pairwise interactions between different commodities, allowing cost and capacity to be balanced at a fine granularity, which is where the quantitative improvement over factor $2$ is ultimately realized. The resulting policies sharply depart from SOSI-style structures and constitute a central algorithmic innovation of this paper.

Section~\ref{sec:po2-sync} concludes our analysis by establishing the Po2-Synchronization Theorem, a new structural result on which the constructions of Section~\ref{sec:suffix_dense} crucially rely. At a high level, we argue that in certain parametric regimes, highly-complex dynamic behavior can be synchronized into a restricted family of replenishment patterns with controlled cost and capacity increments. This result provides the missing analytical bridge between arbitrary dynamic policies and the tractable objects manipulated by our algorithms. Together, Sections~\ref{sec:suffix_dense} and~\ref{sec:po2-sync} deliver the core technical insights of this paper and demonstrate that fully dynamic policies can be exploited --- rather than avoided--- to break the classical $2$-approximation barrier. 
\section{Background: Revisiting the Classical Approximation Framework} \label{sec:anily_approx}

In this section, we revisit some of the technical ideas developed by \citet{Anily91} and \citet{GallegoQS96} for computing a capacity-feasible replenishment policy whose long-run average cost is within factor $2$ of optimal, with the explicit goal of isolating the analytical mechanisms underlying this long-standing barrier. It is important to emphasize that, even though the upcoming contents are slightly different from their conventional presentation, the intrinsic ideas should be fully attributed to these authors. As discussed below, rather than treating their results as black-box benchmarks, we examine basic technical arguments, to distinguish problem-inherent obstacles from those arising due to SOSI-based methodology. This perspective lays the analytical groundwork behind several algorithmic developments in Section~\ref{sec:sub2-approx} and motivates the need for fundamentally different ideas.

\subsection{The average-space bound and its induced relaxation} \label{subsec:average_bound_relaxation}

\paragraph{The average-space bound.} In what follows, we elaborate on the well-known average-space bound, linking the overall capacity ${\cal V}$ and the average inventory levels of the underlying commodities with respect to any capacity-feasible cyclic policy ${\cal P}$. Such policies can be viewed as those repeating precisely the same actions across a bounded-length cycle, say $[0, \tau_\mycyc)$. To arrive at the desired bound, suppose we sample a uniformly distributed random variable $\Theta \sim U[0,\tau_\mycyc)$. Then, over the randomness in $\Theta$, the expected space occupied by ${\cal P}$ can be written as
\begin{align}
\exsub{ \Theta }{V( {\cal P}, \Theta)} & ~~=~~ \frac{ 1 }{ \tau_\mycyc } \cdot \int_{[0,\tau_\mycyc)} V( {\cal P}, t ) \mathrm{d} t \nonumber \\
& ~~=~~ \frac{ 1 }{ \tau_\mycyc } \cdot \sum_{i \in [n]} \gamma_i \cdot \int_{[0,\tau_\mycyc)} I( {\cal P}_i, t ) \mathrm{d} t \nonumber \\
& ~~=~~ \sum_{i \in [n]} \gamma_i \cdot \bar{I}( {\cal P}_i ) \ , \label{eqn:expect_uniform_sample}
\end{align}
with the convention that $\bar{I}( {\cal P}_i ) = \frac{ 1 }{ \tau_\mycyc } \cdot \int_{[0,\tau_\mycyc)} I( {\cal P}_i, t ) \mathrm{d} t$ denotes the expected inventory level of commodity $i$ across a single cycle. Now, since ${\cal P}$ is capacity-feasible, $V( {\cal P}, \Theta ) \leq {\cal V}$ almost surely, implying in conjunction with representation~\eqref{eqn:expect_uniform_sample} that
\begin{equation} \label{eqn:average-space-bound}
\sum_{i \in [n]} \gamma_i \cdot \bar{I}( {\cal P}_i ) ~~\leq~~ {\cal V} \ .
\end{equation}

\paragraph{The resulting relaxation.} Now, suppose we wish to compute a cyclic replenishment policy ${\cal P}$ of minimum long-run average cost, when the capacity constraint $V_{\max}( {\cal P} ) \leq {\cal V}$ is replaced by inequality~\eqref{eqn:average-space-bound}. In other words, we consider the next formulation:
\begin{align} \label{eqn:relax_warehouse}
\tag{$\tilde{\Pi}_{\mycyc}$}
\begin{array}{lll}
{\displaystyle \inf_{{\cal P} \text{ cyclic}}} & C( {\cal P} ) \\
\text{s.t.} & {\displaystyle \sum_{i \in [n]} \gamma_i \cdot \bar{I}( {\cal P}_i ) \leq {\cal V}}
\end{array}
\end{align}
Based on the preceding discussion, it follows that any capacity-feasible cyclic policy for our original problem~\eqref{eqn:model_warehouse} is a feasible solution to~\eqref{eqn:relax_warehouse}. Moreover, it is well known that the non-necessarily-cyclic optimum, $\opt\eqref{eqn:model_warehouse}$, can be approximated by a cyclic policy within any degree of accuracy. In other words, for any $\eps > 0$, there exists a capacity-feasible cyclic policy ${\cal P}^{\eps}$ with a long-run average cost of $C( {\cal P}^{\eps} ) \leq (1 + \eps) \cdot \opt\eqref{eqn:model_warehouse}$. The latter claim can be ascertained, for example, by consulting our earlier work on economic warehouse lot scheduling~\citep{Segev26EWLSP_PTAS}, in which Theorem~2 actually proves a stronger result, with specific limits on the cycle length and number of orders per commodity. Consequently, it follows that~\eqref{eqn:relax_warehouse} forms a relaxation of~\eqref{eqn:model_warehouse}.

\begin{corollary} \label{cor:rel_vs_opt_ware}
$\opt\eqref{eqn:relax_warehouse} \leq \opt\eqref{eqn:model_warehouse}$.
\end{corollary}

\paragraph{Simplifying {\boldmath $\eqref{eqn:relax_warehouse}$}.} Next, let us temporarily restrict relaxation~\eqref{eqn:relax_warehouse} to stationary order sizes and stationary intervals (SOSI) policies. As explained in Section~\ref{subsec:model_definition}, any such policy is fully specified by a given set of ordering intervals, $T_1, \ldots, T_n$. In turn, for each commodity $i \in [n]$, its long-run average inventory level is precisely $\bar{I}( {\cal P}_i ) = \frac{ T_i }{ 2 }$; similarly, its long-run average cost is $C_{\myeoq,i}( T_i ) = \frac{ K_i }{ T_i } + H_i T_i$. Consequently, when specializing formulation~\eqref{eqn:relax_warehouse} to SOSI policies, our resulting problem can be written as:
\begin{align} \label{eqn:modify_relax_warehouse}
\tag{$\tilde{\Pi}_{\text{SOSI}}$}
\begin{array}{lll}
{\displaystyle \min_T} & {\displaystyle \sum_{i \in [n] } C_{\myeoq,i}( T_i )} \\
\text{s.t.} & {\displaystyle \sum_{i \in [n]} \gamma_i T_i \leq 2{\cal V}}
\end{array}
\end{align}
Here, one can easily show that the optimum value is indeed attained via basic convexity arguments. Interestingly, as shown in Lemma~\ref{lem:sosi_optimal} below, by limiting attention to this class of policies, we are not sacrificing optimality in any way. The proof of this result appears in Section~\ref{subsec:proof_lem_sosi_optimal}; as previously mentioned, these details will be important for arguing about several parametric regimes in Section~\ref{sec:sub2-approx}.

\begin{lemma} \label{lem:sosi_optimal}
Problem~\eqref{eqn:relax_warehouse} admits a SOSI optimal replenishment policy, obtained by solving~\eqref{eqn:modify_relax_warehouse}.
\end{lemma}

\paragraph{Optimal solution to~\boldmath{$\eqref{eqn:modify_relax_warehouse}$}?} One approach to solving this relaxation in polynomial time is straightforward, since it is a convex optimization problem. Alternatively, it is not difficult to verify that we can compute a $(1+\eps)$-approximate policy for~\eqref{eqn:modify_relax_warehouse} in $O( \frac{ n^3 }{ \eps } )$ time by means of dynamic programming. In essence, by observing that commodities are only linked together by the additive constraint $\sum_{i \in [n]} \gamma_i T_i \leq 2{\cal V}$, we can discretize the basic units of this budget to integer multiples of $\frac{ \eps }{ n } \cdot {\cal V}$. As such, a knapsack-like dynamic program will process one commodity after the other, solving in each step a single-variable subproblem of the form ``minimize $C_{\myeoq,i}(T_i)$ subject to an upper bound on $T_i$'', which admits a closed-form solution (see, for example, Section~\ref{subsec:useful_partition_dense}).

\subsection{The \texorpdfstring{\boldmath{$2$}}{}-approximate policy} \label{subsec:final_2app_anily}

Let $T^* = (T_1^*, \ldots, T_n^*)$ be an optimal solution to~\eqref{eqn:modify_relax_warehouse}, noting that this policy is generally not capacity-feasible, since its peak space requirement is $V_{\max}( T^* ) = \sum_{i \in [n]} \gamma_i T_i^*$, which could be greater than ${\cal V}$. To correct this issue, consider the scaled-down SOSI policy $\hat{T} = (\hat{T}_1, \ldots, \hat{T}_n)$, defined by setting $\hat{T}_i = \frac{ T_i^* }{ 2 }$ for every commodity $i \in [n]$. As such, $\hat{T}$ is capacity-feasible, since
\begin{equation} \label{eqn:fix_cap_scale}
V_{\max}( \hat{T} ) ~~=~~ \sum_{i \in [n]} \gamma_i \hat{T}_i ~~=~~ \frac{ 1 }{ 2 } \cdot \sum_{i \in [n]} \gamma_i T_i^* ~~\leq~~ {\cal V} \ ,
\end{equation}
where the last inequality holds since $T^*$ is in particular a feasible solution to~\eqref{eqn:modify_relax_warehouse}, implying that $\sum_{i \in [n]} \gamma_i T_i^* \leq 2{\cal V}$. In terms of cost,
\begin{align}
C( \hat{T} ) & ~~=~~ \sum_{i \in [n] } \left( \frac{ K_i }{ \hat{T}_i } + H_i \hat{T}_i \right) \nonumber \\
& ~~=~~ \sum_{i \in [n] } \left( \frac{ 2K_i }{ T_i^* } + \frac{ H_i T_i^* }{ 2 } \right) \nonumber \\
& ~~\leq~~ 2 \cdot \sum_{i \in [n] } C_{\myeoq,i}( T_i^* ) \nonumber \\
& ~~=~~ 2 \cdot \opt\eqref{eqn:modify_relax_warehouse} \label{eqn:final_2app_eq1} \\
& ~~=~~ 2 \cdot \opt\eqref{eqn:relax_warehouse} \label{eqn:final_2app_eq2} \\
& ~~\leq~~ 2 \cdot \opt\eqref{eqn:model_warehouse} \label{eqn:final_2app_eq3} \ .
\end{align}
Here, inequality~\eqref{eqn:final_2app_eq1} holds since $T^*$ is an optimal solution to~\eqref{eqn:modify_relax_warehouse}. Inequality~\eqref{eqn:final_2app_eq2} is obtained by recalling that formulations~\eqref{eqn:relax_warehouse} and~\eqref{eqn:modify_relax_warehouse} are equivalent, as explained in Section~\ref{subsec:average_bound_relaxation}. Finally, inequality~\eqref{eqn:final_2app_eq3} follows by noting that~\eqref{eqn:relax_warehouse} is a relaxation of~\eqref{eqn:model_warehouse}, as shown in Corollary~\ref{cor:rel_vs_opt_ware}.

\subsection{Proof of Lemma~\ref{lem:sosi_optimal}} \label{subsec:proof_lem_sosi_optimal}

To establish the desired claim, let ${\cal P}^{\mycyc}$ be a feasible policy with respect to formulation~\eqref{eqn:relax_warehouse}, noting once again that this problem is restricted to considering cyclic policies. Our proof will show that ${\cal P}^{\mycyc}$ can be iteratively converted to a SOSI policy, feasible with respect to~\eqref{eqn:modify_relax_warehouse}, without increasing its long-run average cost. To this end, we examine how ${\cal P}^{\mycyc}$ operates along a single cycle, $[0, \tau_\mycyc)$. One can easily verify that, for any commodity $i \in [n]$, we may assume without loss of generality that ${\cal P}^{\mycyc}_i$ is a zero-inventory ordering (ZIO) policy, ending with zero inventory at time $\tau_\mycyc$. As such, let us separately consider each commodity $i \in [n]$.

Suppose that ${\cal P}^{\mycyc}_i$ places $m$ orders along $[0,\tau_\mycyc)$, whose durations are $\Delta_1, \ldots, \Delta_m$. Then, we will convert ${\cal P}^{\mycyc}_i$ into a SOSI policy $\hat{\cal P}_i$ for commodity $i$ such that, by defining
\[ \hat{\cal P} ~~=~~ ( \underbrace{{\cal P}^{\mycyc}_1, \ldots, {\cal P}^{\mycyc}_{i-1}}_{ \text{unchanged} }, \underbrace{ \vphantom{{\cal P}^{\mycyc}_{i-1}} \hat{\cal P}_i }_{ \text{new} }, \underbrace{ {\cal P}^{\mycyc}_{i+1}, \ldots, {\cal P}^{\mycyc}_n }_{ \text{unchanged} }) \ , \]
we obtain a feasible solution to~\eqref{eqn:relax_warehouse}, with $C(\hat{\cal P}) \leq C({\cal P}^{\mycyc})$. For this purpose, $\hat{\cal P}_i$ will simply be the policy where we place our $m$ orders at $0, \frac{ \tau_\mycyc }{ m }, \frac{ 2\tau_\mycyc }{ m }, \ldots$; namely, their durations are identical, with $\hat{\Delta} = \cdots = \hat{\Delta}_m = \frac{ \tau_\mycyc }{ m }$.

\paragraph{Feasibility.} We first show that $\sum_{j \in [n]} \gamma_j \cdot \bar{I}( \hat{\cal P}_j ) \leq {\cal V}$. To this end, it suffices to argue that $\bar{I}( \hat{\cal P}_i ) \leq \bar{I}( {\cal P}^{\mycyc}_i )$, which is indeed the case since
\begin{align}
\bar{I}( {\cal P}^{\mycyc}_i ) & ~~=~~ \frac{ 1 }{ 2\tau_\mycyc } \cdot \sum_{ \mu \in [m]} \Delta_{ \mu }^2 \nonumber \\
& ~~\geq~~ \frac{ 1 }{ 2\tau_\mycyc } \cdot \min \left\{ \sum_{ \mu \in [m]} \bar{\Delta}_{ \mu }^2 : \sum_{ \mu \in [m]} \bar{\Delta}_{ \mu } = \tau_\mycyc, \bar{\Delta} \in \bbR^m_+ \right\} \nonumber \\
& ~~=~~ \frac{ 1 }{ 2\tau_\mycyc } \cdot \frac{ \tau_\mycyc^2 }{m} \nonumber \\
& ~~=~~ \frac{ 1 }{ 2\tau_\mycyc } \cdot \sum_{ \mu \in [m]} \hat{\Delta}_{ \mu }^2 \nonumber \\
& ~~=~~ \bar{I}( \hat{\cal P}_i ) \ . \label{eqn:proof_sosi_optimal_1}
\end{align}

\paragraph{Objective value.} In terms of ordering costs, since $\hat{\cal P}_i$ and ${\cal P}^{\mycyc}_i$ place the same number of orders across $[0,\tau_\mycyc)$, this component remains unchanged, that is, ${\cal K}( \hat{\cal P}_i, [0,\tau_\mycyc)) = {\cal K}( {\cal P}^{\mycyc}_i, [0,\tau_\mycyc))$. In terms of holding costs,
\[ {\cal H}( \hat{\cal P}_i, [0,\tau_\mycyc)) ~~=~~ 2H_i \tau_\mycyc \cdot \bar{I}( \hat{\cal P}_i ) ~~\leq~~ 2H_i \tau_\mycyc \cdot \bar{I}( {\cal P}^{\mycyc}_i ) ~~=~~ {\cal H}( {\cal P}^{\mycyc}_i, [0,\tau_\mycyc)) \ , \]
where the inequality above holds since $\bar{I}( \hat{\cal P}_i ) \leq \bar{I}( {\cal P}^{\mycyc}_i )$, as shown in~\eqref{eqn:proof_sosi_optimal_1}. 
\section{Structural Decomposition and Algorithmic Roadmap} \label{sec:sub2-approx}

In what follows, we develop our structural framework for devising an $O( | {\cal I} |^{\tilde{O}( 1/\eps^5)} \cdot 2^{ \tilde{O}( 1 / \eps^{35} ) } )$-time construction of a random capacity-feasible policy whose expected long-run average cost is within factor $2-\frac{17}{5000}+ \eps$ of optimal, as formally stated in Theorem~\ref{thm:2_minus_delta}. For this purpose, Section~\ref{subsec:alg_outline_2minus} presents a high-level outline of these developments, by introducing a carefully chosen near-optimal cyclic benchmark and the notion of ``volume classes'', which partition commodities according to their warehouse capacity contributions. Section~\ref{subsec:high_level_easy} addresses a relatively manageable parametric regime, in which a large fraction of the overall capacity is carried by sparse volume classes, showing that this scenario can be handled by fusing the algorithmic ideas of Section~\ref{sec:anily_approx} and our recent approximation scheme for constantly-many commodities \citep{Segev26EWLSP_PTAS}. The remaining and genuinely challenging regime is isolated in Section~\ref{subsec:high_level_difficult}, where dense volume classes dominate capacity usage, capturing the principal obstruction to prior approaches. The analysis of this regime will heavily rely on a novel structural result, referred to as the Po2-Synchronization Theorem, whose specifics are discussed in Sections~\ref{sec:suffix_dense} and~\ref{sec:po2-sync}.

\subsection{High-level overview} \label{subsec:alg_outline_2minus}

Moving forward, rather than directly comparing ourselves to an optimal replenishment policy, which could be very convoluted, it will be convenient to have a cyclic policy as an intermediate benchmark. To this end, as explained in Section~\ref{subsec:average_bound_relaxation}, we know that for any $\eps > 0$, there exists a capacity-feasible cyclic policy ${\cal P}$ with a long-run average cost of $C( {\cal P} ) \leq (1 + \eps) \cdot \opt\eqref{eqn:model_warehouse}$. From this point on, ${\cal P}^{\eps}$ will stand for one such policy, and we proceed by introducing the basics of volume classes and their properties.

\paragraph{Volume classes.} By writing the average space bound~\eqref{eqn:average-space-bound} in terms of ${\cal P}^{\eps}$, we infer that its average occupied space must be upper-bounded by our overall capacity, i.e., $\sum_{i \in [n]} \gamma_i \cdot \bar{I}( {\cal P}_i^{\eps} ) \leq {\cal V}$. Motivated by this bound, for purposes of analysis, let us partition the underlying set of commodities into $O( \frac{ 1 }{ \eps } \log \frac{ n }{ \eps } )$ volume classes ${\cal V}_1^{\eps}, \ldots, {\cal V}_L^{\eps}, {\cal V}_{\infty}^{\eps}$ based on their average occupied space $\{ \gamma_i \cdot \bar{I}( {\cal P}^{\eps}_i ) \}_{i \in [n]}$. This partition is defined as follows:
\begin{itemize}
    \item The class ${\cal V}_1^{\eps}$ consists of commodities with $\gamma_i \cdot \bar{I}( {\cal P}^{\eps}_i ) \in ( \frac{1}{1+\eps} \cdot {\cal V}, {\cal V}]$.

    \item The class ${\cal V}_2^{\eps}$ consists of those with $\gamma_i \cdot \bar{I}( {\cal P}^{\eps}_i ) \in ( \frac{1}{(1+\eps)^2} \cdot {\cal V}, \frac{1}{1+\eps} \cdot{\cal V}]$.

    \item So on and so forth, up to ${\cal V}_L^{\eps}$, corresponding to commodities with $\gamma_i \cdot \bar{I}( {\cal P}^{\eps}_i ) \in ( \frac{1}{(1+\eps)^L} \cdot {\cal V}, \frac{1}{(1+\eps)^{L-1}} \cdot{\cal V}]$. Here, we set $L = \lceil \log_{1+\eps} (\frac{ n }{ \eps }) \rceil$, meaning that $\frac{ 1 }{ (1+\eps)^L } \leq \frac{ \eps }{ n }$.

    \item In addition, the remaining class ${\cal V}_{\infty}^{\eps}$ is composed of commodities with $\gamma_i \cdot \bar{I}( {\cal P}^{\eps}_i ) \leq \frac{1}{(1+\eps)^L} \cdot {\cal V}$.
\end{itemize}

\paragraph{Sparsity and density.}
For every $\ell \in [L]_{\infty}$, we say that class ${\cal V}_{\ell}^{\eps}$ is sparse when it consists of relatively few commodities, concretely meaning that $| {\cal V}_{\ell}^{\eps} | \leq \frac{ 100 \ln (1/\eps) }{ \eps^4 }$; otherwise, this class is called dense. We make use of ${\cal S}$ and ${\cal D}$ to denote the index sets of sparse and dense classes. Now, suppose that ${\cal S} = \{ \ell_1, \ldots, \ell_M \}$, with the convention that indices are listed in increasing order. We proceed by decomposing this set into
\[ \underbrace{ \ell_1 \quad , \quad \ldots \quad , \quad \ell_{\mymid} }_{ {\cal S}_{\mypref} } \quad , \quad \underbrace{ \ell_{\mymid+1} \quad \ldots \quad , \quad \ell_M }_{ {\cal S}_{\mysuff} } \ . \]
Here, $\ell_{\mymid}$ is the unique index up to which $\Delta = \lceil \log_{1 + \eps} ( \frac{ 125 }{ \eps^7 } ) \rceil$ of the sparse classes ${\cal V}^{\eps}_{\ell_1}, \ldots, {\cal V}^{\eps}_{\ell_{\mymid}}$ are not empty; when there are fewer than $\Delta$ such classes, $\ell_{\mymid} = \ell_M$.

\paragraph{Guessing procedure.} As our first algorithmic step, we guess the next few parameters, related to the unknown $\eps$-optimal policy ${\cal P}^{\eps}$:
\begin{itemize}
    \item {\em Class types}: For every $\ell \in [L]_{\infty}$, we guess whether class ${\cal V}_{\ell}^{\eps}$ is prefix-sparse, suffix-sparse, or dense by enumerating over all $3^{L+1} = O(n^{ \tilde{O}(1 / \eps) })$ options. From this point on, the sets ${\cal S}_{\mypref}$, ${\cal S}_{\mysuff}$, and ${\cal D}$ will be assumed to be known.

    \item {\em Commodities within prefix-sparse classes}: For every $\ell \in {\cal S}_{\mypref}$, we guess the precise identity of ${\cal V}_{\ell}^{\eps}$. Noting that ${\cal S}_{\mypref}$ consists of at most $\Delta = O( \frac{ 1 }{ \eps } \log \frac{ 1 }{ \eps } )$ non-empty classes, and that any such class contains at most $\frac{ 100 \ln (1/\eps) }{ \eps^4 }$ commodities, the total number of guesses across all prefix-sparse classes is $O(n^{ \tilde{O}( 1/\eps^5 ) })$.

    \item {\em Size of suffix-sparse classes}: For every $\ell \in {\cal S}_{\mysuff}$, we guess the cardinality $| {\cal V}_{\ell}^{\eps} |$ of this class. Here, there are at most $(\frac{ 100 \ln (1/\eps) }{ \eps^4 }+1)^{|{\cal S}_{\mysuff}|} = O( n^{ \tilde{O}( 1/ \eps ) } )$ options to be considered.
\end{itemize}

\paragraph{The easy scenario: High volume of sparse classes.} For simplicity of notation, let $\bar{V}^{\eps}_{\cal S} = \sum_{\ell \in {\cal S}} \sum_{i \in {\cal V}_{\ell}^{\eps}} \gamma_i \cdot \bar{I}( {\cal P}_i^{\eps} )$ and $\bar{V}^{\eps}_{\cal D} = \sum_{\ell \in {\cal D}} \sum_{i \in {\cal V}_{\ell}^{\eps}} \gamma_i \cdot \bar{I}( {\cal P}_i^{\eps} )$ be the total average space occupied by the commodities in sparse and dense classes, both with respect to the $\eps$-optimal policy ${\cal P}^{\eps}$. In Section~\ref{subsec:high_level_easy}, we consider the retrospectively simpler scenario, where the average space $\bar{V}^{\eps}_{\cal S}$ of sparse commodities forms a sufficiently large fraction of the capacity bound ${\cal V}$, specifically meaning that $\bar{V}^{\eps}_{\cal S} \geq (\frac{1}{2} + \delta) \cdot {\cal V}$; here, $\delta \in (0, \frac{ 1 }{ 2 })$ is an absolute constant whose value will be determined later on. As formally stated below, by judiciously combining the algorithmic ideas of Section~\ref{sec:anily_approx} and our approximation scheme for constantly-many commodities \citep{Segev26EWLSP_PTAS}, this scenario will allow us to approach the optimal long-run average cost within a factor of essentially $2(1-\delta)$.

\begin{lemma} \label{lem:large_V_sparse}
Suppose that $\bar{V}^{\eps}_{\cal S} \geq (\frac{1}{2} + \delta) \cdot {\cal V}$. Then, we can compute in $O( | {\cal I} |^{\tilde{O}( 1/\eps^5)} \cdot 2^{ \tilde{O}( 1 / \eps^{35} ) } )$ time a deterministic capacity-feasible policy ${\cal P}$ with
$C( {\cal P} ) \leq (2 - 2\delta + 8\eps) \cdot C( {\cal P}^{\eps} )$.
\end{lemma}

\paragraph{The difficult scenario: Low volume of sparse classes.} In Section~\ref{subsec:high_level_difficult}, we will operate in the complementary scenario, where $\bar{V}^{\eps}_{\cal S} < (\frac{1}{2} + \delta) \cdot {\cal V}$, which captures the essence of why sub-$2$-approximations do not seem plausible via existing methods. Since our approach in this context involves quite a few moving parts, its specifics will be discussed in Sections~\ref{sec:suffix_dense} and~\ref{sec:po2-sync}. Technically speaking, up to $\eps$-dependent terms, these ideas will culminate in a polynomial-time algorithm for approximating the optimal long-run average cost within factor
$\max \{ 2(1-\delta), 1.9932 + 2\delta \}$.

\begin{lemma} \label{lem:large_V_dense}
Suppose that $\bar{V}^{\eps}_{\cal S} < (\frac{1}{2} + \delta) \cdot {\cal V}$. Then, we can compute in $O( n^{ \tilde{O}( 1 / \eps^2 ) } )$ time a random capacity-feasible policy ${\cal P}$ with an expected cost of
\[ \ex{ C({\cal P}) } ~~\leq~~ \max \{ 2 - 2\delta + 2\eps, 1.9932 + 2\delta + 3\eps \} \cdot C( {\cal P}^{\eps} ) \ . \]
\end{lemma}

\paragraph{Resulting approximation guarantee.} Clearly, we have no control over whether any given instance falls within the easy scenario or within the difficult one. Thus, for the purpose of minimizing the worst-case performance guarantee over these two scenarios, we set our yet-unspecified threshold parameter at $\delta = \frac{17}{10000}$. As such, the approximation ratios stated in Lemmas~\ref{lem:large_V_sparse} and~\ref{lem:large_V_dense} respectively become $2 - \frac{17}{5000} + 8\eps$ and $2 - \frac{17}{5000} + 3\eps$, thereby concluding the proof of Theorem~\ref{thm:2_minus_delta}.

\subsection{The easy scenario: \texorpdfstring{\boldmath{${\bar{V}^{\eps}_{\cal S} \geq (\frac{1}{2} + \delta) \cdot {\cal V}}$}}{}}\label{subsec:high_level_easy}

\paragraph{Policy for prefix-sparse classes.} Following the guessing procedure of Section~\ref{subsec:alg_outline_2minus}, we already know the index set ${\cal S}_{\mypref}$ of prefix-sparse classes, as well as the set of commodities that belong to each such class. We begin by recalling that, while ${\cal S}_{\mypref}$ may be comprised of $\Omega( \frac{ 1 }{ \eps } \log \frac{ n }{ \eps } )$ classes, at most $\Delta = O( \frac{ 1 }{ \eps} \log \frac{ 1 }{ \eps })$ of these classes are non-empty. Since each such class contains up to $\frac{ 100 \ln (1/\eps) }{ \eps^4 }$ commodities, we have $\sum_{\ell \in {\cal S}_{\mypref}} | {\cal V}^{\eps}_{ \ell } | = O( \frac{ 1 }{ \eps^5 } \log^2 (\frac{ 1 }{ \eps }) ) $. As mentioned in Section~\ref{subsec:related_work}, with respect to $N$ underlying commodities, our recent approximation scheme can be implemented in $O( | {\cal I} |^{O(N)} \cdot 2^{ O( N^{6} / \eps^{5} ) } )$ time~\citep[Thm.~1]{Segev26EWLSP_PTAS}. Therefore, due to currently considering $N = \tilde{O}( \frac{ 1 }{ \eps^5 } )$ commodities, we can compute a $(1+\eps)$-approximate capacity-feasible policy ${\cal P}_{\mypref}$ for these commodities in $O( | {\cal I} |^{\tilde{O}( 1/\eps^5)} \cdot 2^{ \tilde{O}( 1 / \eps^{35} ) } )$ time. In terms of cost, since restricting the policy ${\cal P}^{\eps}$ to these commodities forms a feasible solution in this context, we have
\begin{equation} \label{eqn:easy_sparse_both}
C( {\cal P}_{\mypref} ) ~~\leq~~ (1+\eps) \cdot \sum_{\ell \in {\cal S}_{\mypref}} \sum_{i \in {\cal V}_{\ell}^{\eps}} C_i( {\cal P}_i^{\eps} ) \ .
\end{equation}

\paragraph{Policy for suffix+dense classes.} Moving forward, it is important to emphasize that our guessing procedure does not inform us about how the set of commodities outside of prefix-sparse classes are partitioned between suffix-sparse classes and dense classes. To circumvent this obstacle, we first argue that the total average space occupied by the commodities in ${\cal S}_{\mysuff}$ is rather negligible. The proof of this result is given in Appendix~\ref{app:proof_lem_avg_space_prefix}.

\begin{lemma} \label{lem:avg_space_prefix}
$\sum_{\ell \in {\cal S}_{\mysuff}} \sum_{i \in {\cal V}_{\ell}^{\eps}} \gamma_i \cdot \bar{I}( {\cal P}_i^{\eps} ) \leq \eps {\cal V}$.
\end{lemma}

Motivated by this observation, for the collection of commodities within the union of suffix-sparse classes and dense classes, we design a joint policy by adapting the algorithmic approach of Section~\ref{sec:anily_approx} as follows:
\begin{itemize}
    \item Recalling that $\bar{V}^{\eps}_{\cal D} = \sum_{\ell \in {\cal D}} \sum_{i \in {\cal V}_{\ell}^{\eps}} \gamma_i \cdot \bar{I}( {\cal P}_i^{\eps} )$, we begin by guessing an overestimate $\tilde{V}_{\cal D}$ for this quantity, such that $\tilde{V}_{\cal D} \in [\bar{V}^{\eps}_{\cal D}, \bar{V}^{\eps}_{\cal D} + \eps {\cal V}]$. For this purpose, it suffices to enumerate over $\eps {\cal V}, 2 \eps {\cal V}, 3 \eps {\cal V}, \ldots$, meaning that there are only $O( \frac{ 1 }{ \eps } )$ values to be tested.

    \item In light of Lemma~\ref{lem:avg_space_prefix}, it follows that by restricting ${\cal P}^{\eps}$ to commodities in suffix-sparse classes and dense classes, we obtain a feasible solution to the next formulation:
    \begin{align} \label{eqn:relax_warehouse_dense}
    \tag{$\tilde{\Pi}^{\mysuff+\mydense}$}
    \begin{array}{lll}
    {\displaystyle \min_{\cal P}} & {\displaystyle \sum_{\ell \in {\cal S}_{\mysuff} \cup {\cal D}} \sum_{i \in {\cal V}_{\ell}^{\eps}} C( {\cal P}_i ) } \\
    \text{s.t.} & {\displaystyle \sum_{\ell \in {\cal S}_{\mysuff} \cup {\cal D}} \sum_{i \in {\cal V}_{\ell}^{\eps}} \gamma_i \cdot \bar{I}( {\cal P}_i ) \leq \tilde{V}_{\cal D}} + \eps {\cal V}
    \end{array}
    \end{align}
    Consequently, $\opt\eqref{eqn:relax_warehouse_dense} \leq \sum_{\ell \in {\cal S}_{\mysuff} \cup {\cal D}} \sum_{i \in {\cal V}_{\ell}^{\eps}} C( {\cal P}^{\eps}_i )$.

    \item By duplicating the proof of Lemma~\ref{lem:sosi_optimal} to the letter, one can verify that \eqref{eqn:relax_warehouse_dense} admits a SOSI optimal replenishment policy, i.e., $\opt\eqref{eqn:relax_warehouse_dense} = \opt\eqref{eqn:modify_relax_warehouse_dense}$, where the latter formulation is given by
    \begin{align} \label{eqn:modify_relax_warehouse_dense} \tag{$\tilde{\Pi}^{\mysuff+\mydense}_{\text{SOSI}}$}
    \begin{array}{lll}
    {\displaystyle \min_T} & {\displaystyle \sum_{\ell \in {\cal S}_{\mysuff} \cup {\cal D}} \sum_{i \in {\cal V}_{\ell}^{\eps}} C_{\myeoq,i}(T_i) } \\
    \text{s.t.} & {\displaystyle \sum_{\ell \in {\cal S}_{\mysuff} \cup {\cal D}} \sum_{i \in {\cal V}_{\ell}^{\eps}} \gamma_i T_i \leq 2 \cdot (\tilde{V}_{\cal D}} + \eps {\cal V})
    \end{array}
    \end{align}

    \item Finally, by computing an optimal solution to this relaxation (see Section~\ref{subsec:average_bound_relaxation}), we obtain a policy ${\cal P}_{\mysuff+\mydense}$ whose peak space requirement is
    \begin{equation} \label{eqn:easy_sparse_space}
    V_{\max}( {\cal P}_{\mysuff+\mydense} ) ~~\leq~~ 2 \cdot (\tilde{V}_{\cal D} + \eps {\cal V}) ~~\leq~~ 2 \bar{V}^{\eps}_{\cal D} + 4\eps {\cal V} \ .
    \end{equation}
    At the same time, the long-run average cost of this policy is
    \begin{align}
    C( {\cal P}_{\mysuff+\mydense} ) & ~~=~~ \opt\eqref{eqn:modify_relax_warehouse_dense} \nonumber \\
    & ~~=~~ \opt\eqref{eqn:relax_warehouse_dense} \nonumber \\
    & ~~\leq~~ \sum_{\ell \in {\cal S}_{\mysuff} \cup {\cal D}} \sum_{i \in {\cal V}_{\ell}^{\eps}} C( {\cal P}^{\eps}_i ) \ . \label{eqn:easy_sparse_cost}
    \end{align}
\end{itemize}

\paragraph{The combined policy.} To obtain a single policy ${\cal P}_{\mycomb}$ for all commodities, we first glue ${\cal P}_{\mypref}$ and ${\cal P}_{\mysuff+\mydense}$ together. Since ${\cal P}_{\mypref}$ is capacity-feasible by itself, in conjunction with inequality~\eqref{eqn:easy_sparse_space}, we infer that the peak space requirement of ${\cal P}_{\mycomb}$ is
\[V_{\max}( {\cal P}_{\mycomb} ) ~~\leq~~ {\cal V} + 2 \bar{V}^{\eps}_{\cal D} + 4\eps {\cal V} ~~\leq~~ (2 - 2\delta + 4\eps) \cdot {\cal V} \ , \]
where the last inequality holds since
$\bar{V}^{\eps}_{\cal D} \leq {\cal V} - \bar{V}^{\eps}_{\cal S} \leq (\frac{1}{2} - \delta) \cdot {\cal V}$, due to operating in the scenario where $\bar{V}^{\eps}_{\cal S} \geq (\frac{1}{2} + \delta) \cdot {\cal V}$. In terms of cost, by inequalities~\eqref{eqn:easy_sparse_both} and~\eqref{eqn:easy_sparse_cost}, we have $C( {\cal P}_{\mycomb} ) \leq (1+\eps) \cdot C( {\cal P}^{\eps} )$. Therefore, scaling down ${\cal P}_{\mycomb}$ by a factor of $2 - 2\delta + 4\eps$, we obtain a capacity-feasible policy $\hat{\cal P}_{\mycomb}$ with
\begin{align*}
C( \hat{\cal P}_{\mycomb} ) & ~~\leq~~ (2 - 2\delta + 4\eps) \cdot C( {\cal P}_{\mycomb} ) \\
& ~~\leq~~ (2 - 2\delta + 4\eps) \cdot (1+\eps) \cdot C( {\cal P}^{\eps} )\\
& ~~\leq~~ (2 - 2\delta + 8\eps) \cdot C( {\cal P}^{\eps} ) \ .
\end{align*}

\subsection{The difficult scenario: \texorpdfstring{\boldmath{${\bar{V}^{\eps}_{\cal S} < (\frac{1}{2} + \delta) \cdot {\cal V}}$}}{}} \label{subsec:high_level_difficult}

\paragraph{Lower bound on \boldmath{${\bar{V}^{\eps}_{\cal D}}$}.} Let us first put aside an easily-addressable case, where $\bar{V}^{\eps}_{\cal D} < (\frac{1}{2} - 2\delta) \cdot {\cal V}$, implying that $\bar{V}^{\eps}_{\cal S} + \bar{V}^{\eps}_{\cal D} \leq (\frac{1}{2} + \delta) \cdot {\cal V} + (\frac{1}{2} - 2\delta) \cdot {\cal V} = (1 - \delta) \cdot {\cal V}$. In this case, similarly to how the policy ${\cal P}_{\mysuff+\mydense}$ was constructed in Section~\ref{subsec:high_level_easy}, we can apply precisely the same arguments for the entire collection of commodities, ending up with a policy ${\cal P}_{\mycomb}$ whose peak space requirement is
\[ V_{\max}( {\cal P}_{\mycomb} ) ~~\leq~~ 2 \cdot (\bar{V}^{\eps}_{\cal S} + \bar{V}^{\eps}_{\cal D}) + 2\eps {\cal V} ~~\leq~~ (2 - 2\delta + 2\eps) \cdot {\cal V} \ , \]
and whose long-run average cost is $C( {\cal P}_{\mycomb} ) \leq C( {\cal P}^{\eps} )$. Subsequently, scaling down ${\cal P}_{\mycomb}$ by a factor of $2 - 2\delta + 2\eps$, we obtain a capacity-feasible policy $\hat{\cal P}_{\mycomb}$ with $C( \hat{\cal P}_{\mycomb} ) \leq (2 - 2\delta + 2\eps) \cdot C( {\cal P}^{\eps} )$. The latter expression is precisely the first term mentioned in Lemma~\ref{lem:large_V_dense}. In the remainder of this section, we consider the regime where $\bar{V}^{\eps}_{\cal D} \geq (\frac{1}{2} - 2\delta) \cdot {\cal V}$.

\paragraph{Policy for prefix-sparse classes.} The treatment of these commodities will be identical to our actions in the suffix+dense case of Section~\ref{subsec:high_level_easy}. By repeating this construction, with ``suffix+dense'' and ${\cal S}_{\mysuff} \cup {\cal D}$ replaced by ``prefix-sparse'' and ${\cal S}_{\mypref}$, we can compute in polynomial time a policy ${\cal P}_{\mypref}$ whose peak space requirement and long-run average cost are
\begin{equation} \label{eqn:diff_sparse_space_cost}
V_{\max}( {\cal P}_{\mypref} ) ~~\leq~~ 2 \bar{V}^{\eps}_{\cal S} + 2\eps {\cal V} \qquad \text{and} \qquad C( {\cal P}_{\mypref} ) ~~\leq~~ \sum_{\ell \in {\cal S}_{\mypref}} \sum_{i \in {\cal V}_{\ell}^{\eps}} C( {\cal P}^{\eps}_i ) \ .
\end{equation}

\paragraph{Policy for suffix+dense classes.} We have just reached the most challenging part of our analysis. Since handling these classes is particularly involved, we dedicate Sections~\ref{sec:suffix_dense} and~\ref{sec:po2-sync} to describe our construction, leading to the next result.

\begin{lemma} \label{lem:diff_case_dense_main}
We can compute in $O( n^{ \tilde{O}( 1 / \eps^2 ) } )$ time a random policy ${\cal P}_{\mysuff+\mydense}$ satisfying the next two properties:
\begin{enumerate}
    \item {\em Space requirement:} $V_{\max}( {\cal P}_{\mysuff+\mydense} ) \leq (1 + 8\eps) \cdot \frac{ 7/4 }{ \sqrt{2} \ln 2} \cdot \bar{V}^{\eps}_{\cal D} + 10\eps {\cal V}$ almost surely.

    \item {\em Expected cost}: $\expar{ C( {\cal P}_{\mysuff+\mydense} ) } \leq ( 1 + \frac{2\eps}{5} ) \cdot \frac{ 32/31 }{ \sqrt{2} \ln 2} \cdot \sum_{\ell \in {\cal S}_{\mysuff} \cup {\cal D}} \sum_{i \in {\cal V}_{\ell}^{\eps}} C_i( {\cal P}_i^{\eps} )$.
\end{enumerate}
\end{lemma}

\paragraph{The combined policy.} To obtain a single policy ${\cal P}_{\mycomb}$ for all commodities, we first glue ${\cal P}_{\mypref}$ and ${\cal P}_{\mysuff+\mydense}$ together. By inequality~\eqref{eqn:diff_sparse_space_cost} and Lemma~\ref{lem:diff_case_dense_main}(2), we infer that the expected long-run average cost of this policy is
\begin{align*}
\ex{ C( {\cal P}_{\mycomb} ) } & ~~=~~ C( {\cal P}_{\mypref} ) + \ex{ C( {\cal P}_{\mysuff+\mydense} ) } \\
& ~~\leq~~ \sum_{\ell \in {\cal S}_{\mypref}} \sum_{i \in {\cal V}_{\ell}^{\eps}} C( {\cal P}^{\eps}_i ) + \left( 1 + \frac{2\eps}{5} \right) \cdot \frac{ 32/31 }{ \sqrt{2} \ln 2} \cdot \sum_{\ell \in {\cal S}_{\mysuff} \cup {\cal D}} \sum_{i \in {\cal V}_{\ell}^{\eps}} C_i( {\cal P}_i^{\eps} ) \\
& ~~\leq~~ \left( 1 + \frac{2\eps}{5} \right) \cdot 1.0531 \cdot C( {\cal P}^{\eps} ) \ .
\end{align*}
At the same time, the peak space requirement of ${\cal P}_{\mycomb}$ is almost surely
\begin{align}
V_{\max}( {\cal P}_{\mycomb} ) & ~~\leq~~ V_{\max}( {\cal P}_{\mypref} ) + V_{\max}( {\cal P}_{\mysuff+\mydense} ) \nonumber \\
& ~~\leq~~ 2 \bar{V}^{\eps}_{\cal S} + 2\eps {\cal V} + (1+8\eps) \cdot \frac{ 7/4 }{ \sqrt{2} \ln 2} \cdot \bar{V}^{\eps}_{\cal D} + 10\eps {\cal V}\label{eqn:diff_combined_vmax_1} \\
& ~~\leq~~(1+8\eps) \cdot \left( 2 \cdot \left( \bar{V}^{\eps}_{\cal S} + \bar{V}^{\eps}_{\cal D} \right) - \left( 2 - \frac{ 7/4 }{ \sqrt{2} \ln 2} \right) \cdot \bar{V}^{\eps}_{\cal D} \right) + 12\eps {\cal V} \nonumber \\
& ~~\leq~~ (1+8\eps) \cdot \left( 2 {\cal V} - \left( 2 - \frac{ 7/4 }{ \sqrt{2} \ln 2} \right) \cdot \bar{V}^{\eps}_{\cal D} \right) + 12\eps {\cal V}\label{eqn:diff_combined_vmax_2} \\
& ~~\leq~~ (1+8\eps) \cdot \left( 1 + \frac{ 7/8 }{ \sqrt{2} \ln 2} + \frac{ \delta }{ 2 } + 12\eps \right) \cdot {\cal V} \ . \label{eqn:diff_combined_vmax_3}
\end{align}
Here, inequality~\eqref{eqn:diff_combined_vmax_1} follows from inequality~\eqref{eqn:diff_sparse_space_cost} and Lemma~\ref{lem:diff_case_dense_main}(1). Inequality~\eqref{eqn:diff_combined_vmax_2} is obtained by recalling that the $\eps$-optimal policy ${\cal P}^{\eps}$ is in particular capacity-feasible, implying that $\bar{V}^{\eps}_{\cal S} + \bar{V}^{\eps}_{\cal D} \leq {\cal V}$ due to the average-space bound (see Section~\ref{subsec:average_bound_relaxation}). Finally, inequality~\eqref{eqn:diff_combined_vmax_3} holds since $\bar{V}^{\eps}_{\cal D} \geq (\frac{1}{2} - 2\delta) \cdot {\cal V}$.

Based on these observations, when scaling down ${\cal P}_{\mycomb}$ by a factor of $(1+8\eps) \cdot ( 1 + \frac{ 7/8 }{ \sqrt{2} \ln 2} + \frac{ \delta }{ 2 } + 12\eps )$, we obtain a capacity-feasible policy $\hat{\cal P}_{\mycomb}$ with an expected long-run average cost of
\begin{align*}
\ex{ C( \hat{\cal P}_{\mycomb} ) } & ~~\leq~~ (1+8\eps) \cdot \left( 1 + \frac{ 7/8 }{ \sqrt{2} \ln 2} + \frac{ \delta }{ 2 } + 12\eps \right) \cdot \ex{ C( {\cal P}_{\mycomb} ) } \\
& ~~\leq~~ (1+9\eps) \cdot \left( 1 + \frac{ 7/8 }{ \sqrt{2} \ln 2} + \frac{ \delta }{ 2 } + 12\eps \right) \cdot \left( 1 + \frac{2\eps}{5} \right) \cdot 1.0531 \cdot C( {\cal P}^{\eps} )\\
& ~~\leq~~ (1.9932 + 2\delta + 3\eps) \cdot C( {\cal P}^{\eps} ) \ .
\end{align*} 
\section{Designing the Suffix+Dense Policy} \label{sec:suffix_dense}

In this section, we present the main technical ideas behind Lemma~\ref{lem:diff_case_dense_main}, showing that in $O( n^{ \tilde{O}( 1 / \eps^2 ) } )$  time, we can compute a random policy ${\cal P}_{\mysuff+\mydense}$ satisfying
\begin{align*}
& V_{\max}( {\cal P}_{\mysuff+\mydense} ) ~~\leq~~ (1 + 8\eps) \cdot \frac{ 7/4 }{ \sqrt{2} \ln 2} \cdot \bar{V}^{\eps}_{\cal D} + 10\eps  {\cal V} \\
& \qquad \qquad \qquad \qquad \text{and} \qquad \ex{ C( {\cal P}_{\mysuff+\mydense} ) } ~~\leq~~ \left( 1 + \frac{2\eps}{5} \right) \cdot \frac{ 32/31 }{ \sqrt{2} \ln 2} \cdot \sum_{\ell \in {\cal S}_{\mysuff} \cup {\cal D}} \sum_{i \in {\cal V}_{\ell}^{\eps}} C_i( {\cal P}_i^{\eps} ) \ ,
\end{align*}
with the former property holding almost surely. For this purpose, Section~\ref{subsec:useful_partition_dense} formalizes the notion of ``mimicking partitions'' that lies at the heart of our construction, and explains how such partitions can be efficiently identified by means of minimum-weight $b$-matchings. Section~\ref{subsec:V_ell_rounding} utilizes this partition to define a randomized replenishment policy, motivating us to introduce the Po2-Synchronization Theorem. Section~\ref{subsec:dense_class_analysis} concludes by deriving explicit bounds on the expected long-run average cost and peak space requirement of this policy.

\subsection{Computing a mimicking partition} \label{subsec:useful_partition_dense}

With this goal in mind, let us recall that following the guessing procedure described in Section~\ref{subsec:alg_outline_2minus}, the index sets ${\cal S}_{\mysuff}$ and ${\cal D}$ of suffix-sparse and dense classes, as well as the commodities $U = \bigcup_{\ell \in {\cal S}_{\mysuff} \cup {\cal D}} {\cal V}_{\ell}^{\eps}$ belonging to these classes, are completely known. However, what remains unspecified is how $U$ is distributed across  individual classes. In what follows, we devise a matching-based method for partitioning these commodities into $\{ \tilde{\cal V}_{ \ell } \}_{\ell \in {\cal S}_{\mysuff} \cup {\cal D}}$, in a way that preserves several high-level structural properties exhibited by the unknown classes $\{ {\cal V}_{ \ell }^{\eps} \}_{\ell \in {\cal S}_{\mysuff} \cup {\cal D}}$, yielding what we refer to as mimicking partitions.

\paragraph{Additional guessing step.} To this end, for every class index $\ell \in {\cal D} \setminus \{ \infty \}$, let $\bar{V}^{\eps}_{\ell} = \sum_{i \in {\cal V}_{\ell}^{\eps}} \gamma_i \cdot \bar{I}( {\cal P}_i^{\eps} )$ be the total average space occupied by ${\cal V}_{\ell}^{\eps}$-commodities with respect to the policy ${\cal P}^{\eps}$. As our final guessing step, we obtain an over-estimate $\tilde{V}^{\eps}_{\ell}$ for this quantity, satisfying $\bar{V}^{\eps}_{\ell} \leq \tilde{V}^{\eps}_{\ell} \leq \bar{V}^{\eps}_{\ell} + \frac{ \eps }{ |{\cal D}| } \cdot {\cal V}$. By observing that $\sum_{\ell \in {\cal D}} \tilde{V}^{\eps}_{\ell} \leq  \bar{V}^{\eps}_{\cal D} + {\cal V} \leq 2{\cal V}$, elementary balls-and-bins counting arguments show that there are only $2^{ O( |{\cal D}| / \eps ) } = O(n^{ \tilde{O}( 1 / \eps^2 ) })$ options to be jointly considered for $\{ \tilde{V}^{\eps}_{\ell} \}_{\ell \in {\cal D}}$. Furthermore, letting $\tilde{\cal N}_{\ell} = \frac{ (1 + \eps)^{\ell} }{ {\cal V} } \cdot \tilde{V}^{\eps}_{\ell}$, we argue that this term provides an upper bound on the cardinality of ${\cal V}_{ \ell }^{\eps}$. Indeed,
\[ \tilde{V}^{\eps}_{\ell} ~~\geq~~  \bar{V}^{\eps}_{\ell} ~~=~~ \sum_{i \in {\cal V}_{\ell}^{\eps}} \gamma_i \cdot \bar{I}( {\cal P}_i^{\eps} )  ~~\geq~~ | {\cal V}_{\ell}^{\eps} | \cdot \frac{\cal V}{(1+\eps)^{\ell}} \ , \]
where the last inequality holds since $\gamma_i \cdot \bar{I}( {\cal P}^{\eps}_i ) \in ( \frac{1}{(1+\eps)^{\ell}} \cdot {\cal V}, \frac{1}{(1+\eps)^{\ell-1}} \cdot{\cal V}]$ for every commodity $i \in {\cal V}_{\ell}^{\eps}$. By rearranging the above inequality, we immediately get $\tilde{\cal N}_{\ell} \geq | {\cal V}_{ \ell }^{\eps} |$.

\paragraph{The mimicking partition.} We are now ready to partition the set of commodities $U = \bigcup_{\ell \in {\cal S}_{\mysuff} \cup {\cal D}} {\cal V}_{\ell}^{\eps}$ into subsets $\{ \tilde{\cal V}_{ \ell } \}_{\ell \in {\cal S}_{\mysuff} \cup {\cal D}}$ and  to identify SOSI replenishment policies $\{ \hat{T}_i \}_{i \in U}$ for these commodities, jointly mimicking the behavior of $\{ {\cal V}_{ \ell }^{ \eps } \}_{\ell \in {\cal S}_{\mysuff} \cup {\cal D}}$ and their respective policies $\{ {\cal P}^{\eps}_i \}_{i \in U}$, in the sense of satisfying the next three properties:
\begin{enumerate}
    \item \label{prop:tildeV_size} {\em Class size}: $|\tilde{\cal V}_{ \ell }| = |{\cal V}_{\ell}^{\eps} |$, for every $\ell \in {\cal S}_{\mysuff}$, and $|\tilde{\cal V}_{ \ell }| \in [\frac{ 100 \ln (1/\eps) }{ \eps^4 }, \tilde{\cal N}_{\ell}]$ for every $\ell \in {\cal D}$.

    \item \label{prop:hatT_cost} {\em Total cost}: $\sum_{i \in U}  C_{\myeoq,i}( \hat{T}_i ) \leq  \sum_{i \in U} C_i( {\cal P}^{\eps}_i )$.

    \item \label{prop:hatT_space} {\em Average space within class}: For every class $\ell \in {\cal S}_{\mysuff} \cup {\cal D}$ and commodity $i \in \tilde{\cal V}_{\ell}$,
    \begin{equation} \label{eqn:occupied_space_constraint}
    \gamma_i \cdot \bar{I}( \hat{T}_i ) ~~\leq~~ \begin{cases}
    \frac{1}{(1+\eps)^{\ell-1}} \cdot{\cal V}, \qquad & \text{when } \ell \in [L] \\
    \frac{ \eps }{ n } \cdot {\cal V}, & \text{when } \ell = \infty
    \end{cases}
    \end{equation}
\end{enumerate}

\paragraph{Matching-based formulation.} Toward this objective, we define an instance ${\cal I}$ of the minimum-weight $b$-matching problem as follows:
\begin{itemize}
    \item {\em Graph}: The underlying graph $G$ is bipartite and complete, with the set of commodities in $U$ on one side, and with the index set ${\cal S}_{\mysuff} \cup {\cal D}$ on the other.

    \item {\em Edge weights}: The weight $w_{i \ell}$ of each edge $(i,\ell)$ is set as the minimum $C_{\myeoq,i}$-cost of a SOSI policy $\hat{T}_{i \ell}$ for commodity $i$ satisfying constraint~\eqref{eqn:occupied_space_constraint}. One can easily notice that $w_{i \ell}$ admits a closed-form expression. Indeed, since $\bar{I}( T_{i \ell} ) = \frac{ T_{i \ell} }{ 2 }$, we are minimizing $C_{\myeoq,i}( T_{i \ell} ) = \frac{ K_i }{ T_{i \ell} } + H_i T_{i \ell}$ subject to an upper bound on $T_{i \ell}$. As such, by consulting Claim~\ref{clm:EOQ_properties}, it is not difficult to verify that
    \begin{equation} \label{eqn:closed_form_Ti}
    \hat{T}_{i \ell} ~~=~~ \begin{cases}
    \min \{ \sqrt{K_i / H_i}, \frac{ 2{\cal V} }{(1+\eps)^{\ell-1} \cdot \gamma_i} \}, \qquad & \text{when } \ell \in [L] \\
    \min \{ \sqrt{K_i / H_i}, \frac{ 2\eps {\cal V} }{n \gamma_i} \}, & \text{when } \ell = \infty
    \end{cases}
    \end{equation}

    \item {\em Degree constraints}: Each commodity-vertex $i \in U$ should have a degree of exactly $1$. On the opposing side, the degree of each class-vertex $\ell \in {\cal D}$ should reside  within $[\frac{ 100 \ln (1/\eps) }{ \eps^4 }, \tilde{\cal N}_{\ell}]$. Finally, each class-vertex  $\ell \in {\cal S}_{\mysuff}$ should have a degree of exactly $|{\cal V}_{ \ell }^{\eps}|$, noting that the latter quantity is known, given the guessing procedure in Section~\ref{subsec:alg_outline_2minus}.
\end{itemize}

With respect to this  formulation, we compute a minimum-weight set of edges ${\cal E}^* \subseteq E(G)$ satisfying the above-mentioned degree constraints. Any polynomial-time bipartite $b$-matching algorithm (see, e.g., \citet[Chap.~21]{Schrijver03} or \citet[Chap.~12]{KorteV18}) works for our purposes, noting that this particular instance is clearly feasible. Indeed, one possible solution can be obtained by connecting each commodity-vertex $i$ to the unique class-vertex $\ell$ for which $i \in {\cal V}_{ \ell }^{\eps}$. Next, to define the subsets of commodities  $\{ \tilde{\cal V}_{ \ell } \}_{\ell \in {\cal S}_{\mysuff} \cup {\cal D}}$, each such set $\tilde{\cal V}_{ \ell }$ will be given by the ${\cal E}^*$-neighbors of the class-vertex $\ell$, i.e., $\tilde{\cal V}_{ \ell } = \{ i \in U: (i,\ell) \in {\cal E}^* \}$. In addition, for each commodity $i \in \tilde{\cal V}_{ \ell }$, its corresponding SOSI policy $\hat{T}_i = \hat{T}_{i \ell}$ is the one computed via the closed-form expression~\eqref{eqn:closed_form_Ti}.

\paragraph{Verifying properties~\ref{prop:tildeV_size}-\ref{prop:hatT_space}.} We begin by observing that $|\tilde{\cal V}_{ \ell }|$ is precisely the degree of each class-vertex $\ell \in {\cal S}_{\mysuff} \cup {\cal D}$ with respect to the edge set ${\cal E}^*$, meaning that property~\ref{prop:tildeV_size} is trivially guaranteed by our degree constraints. Property~\ref{prop:hatT_space} is satisfied as well, simply by definition of $\hat{T}_i$. The non-trivial argument is regarding the total cost of the policies $\{ \hat{T}_i \}_{i \in U}$, corresponding to property~\ref{prop:hatT_cost}. The next claim, whose proof is provided in Appendix~\ref{app:proof_lem_cost_U_dense}, establishes this property.

\begin{lemma} \label{lem:cost_U_dense}
$\sum_{i \in U}  C_{\myeoq,i}( \hat{T}_i ) \leq  \sum_{i \in U} C_i( {\cal P}^{\eps}_i )$.
\end{lemma}

\subsection{Policy construction} \label{subsec:V_ell_rounding}

Let $\{ \tilde{\cal V}_{ \ell } \}_{\ell \in {\cal S}_{\mysuff} \cup {\cal D}}$ be the mimicking partition we obtain, and let $\{ \hat{T}_i \}_{i \in U}$ be the SOSI policies associated with its underlying set of commodities $U = \bigcup_{\ell \in {\cal S}_{\mysuff} \cup {\cal D}} {\cal V}_{\ell}^{\eps} = \bigcup_{\ell \in {\cal S}_{\mysuff} \cup {\cal D}} \tilde{\cal V}_{ \ell }$, jointly satisfying properties~\ref{prop:tildeV_size}-\ref{prop:hatT_space}. In what follows, we convert these objects into a randomized replenishment policy, with distinct treatments of suffix-sparse and dense classes.

\paragraph{Policy for suffix-sparse classes.} Starting with the easy task, our replenishment policy for the commodities in $\{ \tilde{\cal V}_{ \ell } \}_{\ell \in {\cal S}_{\mysuff}}$ is very simple: For each such commodity $i$, we directly make use of its corresponding SOSI policy, $\hat{T}_i$. The next claim, whose proof is provided in Appendix~\ref{app:proof_lem_bound_space_suffix}, shows that the peak space requirement of this policy is negligible.

\begin{lemma} \label{lem:bound_space_suffix}
Let ${\cal P}_{\mysuff}$ be the deterministic  policy where ${\cal P}_{\mysuff,i} = \hat{T}_i$ for every $i \in \bigcup_{\ell \in {\cal S}_{\mysuff}} \tilde{\cal V}_{ \ell }$. Then, $V_{\max}( {\cal P}_{\mysuff} ) \leq 4 \eps {\cal V}$ and $C( {\cal P}_{\mysuff} ) = \sum_{\ell \in {\cal S}_{\mysuff}} \sum_{i \in \tilde{\cal V}_{ \ell } } C_{\myeoq,i}( \hat{T}_i )$.
\end{lemma}

\paragraph{Policy for dense classes.} In contrast, our policy for the commodities in $\{ \tilde{\cal V}_{ \ell } \}_{\ell \in {\cal D}}$ is significantly more involved. In what follows, for every $\ell \in {\cal D}$, we explain how to efficiently construct a random replenishment policy $\tilde{\cal P}^{\ell}$ for $\tilde{\cal V}_{ \ell }$-commodities that ``approximates'' the behavior of the $\eps$-optimal policy ${\cal P}^{\eps}$ with respect to ${\cal V}_{\ell}^{\eps}$ in a very specific sense. To make this objective more concrete, we say that $\tilde{\cal P}^{\ell}$ guarantees an $(\alpha,\beta)$-ratio when
\[ V_{\max}( \tilde{\cal P}^{\ell} ) ~~\leq~~ \alpha \cdot  | \tilde{\cal V}_{\ell} | \cdot \frac{1}{(1+\eps)^{\ell-1}} \cdot{\cal V} \qquad \text{and} \qquad \expar{ C( \tilde{\cal P}^{\ell} ) } ~~\leq~~ \beta \cdot \sum_{i \in \tilde{\cal V}_{ \ell } } C_{\myeoq,i}( \hat{T}_i ) \ , \]
with the former condition holding almost surely. We proceed by arguing that, up to $\eps$-dependent terms, a ratio of $(\frac{ 7/4 }{ \sqrt{2} \ln 2} , \frac{ 32/31 }{ \sqrt{2} \ln 2})$ is attainable for the general case, where $\ell \in {\cal D} \setminus \{ \infty \}$. Subsequently, the special case of $\ell = \infty$ will be handled via a separate argument.

\paragraph{Light and heavy commodities.} Starting with the scenario where $\ell \in {\cal D} \setminus \{ \infty \}$, let us recall that by property~\ref{prop:hatT_space}, the SOSI policy $\hat{T}_i$ corresponding to each commodity $i \in \tilde{\cal V}_{ \ell }$  satisfies constraint~\eqref{eqn:occupied_space_constraint}. Namely, $\gamma_i \cdot \bar{I}( \hat{T}_i ) \leq \frac{1}{(1+\eps)^{\ell-1}} \cdot{\cal V}$, and as a result, we can decompose $\tilde{\cal V}_{ \ell }$ into commodities of two possible types: Commodity $i \in \tilde{\cal V}_{ \ell }$ is called light when $\gamma_i \cdot \bar{I}( \hat{T}_i ) \leq \frac{ 3 }{ 4 } \cdot \frac{1}{(1+\eps)^{\ell-1}} \cdot{\cal V}$; otherwise, $\gamma_i \cdot \bar{I}( \hat{T}_i ) \in (\frac{ 3 }{ 4 } \cdot \frac{1}{(1+\eps)^{\ell-1}} \cdot{\cal V}, \frac{1}{(1+\eps)^{\ell-1}} \cdot{\cal V}]$, and this commodity is called heavy. The collections of light and heavy commodities within $\tilde{\cal V}_{ \ell }$ will be denoted by ${\cal L}_{ \ell }$ and ${\cal H}_{ \ell }$, respectively. We proceed to construct a replenishment policy $\tilde{\cal P}^{\ell}$ for $\tilde{\cal V}_{ \ell }$-commodities, independently of any other class, by considering two cases, depending on the cardinality of ${\cal L}_{ \ell }$ and ${\cal H}_{ \ell }$.

\paragraph{The light-majority case: \boldmath{${ | {\cal L}_{ \ell } | \geq \frac{ | \tilde{\cal V}_{ \ell } | }{ 2 }}$}.} We first consider the rather straightforward scenario, where at least half of the $\tilde{\cal V}_{ \ell }$-commodities are light. In this case, due to having a majority of light commodities, the SOSI policies $\{ \hat{T}_i \}_{i \in \tilde{\cal V}_{ \ell }}$ themselves lead to a $(\frac{ 7 }{ 4 }, 1)$-ratio, which is even better than the $(\frac{ 7/4 }{ \sqrt{2} \ln 2} , \frac{ 32/31 }{ \sqrt{2} \ln 2})$-ratio we are aiming for. The next claim, whose proof is provided in Appendix~\ref{app:proof_lem_main_result_Vell_light}, formalizes this statement.

\begin{lemma} \label{lem:main_result_Vell_light}
Let $\tilde{\cal P}^{\ell} = \{ \tilde{\cal P}^{\ell}_i \}_{i \in \tilde{\cal V}_{ \ell }}$ be the deterministic policy where $\tilde{\cal P}^{\ell}_i = \hat{T}_i$ for every $i \in \tilde{\cal V}_{ \ell }$. When $| {\cal L}_{ \ell } | \geq \frac{ | \tilde{\cal V}_{ \ell } | }{ 2 }$, this policy guarantees a $(\frac{ 7 }{ 4 }, 1)$-ratio.
\end{lemma}

\paragraph{The heavy-majority case: \boldmath{${ | {\cal H}_{ \ell } | > \frac{ | \tilde{\cal V}_{ \ell } | }{ 2 }}$}.} As it turns out, handling the opposite scenario where ${ | {\cal H}_{ \ell } | > \frac{ | \tilde{\cal V}_{ \ell } | }{ 2 }}$ is much more challenging, necessitating the development of new algorithmic tools and analytical ideas. For ease of exposition, these contents are discussed in Section~\ref{sec:po2-sync}, where we prove the following result, referred to as the Po2-Synchronization Theorem.

\begin{theorem}[Po2-Synchronization] \label{thm:main_result_Vell_heavy}
When ${ | {\cal H}_{ \ell } | > \frac{ | \tilde{\cal V}_{ \ell } | }{ 2 }}$, we can construct in polynomial time a random  policy $\tilde{\cal P}^{\ell} = \{ \tilde{\cal P}^{\ell}_i \}_{i \in \tilde{\cal V}_{ \ell }}$ that guarantees a $((1 + 6\eps) \cdot \frac{ 7/4 }{ \sqrt{2} \ln 2},(1 + \frac{2\eps}{5}) \cdot \frac{ 32/31 }{ \sqrt{2} \ln 2})$-ratio.
\end{theorem}

\paragraph{Constructing \boldmath{${\tilde{\cal P}^{\ell}}$} when \boldmath{${\ell = \infty}$}.} In this special case, it is unclear how to efficiently attain a $(\frac{ 7/4 }{ \sqrt{2} \ln 2} , \frac{ 32/31 }{ \sqrt{2} \ln 2})$-ratio, or any other useful ratio for that matter. However, by property~\ref{prop:hatT_space}, we know  that $\gamma_i \cdot \bar{I}( \hat{T}_i ) \leq \frac{ \eps }{ n } \cdot {\cal V}$ for every commodity $i \in \tilde{\cal V}_{\infty}$, implying that the SOSI policies $\{ \hat{T}_i \}_{i \in \tilde{\cal V}_{ \infty }}$ themselves carry a small additive error in terms of their space requirement, since $\sum_{i \in \tilde{\cal V}_{ \infty }} \gamma_i \hat{T}_i \leq | \tilde{\cal V}_{ \infty } | \cdot \frac{ 2\eps }{ n } \cdot {\cal V}  \leq 2\eps  {\cal V}$. As an immediate consequence, we obtain the next claim.

\begin{observation} \label{obs:main_result_Vinfty}
Let $\tilde{\cal P}^{\infty} = \{ \tilde{\cal P}^{\infty}_i \}_{i \in \tilde{\cal V}_{ \infty }}$ be the deterministic  policy where $\tilde{\cal P}^{\infty}_i = \hat{T}_i$ for every $i \in \tilde{\cal V}_{ \infty }$. Then, $V_{\max}( \tilde{\cal P}^{\infty} ) \leq 2\eps  {\cal V}$ and $C( \tilde{\cal P}^{\infty} ) = \sum_{i \in \tilde{\cal V}_{ \infty } } C_{\myeoq,i}( \hat{T}_i )$.
\end{observation}

\paragraph{Defining a combined policy.} Having just dealt with all possible cases, it remains to propose a single policy ${\cal P}_{\mysuff+\mydense}$ for the entire collection of commodities in $\{ \tilde{\cal V}_{ \ell } \}_{\ell \in {\cal S}_{\mysuff} \cup {\cal D}}$. To this end, we simply glue together ${\cal P}_{\mysuff}$ and ${\cal P}_{\mydense}$, where the latter is comprised of the policies $\{ \tilde{\cal P}^{ \ell } \}_{\ell \in {\cal D}}$ mentioned above. It is worth noting that ${\cal P}_{\mydense}$ is generally a random policy, since some of $\{ \tilde{\cal P}^{ \ell } \}_{\ell \in {\cal D}}$ are deterministic and some are random, depending on whether $\ell = \infty$ or not and on whether ${ | {\cal L}_{ \ell } | \geq \frac{ | \tilde{\cal V}_{ \ell } | }{ 2 }}$ or ${ | {\cal H}_{ \ell } | > \frac{ | \tilde{\cal V}_{ \ell } | }{ 2 }}$.

\subsection{Analysis} \label{subsec:dense_class_analysis}

We conclude the proof of Lemma~\ref{lem:diff_case_dense_main} by deriving the next two claims, with the desired bounds on the peak space requirement and the expected long-run average cost of our combined policy ${\cal P}_{\mysuff+\mydense}$.

\begin{lemma}
$V_{\max}( {\cal P}_{\mysuff+\mydense} ) \leq (1 + 8\eps) \cdot \frac{ 7/4 }{ \sqrt{2} \ln 2} \cdot \bar{V}^{\eps}_{\cal D} + 10\eps  {\cal V}$, almost surely.
\end{lemma}
\begin{proof}
Based on the preceding discussion, we almost surely have
\begin{align}
V_{\max}( {\cal P}_{\mysuff+\mydense} ) & ~~\leq~~ \sum_{ \MyAbove{ \ell \in {\cal D} \setminus \{ \infty \} : }{ | {\cal L}_{ \ell } | \geq | \tilde{\cal V}_{ \ell } | / 2 } } V_{\max}( \tilde{\cal P}^{\ell} ) + \sum_{ \MyAbove{ \ell \in {\cal D} \setminus \{ \infty \} : }{ | {\cal H}_{ \ell } | > | \tilde{\cal V}_{ \ell } | / 2 } } V_{\max}( \tilde{\cal P}^{\ell} ) + V_{\max}( \tilde{\cal P}^{\infty} ) + V_{\max}( {\cal P}_{\mysuff} ) \nonumber \\
& ~~\leq~~ \frac{ 7 }{ 4 } \cdot \sum_{ \MyAbove{ \ell \in {\cal D} \setminus \{ \infty \} : }{ | {\cal L}_{ \ell } | \geq | \tilde{\cal V}_{ \ell } |  / 2  } } | \tilde{\cal V}_{\ell} | \cdot \frac{1}{(1+\eps)^{\ell-1}} \cdot{\cal V} \nonumber \\
& \phantom{~~\leq~~} \mbox{} + (1 + 6\eps) \cdot \frac{ 7/4 }{ \sqrt{2} \ln 2} \cdot \sum_{ \MyAbove{ \ell \in {\cal D} \setminus \{ \infty \} : }{ | {\cal H}_{ \ell } | > | \tilde{\cal V}_{ \ell } | / 2  } }  | \tilde{\cal V}_{\ell} | \cdot \frac{1}{(1+\eps)^{\ell-1}} \cdot{\cal V} + 6\eps  {\cal V} \label{eqn:UB_Pdense_space_1} \\
& ~~\leq~~ (1 + 6\eps) \cdot \frac{ 7/4 }{ \sqrt{2} \ln 2} \cdot \sum_{\ell \in {\cal D} \setminus \{ \infty \}} | \tilde{\cal V}_{\ell} | \cdot \frac{1}{(1+\eps)^{\ell-1}} \cdot{\cal V} + 6\eps  {\cal V} \nonumber \\
& ~~\leq~~ (1 + 6\eps) \cdot \frac{ 7/4 }{ \sqrt{2} \ln 2} \cdot (1 + \eps) \cdot \sum_{\ell \in {\cal D} \setminus \{ \infty \}} \tilde{V}^{\eps}_{\ell} + 6\eps  {\cal V} \label{eqn:UB_Pdense_space_2}  \\
& ~~\leq~~ (1 + 8\eps) \cdot \frac{ 7/4 }{ \sqrt{2} \ln 2} \cdot \sum_{\ell \in {\cal D} \setminus \{ \infty \}}  \left( \bar{V}^{\eps}_{\ell} + \frac{ \eps }{ |{\cal D}| } \cdot {\cal V}\right) + 6\eps  {\cal V} \label{eqn:UB_Pdense_space_3} \\
& ~~\leq~~ (1 + 8\eps) \cdot \frac{ 7/4 }{ \sqrt{2} \ln 2} \cdot \bar{V}^{\eps}_{\cal D} + 10\eps  {\cal V} \ . \label{eqn:UB_Pdense_space_4}
\end{align}
Here, inequality~\eqref{eqn:UB_Pdense_space_1} is obtained by combining Lemmas~\ref{lem:bound_space_suffix} and~\ref{lem:main_result_Vell_light}, Theorem~\ref{thm:main_result_Vell_heavy}, and Observation~\ref{obs:main_result_Vinfty}. Inequality~\eqref{eqn:UB_Pdense_space_2} follows from property~\ref{prop:tildeV_size}, stating in particular that $|\tilde{\cal V}_{ \ell }| \leq  \tilde{\cal N}_{\ell} = \frac{ (1 + \eps)^{\ell} }{ {\cal V} } \cdot \tilde{V}^{\eps}_{\ell}$ for every $\ell \in {\cal D}$. Inequality~\eqref{eqn:UB_Pdense_space_3} holds since $\tilde{V}^{\eps}_{\ell} \leq \bar{V}^{\eps}_{\ell} + \frac{ \eps }{ |{\cal D}| } \cdot {\cal V}$, as explained in Section~\ref{subsec:useful_partition_dense}. Finally, we arrive at inequality~\eqref{eqn:UB_Pdense_space_4} by recalling that  $\bar{V}^{\eps}_{\cal D} = \sum_{\ell \in {\cal D}} \bar{V}^{\eps}_{\ell}$.
\end{proof}

\begin{lemma}
$\expar{ C( {\cal P}_{\mysuff+\mydense} ) } \leq ( 1 + \frac{2\eps}{5} ) \cdot \frac{ 32/31 }{ \sqrt{2} \ln 2} \cdot \sum_{\ell \in {\cal S}_{\mysuff} \cup {\cal D}} \sum_{i \in {\cal V}_{\ell}^{\eps}} C_i( {\cal P}_i^{\eps} )$.
\end{lemma}
\begin{proof}
Similarly to the proof of the previous lemma, we have
\begin{align}
&\ex{ C( {\cal P}_{\mysuff+\mydense} ) } \nonumber \\
& \qquad =~~ \sum_{ \MyAbove{ \ell \in {\cal D} \setminus \{ \infty \} : }{ | {\cal L}_{ \ell } | \geq | \tilde{\cal V}_{ \ell } | / 2 } }  C( \tilde{\cal P}^{\ell} ) + \sum_{ \MyAbove{ \ell \in {\cal D} \setminus \{ \infty \} : }{ | {\cal H}_{ \ell } | > | \tilde{\cal V}_{ \ell } | / 2 } } \ex{ C( \tilde{\cal P}^{\ell} ) } + C( \tilde{\cal P}^{\infty} ) +C( {\cal P}_{\mysuff} )
\nonumber \\
& \qquad \leq~~ \sum_{ \MyAbove{ \ell \in {\cal D} \setminus \{ \infty \} : }{ | {\cal L}_{ \ell } | \geq | \tilde{\cal V}_{ \ell } | / 2 } } \sum_{i \in \tilde{\cal V}_{ \ell } } C_{\myeoq,i}( \hat{T}_i ) + \left( 1 + \frac{2\eps}{5} \right) \cdot \frac{ 32/31 }{ \sqrt{2} \ln 2} \cdot \sum_{ \MyAbove{ \ell \in {\cal D} \setminus \{ \infty \} : }{ | {\cal H}_{ \ell } | > | \tilde{\cal V}_{ \ell } | / 2 } }  \sum_{i \in \tilde{\cal V}_{ \ell } } C_{\myeoq,i}( \hat{T}_i ) \nonumber \\
& \qquad \qquad  \mbox{} + \sum_{i \in \tilde{\cal V}_{ \infty } } C_{\myeoq,i}( \hat{T}_i ) + \sum_{ \ell \in {\cal S}_{\mysuff} } \sum_{i \in \tilde{\cal V}_{ \ell } } C_{\myeoq,i}( \hat{T}_i ) \label{eqn:UB_Pdense_cost_1} \\
& \qquad \leq~~ \left( 1 + \frac{2\eps}{5} \right) \cdot \frac{ 32/31 }{ \sqrt{2} \ln 2} \cdot \sum_{\ell \in {\cal S}_{\mysuff} \cup {\cal D}} \sum_{i \in {\cal V}_{\ell}^{\eps}} C_{\myeoq,i}( \hat{T}_i ) \nonumber \\
& \qquad \leq~~ \left( 1 + \frac{2\eps}{5} \right) \cdot \frac{ 32/31 }{ \sqrt{2} \ln 2} \cdot \sum_{\ell \in {\cal S}_{\mysuff} \cup {\cal D}} \sum_{i \in {\cal V}_{\ell}^{\eps}} C_i( {\cal P}_i^{\eps} ) \ . \label{eqn:UB_Pdense_cost_2}
\end{align}
Here, inequality~\eqref{eqn:UB_Pdense_cost_1} is obtained by combining Lemmas~\ref{lem:bound_space_suffix} and~\ref{lem:main_result_Vell_light}, Theorem~\ref{thm:main_result_Vell_heavy}, and Observation~\ref{obs:main_result_Vinfty}. Inequality~\eqref{eqn:UB_Pdense_cost_2} holds since $\bigcup_{\ell \in {\cal S}_{\mysuff} \cup {\cal D}} {\cal V}_{\ell}^{\eps} = U$ and  $\sum_{i \in U}  C_{\myeoq,i}( \hat{T}_i ) \leq  \sum_{i \in U} C_i( {\cal P}^{\eps}_i )$, by property~\ref{prop:hatT_cost}.
\end{proof}

\section{Proof of the Po2-Synchronization Theorem} \label{sec:po2-sync}

The main objective of this section is to establish Theorem~\ref{thm:main_result_Vell_heavy}, stating that for every dense class $\ell$ with ${ | {\cal H}_{ \ell } | > \frac{ | \tilde{\cal V}_{ \ell } | }{ 2 }}$, we can efficiently construct  a random policy $\tilde{\cal P}^{\ell} = \{ \tilde{\cal P}^{\ell}_i \}_{i \in \tilde{\cal V}_{ \ell }}$ satisfying
\begin{align*}
& V_{\max}( \tilde{\cal P}^{\ell} ) ~~\leq~~ (1 + 6\eps) \cdot \frac{ 7/4 }{ \sqrt{2} \ln 2} \cdot  | \tilde{\cal V}_{\ell} | \cdot \frac{1}{(1+\eps)^{\ell-1}} \cdot{\cal V} \\
& \qquad \qquad \qquad \qquad \qquad \text{and} \qquad \expar{ C( \tilde{\cal P}^{\ell} ) } ~~\leq~~ \left( 1 + \frac{2\eps}{5} \right) \cdot \frac{ 32/31 }{ \sqrt{2} \ln 2} \cdot \sum_{i \in \tilde{\cal V}_{ \ell } } C_{\myeoq,i}( \hat{T}_i ) \ ,
\end{align*}
with the former property holding almost surely. To this end, Section~\ref{subsec:sync_hack} uncovers a basic obstacle on the way to improving the approximation guarantees of \citet{Anily91} and \citet{GallegoQS96}, identifying a seemingly stylized setting where the latter barrier can be breached. Sections~\ref{subsec:pow-rounding} and~\ref{subsec:near_far_pairs} describe a probabilistic reduction by which we make this stylized setting applicable in proving Theorem~\ref{thm:main_result_Vell_heavy}. Finally, Sections~\ref{subsec:PO2_sync_policy} and~\ref{subsec:analysis_Aell_space_cost} explain how our randomized policy is created, with a detailed analysis of its performance guarantees.

\subsection{The pairwise synchronization gadget} \label{subsec:sync_hack}

\paragraph{The \boldmath{$2$}-approximation barrier.} Circling back to Section~\ref{sec:anily_approx}, let us develop some basic understanding of why improving on its approximation guarantees seems implausible without additional advancements. In essence, letting $T^* = (T_1^*, \ldots, T_n^*)$ be an optimal vector of SOSI policies with respect to formulation~\eqref{eqn:modify_relax_warehouse}, we noticed that these policies are generally not capacity-feasible, since their  peak space requirement $V_{\max}( T^* ) = \sum_{i \in [n]} \gamma_i T_i^*$ could be as large as $2{\cal V}$. To correct this issue, we scaled down each SOSI policy by a factor of $2$. As inequalities~\eqref{eqn:fix_cap_scale} and~\eqref{eqn:final_2app_eq3} show, the resulting policy becomes capacity-feasible, with the downside of potentially increasing its long-run average cost to $2 \cdot \opt\eqref{eqn:model_warehouse}$.

Along these lines, one may wish to be more sophisticated, in the sense of scaling each policy $T_i^*$ by a factor of $\alpha_i > 0$ and including a horizontal shift of $\tau_i \in [0, \alpha_i T_i^*)$. Such an alteration scales the peak space utilization of this commodity by $\alpha_i$; however, its cost could blow-up by $\max \{ \alpha_i, \frac{ 1 }{ \alpha_i } \}$. As a result, since $\alpha_i \cdot \max \{ \alpha_i, \frac{ 1 }{ \alpha_i } \} \geq 1$, each commodity by itself appears to be in worse shape, and it is  unclear whether there is a way to pick $( \alpha_i, \tau_i )_{i \in [n]}$ such that we are creating a capacity-feasible policy whose  cost is at most $(2 - \delta) \cdot \opt\eqref{eqn:model_warehouse}$, for some absolute constant $\delta > 0$.

\paragraph{Additional structure and dynamic policies?} That said, the crux of our approach towards deriving the Po2-Synchronization Theorem resides in identifying a stylized setting where one could beat the space-cost tradeoff discussed above on a pairwise basis. Specifically, suppose we are given a pair of commodities, $A$ and $B$, along with corresponding SOSI policies $T_A$ and $T_B$, that satisfy the next two properties:
\begin{enumerate}
    \item \label{prop:sub1_couple_space} The peak space utilization of $T_A$ and $T_B$ are nearly equal, with $\gamma_A T_A \in (1 \pm \eps) \cdot \gamma_B T_B$.

    \item \label{prop:sub1_couple_ratio} The ratio between $T_A$ and $T_B$ is an integer power of $2$.
\end{enumerate}
In this case, we say that $(A,T_A)$ and $(B, T_B)$ jointly form a sub-$1$ couple, shortly justifying this terminology.

The following result, whose full proof is provided in Appendix~\ref{app:proof_lem_construction_pair}, argues that by judiciously synchronizing the replenishment cycles of a sub-1 couple, we can indeed overcome the space-cost limitations inherent to  SOSI policies. Our proof relies on constructing a joint dynamic schedule, where the two replenishment cycles are synchronized and horizontally-shifted to minimize peak overlap. By carefully interleaving orders, we ensure that their aggregate space consumption remains stable, avoiding the high peaks typical of independent SOSI policies.

\begin{lemma} \label{lem:construction_pair}
Given any sub-$1$ couple, $(A,T_A)$ and $(B, T_B)$, we can construct in polynomial time dynamic replenishment policies ${\cal P}_A$ and ${\cal P}_B$ such that:
\begin{enumerate}
    \item {\em Joint occupied space:} $V_{\max}( {\cal P}_A, {\cal P}_B ) \leq (1 + \eps) \cdot \frac{ 7 }{ 8 } \cdot ( \gamma_A T_A  + \gamma_B T_B )$.

    \item {\em Cost blow-up:} $\max \{ \frac{ C_A( {\cal P}_A ) }{ C_{\myeoq,A}( T_A )}, \frac{ C_B( {\cal P}_B ) }{ C_{\myeoq,B}( T_B)} \} \leq \frac{ 32 }{ 31}$.
\end{enumerate}
\end{lemma}

To interpret this result, let us first point out the obvious: We take explicit advantage of our freedom to design dynamic policies, unlike the approach of Section~\ref{sec:anily_approx}, which solely focuses on SOSI policies. To unveil the not-so-obvious feature of this construction, note that by item~1, the peak space requirement of gluing ${\cal P}_A$ and ${\cal P}_B$ together is at most $ (1 + \eps) \cdot\frac{ 7 }{ 8 }$ times that of $(T_A, T_B)$. In addition, by item~2, the combined cost $C_A({\cal P}_A) + C_B( {\cal P}_B )$ is within factor $\frac{ 32 }{ 31 }$ of $C_{\myeoq,A}( T_A ) + C_{\myeoq,B}( T_B )$. Consequently, while we have no guarantees on the space-cost tradeoff of commodities $A$ and $B$ by themselves, their pairwise tradeoff is upper-bounded by $ (1 + \eps) \cdot \frac{ 28 }{ 31 }$, motivating the ``sub-$1$ couple'' terminology.

\paragraph{Why is this setting useful?} At the moment, readers can only wonder about basic questions such as: With respect to our current SOSI policies, $\{ \hat{T}_i \}_{i \in \tilde{\cal V}_{ \ell }}$, why are we guaranteed to have sufficiently-many sub-$1$ couples? What is their combined contribution toward the overall space requirement and long-run average cost? How are we going to handle commodities outside of sub-$1$ couples? In what follows, we address these issues by probabilistically modifying the policies $\{ \hat{T}_i \}_{i \in \tilde{\cal V}_{ \ell }}$ into having many sub-$1$ couples, without overly perturbing their space requirement and cost. Subsequently, combining Lemma~\ref{lem:construction_pair} with additional ideas will allow us to derive the Po2-Synchronization Theorem.

\subsection{Power-of-2 rounding} \label{subsec:pow-rounding}

To explain our method for probabilistically altering $\{ \hat{T}_i \}_{i \in \tilde{\cal V}_{ \ell }}$, we remind the reader that $\tilde{\cal V}_{ \ell }$ was partitioned in Section~\ref{subsec:V_ell_rounding} into heavy and light commodities, corresponding to ${\cal H}_{\ell}$ and ${\cal L}_{\ell}$. Let us arbitrarily partition ${\cal H}_{ \ell }$ into $Q = \frac{ 20 \ln (1/\eps) }{ \eps^2 }$ subsets ${\cal H}_{ \ell , 1}, \ldots, {\cal H}_{\ell,Q}$, each of size either $\lfloor \frac{ |{\cal H}_{ \ell }| }{ Q } \rfloor$ or $\lceil  \frac{ |{\cal H}_{ \ell }| }{ Q } \rceil$, noting that
\begin{equation} \label{eqn:ratio_HQ_eps}
\left\lfloor \frac{ |{\cal H}_{ \ell }| }{ Q } \right\rfloor ~~>~~ \left\lfloor \frac{ | \tilde{\cal V}_{ \ell } |/2 }{ Q } \right\rfloor  ~~\geq~~ \left\lfloor \frac{ 50\ln (1/\eps) }{ Q \eps^4 } \right\rfloor ~~\geq~~ \frac{ 2 }{ \eps^2 } \ .
\end{equation}
Here, the first inequality holds since  ${ | {\cal H}_{ \ell } | > \frac{ | \tilde{\cal V}_{ \ell } | }{ 2 }}$, by the hypothesis of Theorem~\ref{thm:main_result_Vell_heavy}, and the second inequality follows by recalling that $|\tilde{\cal V}_{ \ell }| \geq \frac{ 100 \ln (1/\eps) }{ \eps^4 }$ for every $\ell \in {\cal D}$, according to property~\ref{prop:tildeV_size}.

Now, for each subset ${\cal H}_{ \ell ,q}$, independently of any other subset, we proceed by power-of-$2$ rounding its SOSI policies $\{ \hat{T}_i \}_{i \in {\cal H}_{ \ell ,q}}$. While there are several different methods in this context (see, e.g., \citet{Roundy85, Roundy86} and \citet{JacksonMM85}), for our particular purposes, we employ  the dependent rounding procedure of \citet{TeoB01} to create the collection of random policies $\{ T_i^{ \Theta_{\ell, q} } \} _{i \in {\cal H}_{ \ell ,q}}$. This procedure operates as follows:
\begin{itemize}
    \item For each commodity $i \in {\cal H}_{ \ell ,q}$, we first express its corresponding SOSI policy as $\hat{T}_i = 2^{\alpha_i + \beta_i} \cdot \hat{T}_{\min,\ell ,q}$, for some integer $\alpha_i \geq 0$ and real $\beta_i \in [0,1)$, where $\hat{T}_{\min,\ell ,q} = \min_{i \in {\cal H}_{ \ell ,q}} \hat{T}_i$.

    \item Next, we generate a  uniform random variable $\Theta_{\ell, q} \sim U[-\frac{1}{2}, \frac{1}{2}]$.

    \item Finally, for every commodity $i \in {\cal H}_{ \ell ,q}$, we set
    \[ T_i^{ \Theta_{\ell, q} } ~~=~~ \begin{cases}
    2^{\alpha_i + \Theta_{\ell, q}} \cdot \hat{T}_{\min,\ell ,q}, & \text{when } \Theta_{\ell, q} \geq \beta_i - \frac{ 1 }{ 2 } \\
    2^{\alpha_i + \Theta_{\ell, q}+1} \cdot \hat{T}_{\min,\ell ,q}, \qquad & \text{when } \Theta_{\ell, q} < \beta_i - \frac{ 1 }{ 2 }
    \end{cases} \]
\end{itemize}
We mention in passing that readers who are thoroughly familiar with the work of \citet[Sec.~2.2]{TeoB01} will notice subtle differences between their procedure and the one described above. Our point in  presenting this slightly modified implementation is that it allows one to easily verify a number of structural properties exhibited by $\{ T_i^{ \Theta_{\ell, q} } \}_{i \in {\cal H}_{ \ell ,q}}$, formally summarized in Claim~\ref{clm:properties_PO2} below. The credit should be fully given to \citeauthor{TeoB01}; we are merely rephrasing their approach.

\begin{claim} \label{clm:properties_PO2}
The random policies $\{ T_i^{ \Theta_{\ell, q} } \} _{i \in {\cal H}_{ \ell ,q}}$ satisfy the next three properties:
\begin{enumerate}
    \item $\exsubpar{ \Theta_{\ell, q} } { T_i^{ \Theta_{\ell, q} } } = \frac{ 1 }{ \sqrt{2} \ln 2 } \cdot \hat{T}_i$ and $\exsubpar{ \Theta_{\ell, q} }{ \frac{ 1 }{ T_i^{ \Theta_{\ell, q} } } } = \frac{ 1 }{ \sqrt{2} \ln 2 } \cdot \frac{ 1 }{ \hat{T}_i }$.

    \item $T_i^{ \Theta_{\ell, q} } \in [ \frac{ \hat{T}_i }{ \sqrt{2} } , \sqrt{2} \hat{T}_i ]$ almost surely.

    \item For every pair of commodities $i_1, i_2 \in {\cal H}_{ \ell ,q}$, the ratio between $T_{i_1}^{ \Theta_{\ell, q} }$ and $T_{i_2}^{ \Theta_{\ell, q} }$ is an integer power of $2$.
\end{enumerate}
\end{claim}

It is worth pointing out that, in terms of the random policies $\{ T_i^{ \Theta_{\ell, q} } \} _{i \in {\cal H}_{ \ell ,q}}$, every pair $(i_1, T_{i_1}^{ \Theta_{\ell, q} })$ and $(i_2, T_{i_2}^{ \Theta_{\ell, q} })$ satisfies characterization~\ref{prop:sub1_couple_ratio} of sub-$1$ couples, by property~3 above.

\subsection{Near and far pairs} \label{subsec:near_far_pairs}

To concurrently instill characterization~\ref{prop:sub1_couple_space} for a sufficiently large collection of couples, let us consider a single subset ${\cal H}_{ \ell ,q}$, whose commodities will be denoted by $1, \ldots, m$. Recalling that the corresponding power-of-$2$ policies $\{ T_i^{ \Theta_{\ell, q} } \} _{i \in {\cal H}_{ \ell ,q}}$ are random, let $\Pi_{ \ell, q } : [m] \to [m]$ be the random permutation for which
\begin{equation} \label{eqn:sequence_gamma_T}
\gamma_{ \Pi_{ \ell, q }(1) } \cdot T_{ \Pi_{ \ell, q }(1)}^{ \Theta_{\ell, q} } ~~\geq~~ \cdots ~~\geq~~ \gamma_{ \Pi_{ \ell, q }(m) } \cdot T_{ \Pi_{ \ell, q }(m)}^{ \Theta_{\ell, q} } \ ,
\end{equation}
where ties are broken by some fixed decision rule, say by minimum index, just to ensure that $\Pi_{ \ell, q }$ is uniquely defined.

\paragraph{Far and near pairs.} Now, suppose we break ${\cal H}_{ \ell ,q}$ into successive pairs $\{ \Pi_{ \ell, q }(1),\Pi_{ \ell, q }(2) \}$, $\{ \Pi_{ \ell, q }(3), \Pi_{ \ell, q }(4) \}$, so on and so forth, leaving $\Pi_{ \ell, q }(m)$ out when $m$ is odd. We say that such a pair $\{ \Pi_{ \ell, q }(2i-1), \Pi_{ \ell, q }(2i) \}$ is far when $\gamma_{ \Pi_{ \ell, q }(2i-1) } \cdot T_{ \Pi_{ \ell, q }(2i-1)}^{ \Theta_{\ell, q} } \geq (1+\eps) \cdot \gamma_{ \Pi_{ \ell, q }(2i) } \cdot T_{ \Pi_{ \ell, q }(2i)}^{ \Theta_{\ell, q} }$. Namely, there is a multiplicative gap of at least $1+\eps$ between the $(2i-1)$-th and $2i$-th terms of the sequence~\eqref{eqn:sequence_gamma_T}. In the opposite case, where $\gamma_{ \Pi_{ \ell, q }(2i-1) } \cdot T_{ \Pi_{ \ell, q }(2i-1)}^{ \Theta_{\ell, q} } < (1+\eps) \cdot \gamma_{ \Pi_{ \ell, q }(2i) } \cdot T_{ \Pi_{ \ell, q }(2i)}^{ \Theta_{\ell, q} }$, we say that this pair is near. We use
${\cal F}_{ \ell ,q}^{ \Pi_{ \ell, q } }$ and ${\cal N}_{ \ell ,q}^{ \Pi_{ \ell, q } }$ to denote the collection of far and near pairs in ${\cal H}_{ \ell ,q}$; these sets are of course random, due to their dependency on $\Pi_{ \ell, q }$. The next claim, whose proof appears in Appendix~\ref{app:proof_lem_bound_on_far_pairs}, shows that the number of far pairs is only $O( \frac{ 1 }{ \eps } )$.

\begin{claim} \label{clm:bound_on_far_pairs}
$|{\cal F}_{ \ell ,q}^{ \Pi_{ \ell, q } }| \leq \frac{ 11 }{ 10\eps }$.
\end{claim}

\paragraph{Near pairs are  sub-\boldmath{$1$} couples.} An immediate consequence of this result is that there are considerably more near pairs than far ones, since
\[ |{\cal N}_{ \ell ,q}^{ \Pi_{ \ell, q } }| ~~=~~ \left\lfloor \frac{ | {\cal H}_{ \ell, q }  | }{ 2 } \right\rfloor - |{\cal F}_{ \ell ,q}^{ \Pi_{ \ell, q } }| ~~\geq~~ \left\lfloor \frac{ \lfloor |{\cal H}_{ \ell }| / Q \rfloor }{ 2 } \right\rfloor - \frac{ 11 }{ 10\eps } ~~\geq~~ \left\lfloor \frac{ 1 }{ \eps^2 } \right\rfloor - \frac{ 11 }{ 10\eps } ~~=~~ \Omega \left( \frac{ 1 }{ \eps^2 } \right) \ , \]
where the second inequality  follows from~\eqref{eqn:ratio_HQ_eps}. The important observation is that for every near pair,  $( \Pi_{ \ell, q }(2i-1), T_{ \Pi_{ \ell, q }(2i-1)}^{ \Theta_{\ell, q} })$ and $( \Pi_{ \ell, q }(2i), T_{ \Pi_{ \ell, q }(2i)}^{ \Theta_{\ell, q} })$ jointly form a sub-$1$ couple. Indeed, by Claim~\ref{clm:properties_PO2}(3), we already know that the ratio between $T_{ \Pi_{ \ell, q }(2i-1)}^{ \Theta_{\ell, q} }$ and $T_{ \Pi_{ \ell, q }(2i)}^{ \Theta_{\ell, q} }$ is an integer power of $2$. In addition, by definition of near pairs, $\gamma_{ \Pi_{ \ell, q }(2i-1) } \cdot T_{ \Pi_{ \ell, q }(2i-1)}^{ \Theta_{\ell, q} } \in [1,1+\eps) \cdot \gamma_{ \Pi_{ \ell, q }(2i) } \cdot T_{ \Pi_{ \ell, q }(2i)}^{ \Theta_{\ell, q} }$. Consequently, by employing Lemma~\ref{lem:construction_pair}, we obtain the next result.

\begin{corollary} \label{cor:construct_sub1_near}
For every $\{ \Pi_{ \ell, q }(2i-1), \Pi_{ \ell, q }(2i) \} \in {\cal N}_{ \ell ,q}^{ \Pi_{ \ell, q } }$, we can construct in polynomial time dynamic replenishment policies ${\cal P}_{ \Pi_{ \ell, q }(2i-1) }^{\ell,q}$ and ${\cal P}_{ \Pi_{ \ell, q }(2i)}^{\ell,q}$ such that:
\begin{enumerate}
    \item {\em Joint occupied space:}
    \[ V_{\max}( {\cal P}_{ \Pi_{ \ell, q }(2i-1) }^{\ell,q}, {\cal P}_{ \Pi_{ \ell, q }(2i)}^{\ell,q} ) ~~\leq~~ (1 + \eps) \cdot \frac{ 7 }{ 8 } \cdot \left( \gamma_{ \Pi_{ \ell, q }(2i-1) } \cdot T_{ \Pi_{ \ell, q }(2i-1)}^{ \Theta_{\ell, q} } + \gamma_{ \Pi_{ \ell, q }(2i) } \cdot T_{ \Pi_{ \ell, q }(2i)}^{ \Theta_{\ell, q} } \right) \ . \]

    \item {\em Cost blow-up:}
    \[ \max \left\{ \frac{ C_{ \Pi_{ \ell, q }(2i-1) }( {\cal P}^{\ell,q}_{ \Pi_{ \ell, q }(2i-1) } ) }{ C_{ \myeoq, \Pi_{ \ell, q }(2i-1) }( T_{ \Pi_{ \ell, q }(2i-1)}^{ \Theta_{\ell, q} } ) }, \frac{ C_{ \Pi_{ \ell, q }(2i) }( {\cal P}^{\ell,q}_{ \Pi_{ \ell, q }(2i) } ) }{ C_{ \myeoq,  \Pi_{ \ell, q }(2i) }( T_{ \Pi_{ \ell, q }(2i)}^{ \Theta_{\ell, q} } ) } \right\} ~~\leq~~ \frac{ 32 }{ 31} \ . \]
\end{enumerate}
\end{corollary}

\subsection{Overall policy construction} \label{subsec:PO2_sync_policy}

\paragraph{Distinction between commodity types.} Based on the preceding discussion, it follows that the commodities in $\tilde{\cal V}_{\ell} = {\cal H}_{\ell} \cup {\cal L}_{\ell}$ can be classified into four types:
\begin{enumerate}
    \item Those belonging to near pairs across ${\cal H}_{ \ell , 1}, \ldots, {\cal H}_{\ell,Q}$.

    \item Those belonging to far pairs across ${\cal H}_{ \ell , 1}, \ldots, {\cal H}_{\ell,Q}$.

    \item Within each of the subsets ${\cal H}_{ \ell , 1}, \ldots, {\cal H}_{\ell,Q}$ with odd cardinality, there is a single commodity that does not belong to any pair.

    \item The collection of light commodities ${\cal L}_{\ell}$.
\end{enumerate}
We denote these types by ${\cal T}_1^{ \Theta_\ell }$, ${\cal T}_2^{\Theta_\ell }$, ${\cal T}_3^{ \Theta_\ell }$, and ${\cal T}_4$, with the superscript $\Theta_\ell$ emphasizing the randomness inherent to the first three types. Here, $\Theta_\ell = ( \Theta_{ \ell, q } )_{q \in [Q]}$ is the random vector resulting from our power-of-$2$ rounding procedure (see Section~\ref{subsec:pow-rounding}), noting that by construction, its components are mutually independent.

\paragraph{The overall policy.} The final policy we return depends on whether an easily-checkable condition is satisfied or not. To formalize this condition, we remind the reader that for every subset ${\cal H}_{ \ell ,q}$, starting with the SOSI policies $\{ \hat{T}_i \}_{i \in {\cal H}_{ \ell ,q}}$, our power-of-$2$ rounding procedure  created the collection of random policies $\{ T_i^{ \Theta_{\ell, q} } \} _{i \in {\cal H}_{ \ell ,q}}$, independently of other subsets.  With respect to these policies, let ${\cal A}_{\ell}$ be the event where
\begin{equation} \label{eqn:definition_A_ell}
\sum_{i \in {\cal H}_{\ell} } \gamma_i   T_i^{ \Theta_{\ell} } ~~\leq~~ (1 + \eps) \cdot \frac{ 1 }{ \sqrt{2} \ln 2 } \cdot \sum_{i \in {\cal H}_{\ell} } \gamma_i   \hat{T}_i \ .
\end{equation}
Here, for ease of notation, we make use of $T_i^{ \Theta_{\ell} } = T_i^{ \Theta_{\ell, q} }$, where $q \in [Q]$ is the unique index for which $i \in {\cal H}_{\ell,q}$. Given this definition, our final policy $\tilde{\cal P}^{ \ell }$ depends on whether ${\cal A}_{ \ell }$ occurs or not:
\begin{itemize}
    \item {\em When ${\cal A}_{\ell}$ occurs}: Our policies for ${\cal T}_1^{ \Theta_\ell }$-commodities are exactly those  obtained via Corollary~\ref{cor:construct_sub1_near}. It is important to point out that, for each such commodity $i \in {\cal T}_1^{ \Theta_\ell }$, due to conditioning on ${\cal A}_{\ell}$, the object we return is the conditional random policy ${\cal P}_i^{\ell+} = [ {\cal P}_i^{\ell} | {\cal A}_{\ell} ]$. For every commodity $i \in {\cal T}_2^{ \Theta_\ell } \cup {\cal T}_3^{ \Theta_\ell }$, its random policy is kept unchanged, meaning that we return ${\cal P}_i^{\ell+} = [T_i^{ \Theta_{\ell} } | {\cal A}_{\ell}]$, again due to conditioning on ${\cal A}_{\ell}$. Finally, for every commodity $i \in {\cal T}_4$, we return its original SOSI policy ${\cal P}_i^{\ell+} = \hat{T}_i$, which is of course deterministic.

    \item {\em When ${\cal A}_{\ell}$ does not occur}: In this case, the policy we employ is a scaled version of $\hat{T}$. Specifically, for every commodity $i \in \tilde{\cal V}_{\ell}$, we return the policy ${\cal P}_i^{\ell-} = \alpha \hat{T}_i$, where $\alpha = \frac{ 7/8 }{ \sqrt{2} \ln 2 }$.
\end{itemize}

\subsection{Analysis} \label{subsec:analysis_Aell_space_cost}

In the remainder of this section, we conclude the proof of Theorem~\ref{thm:main_result_Vell_heavy} by arguing that our random replenishment policy $\tilde{\cal P}^{\ell} = \{ \tilde{\cal P}^{\ell}_i \}_{i \in \tilde{\cal V}_{ \ell }}$ guarantees a $((1 + 6\eps) \cdot \frac{ 7/4 }{ \sqrt{2} \ln 2},(1 + \frac{2\eps}{5}) \cdot \frac{ 32/31 }{ \sqrt{2} \ln 2})$-ratio. In other words, we will prove that the peak space requirement of this policy is almost surely
\[ V_{\max}( \tilde{\cal P}^{\ell} ) ~~\leq~~ (1 + 6\eps) \cdot \frac{ 7/4 }{ \sqrt{2} \ln 2 } \cdot | \tilde{\cal V}_{\ell} | \cdot \frac{1}{(1+\eps)^{\ell-1}} \cdot{\cal V} \ , \]
and that its expected long-run average cost is
\[ \exsub{ \Theta_{\ell} }{ C( \tilde{\cal P}^{\ell} ) } ~~\leq~~ \left( 1 + \frac{2\eps}{5}  \right) \cdot \frac{ 32/31 }{ \sqrt{2} \ln 2} \cdot \sum_{i \in \tilde{\cal V}_{ \ell } } C_{\myeoq,i}( \hat{T}_i ) \ . \]

\paragraph{The likelihood of \boldmath{${{\cal A}_{\ell}}$}.} To go about deriving these bounds, we first show that the event ${\cal A}_{\ell}$ occurs with probability $1-O(\eps)$,  as stated in Lemma~\ref{lem:success_Aell} below. To understand the intuition behind this result, let us observe that by Claim~\ref{clm:properties_PO2}(1), we know that $\exsubpar{ \Theta_{\ell} } { T_i^{ \Theta_{\ell} } } = \frac{ 1 }{ \sqrt{2} \ln 2 } \cdot \hat{T}_i$ for every commodity $i \in \tilde{\cal V}_{\ell}$. Therefore, by defining ${\cal A}_{\ell}$ via condition~\eqref{eqn:definition_A_ell}, this event essentially states that the random variable $\sum_{i \in {\cal H}_{\ell} } \gamma_i T_i^{ \Theta_{\ell} }$ does not deviate much above its expectation, $\frac{ 1 }{ \sqrt{2} \ln 2 } \cdot \sum_{i \in {\cal H}_{\ell} } \gamma_i  \hat{T}_i$. While the rather involved proof of this claim is deferred to Appendix~\ref{app:proof_lem_success_Aell}, it will assist readers in realizing that various choices up until now were meant to guarantee that  appropriate concentration inequalities will be applicable. Among others, these choices include making use of power-of-$2$ rounding, independently drawing $( \Theta_{ \ell, q } )_{q \in [Q]}$, and ensuring that each subset ${\cal H}_{\ell,q}$ is sufficiently large.

\begin{lemma} \label{lem:success_Aell}
$\prsubpar{ \Theta_\ell }{ {\cal A}_{\ell} } \geq 1 - \frac{ \eps }{ 10 }$.
\end{lemma}

\paragraph{Space requirement.} We are now ready to upper-bound the peak space requirement of our random policy $\tilde{\cal P}^{\ell}$. Once again, the proof of this result is quite technical, and we therefore present its finer details in Appendix~\ref{app:proof_lem_PO2_bound_Vmax}. At a high level, when ${\cal A}_{\ell}$ does not occur, it will turn out that scaling $\hat{T}$ by a factor of $\alpha = \frac{ 7/8 }{ \sqrt{2} \ln 2 }$ indeed suffices. The complementary case, where ${\cal A}_{\ell}$ occurs, is significantly more involved, and our approach consists of separately bounding each of the commodity types ${\cal T}_1^{ \Theta_\ell }$, ${\cal T}_2^{\Theta_\ell }$, ${\cal T}_3^{ \Theta_\ell }$, and ${\cal T}_4$. Here, the observation that there are considerably more near pairs than far ones (see Section~\ref{subsec:near_far_pairs}) will play an important role.

\begin{lemma} \label{lem:PO2_bound_Vmax}
$V_{\max}( \tilde{\cal P}^{\ell} ) \leq (1 + 6\eps) \cdot \frac{ 7/4 }{ \sqrt{2} \ln 2 } \cdot | \tilde{\cal V}_{\ell} | \cdot \frac{1}{(1+\eps)^{\ell-1}} \cdot{\cal V}$ almost surely.
\end{lemma}

\paragraph{Expected cost.} The final component of our analysis resides in upper-bounding the expected long-run average cost of $\tilde{\cal P}^{\ell}$, as formally stated in Lemma~\ref{lem:PO2_bound_cost} below. The complete proof of this result is provided in Appendix~\ref{app:proof_lem_PO2_bound_cost}. Toward obtaining the desired bound, we will first observe that when ${\cal A}_{\ell}$ does not occur, the specific policy we pick has negligible effects on the expected cost, since $\prsubpar{ \Theta_\ell }{ \bar{\cal A}_{\ell} } = O(\eps)$. As before, the difficult case to analyze is the one where ${\cal A}_{\ell}$ occurs. Here, our approach will provide separate arguments for the four commodity types, showing that even when conditioning on ${\cal A}_{\ell}$, the guarantees of Corollary~\ref{cor:construct_sub1_near} remain essentially unchanged.

\begin{lemma} \label{lem:PO2_bound_cost}
$\exsubpar{ \Theta_{\ell} }{ C( \tilde{\cal P}^{\ell} ) } \leq ( 1 + \frac{ 2\eps }{ 5 }  ) \cdot \frac{ 32/31 }{ \sqrt{2} \ln 2} \cdot \sum_{i \in \tilde{\cal V}_{ \ell } } C_i( \hat{T}_i )$.
\end{lemma}

\section{Concluding Remarks}

In this paper, we addressed several long-standing algorithmic questions surrounding the economic warehouse lot scheduling problem, for which  the best-known approximation guarantees have remained unchanged since the mid-1990s. Beyond breaking this deadlock, our work opens up broad directions for future research, in terms of both algorithmic design and structural understanding of dynamic replenishment policies. The remainder of this section highlights a selection of such directions, with the aim of stimulating further investigation into the computational and analytical challenges posed by capacity-constrained inventory systems.

\paragraph{The strategic version.}
As observed by \cite*[p.~51]{GallegoSS92} and corroborated by subsequent work, the vast majority of academic research around economic warehouse lot scheduling has focused on its ``tactical'' version, which is precisely the one
studied in this paper. At the same time, a related ``strategic'' version has attracted sustained attention (see, e.g., \cite{HodgsonL82, ParkY85, Hall88, RosenblattR90}), in which warehouse space is incorporated directly into the objective function rather than imposed as a hard constraint, aiming to minimize the combined cost $C({\cal P}) + V_{\max}({\cal P})$. It is not difficult to show that, by guessing the peak inventory level $V_{\max}({\cal P}^*)$ of an optimal policy ${\cal P}^*$ and enforcing it as a capacity constraint, any tactical $\alpha$-approximation  can be transformed into a strategic $(\sqrt{\alpha}+\epsilon)$-approximation. As a consequence, our work immediately yields a polynomial-time $(\sqrt{2-\frac{17}{5000}}+\epsilon)$-approximation for general strategic instances,  improving upon the classical $\sqrt{2}$-approximation due to \citet{Anily91} and \citet{GallegoQS96}. It would be particularly interesting to investigate whether the additional flexibility afforded by penalizing $V_{\max}({\cal P})$ in the objective, rather than explicitly constraining it, can be further exploited to obtain stronger approximation guarantees via refinements of our approach.

\paragraph{Efficient implementation?}
As formally stated in Theorem~\ref{thm:2_minus_delta}, our main algorithmic finding for general problem instances consists in devising an $O( | {\cal I} |^{\tilde{O}( 1/\eps^5)} \cdot 2^{ \tilde{O}( 1 / \eps^{35} ) } )$-time construction of a random capacity-feasible policy whose expected long-run average cost is within factor $2-\frac{17}{5000} + \eps$ of optimal. Here, it is important to point out that much of the resulting exponents arise from deliberately prioritizing conceptual clarity over an optimized implementation of our current algorithmic ideas and  preceding approximation scheme for constantly-many commodities~\citep{Segev26EWLSP_PTAS}. We believe that these exponents are by no means intrinsic, and could be substantially reduced through a careful integration of the existing algorithmic components. We view the development of more efficient implementations as a promising direction for future work, even though the potential inaccessibility of efforts along these lines is still unknown.

\paragraph{Derandomization?} On a different front, devising a deterministic construction of sub-$2$-approximate policies is a compelling open question to be tackled as part of future work. To this end, in relation to  Section~\ref{subsec:pow-rounding}, let us point out a well-known observation regarding the power-of-$2$ rounding procedure of \citet{TeoB01}: Up to scaling, the resulting random policies $\{ T_i^{ \Theta_{\ell, q} } \} _{i \in {\cal H}_{ \ell ,q}}$ jointly have only $O(n)$ possible realizations. That said, we are currently not seeing how a given set of realizations can be efficiently scaled while keeping the Po2-Synchronization Theorem intact. It would be interesting to investigate whether this direction could lead to derandomizing our approach, or perhaps, whether completely different ideas are required. With the scarcity of rigorous results around economic warehouse lot scheduling, we expect this direction to be particularly challenging.

\phantomsection
\addcontentsline{toc}{section}{References}
\bibliographystyle{plainnat}
\bibliography{BIB-Lot-Sizing}

\addtocontents{toc}{\protect\setcounter{tocdepth}{1}} 
\appendix
\section{Additional Proofs from Section~\ref{sec:sub2-approx}}

\subsection{Proof of Lemma~\ref{lem:avg_space_prefix}} \label{app:proof_lem_avg_space_prefix}

Our proof is based on a charging argument, comparing the total average space occupied by the commodities in ${\cal S}_{\mysuff}$ to the analogous quantity with respect to ${\cal V}_{\hat{\ell}}^{\eps}$, where $\hat{\ell}$ is the smallest index of a non-empty class in ${\cal S}_{\mypref}$. For this purpose, regarding the latter term, we observe  that
\begin{equation} \label{eqn:LB_sum_gammaI_l1}
\sum_{i \in {\cal V}_{\hat{\ell}}^{\eps}} \gamma_i \cdot \bar{I}( {\cal P}_i^{\eps} ) ~~\geq~~ \frac{ 1 }{ (1 + \eps)^{\hat{\ell} } } \cdot {\cal V} \ ,
\end{equation}
since $\gamma_i \cdot \bar{I}( {\cal P}^{\eps}_i ) \in ( \frac{1}{(1+\eps)^{\hat{\ell}}} \cdot {\cal V}, \frac{1}{(1+\eps)^{\hat{\ell}-1}} \cdot{\cal V}]$ for every commodity $i \in {\cal V}^{\eps}_{ \hat{\ell} }$, and since ${\cal V}^{\eps}_{ \hat{\ell} } \neq \emptyset$. On the other hand, to upper-bound the total average space occupied by ${\cal S}_{\mysuff}$-commodities, note that
\begin{align}
\sum_{\ell \in {\cal S}_{\mysuff}} \sum_{i \in {\cal V}_{\ell}^{\eps}} \gamma_i \cdot \bar{I}( {\cal P}_i^{\eps} )
& ~~\leq~~ \sum_{\ell \in {\cal S}_{\mysuff}} | {\cal V}_{\ell}^{\eps} | \cdot \frac{ 1 }{ (1 + \eps)^{ \ell - 1}} \cdot {\cal V} \label{eqn:proof_lem_avg_space_prefix_0} \\
& ~~\leq~~ \frac{ 100 }{ \eps^5 } \cdot \sum_{\ell \in {\cal S}_{\mysuff}} \frac{ 1 }{ (1 + \eps)^{ \ell - 1}} \cdot {\cal V} \nonumber \\
& ~~\leq~~ \frac{ 100 }{ \eps^5 } \cdot \sum_{\ell \geq \ell_{\mymid}+1}^{\infty} \frac{ 1 }{ (1 + \eps)^{ \ell - 1}} \cdot {\cal V} \label{eqn:proof_lem_avg_space_prefix_1} \\
& ~~=~~ \frac{ 100 }{ \eps^5 } \cdot \frac{ 1 + \eps }{ \eps } \cdot \frac{ 1 }{ (1 + \eps)^{ \ell_{\mymid} }} \cdot {\cal V} \nonumber \\
& ~~\leq~~ \frac{ 125 }{ \eps^6 } \cdot \frac{ 1 }{ (1 + \eps)^{ \Delta } }  \cdot \frac{ 1 }{ (1 + \eps)^{ \hat{\ell}  }} \cdot {\cal V} \label{eqn:proof_lem_avg_space_prefix_2}  \\
& ~~\leq~~ \eps \cdot \sum_{i \in {\cal V}_{\hat{\ell}}^{\eps}} \gamma_i \cdot \bar{I}( {\cal P}_i^{\eps} ) \label{eqn:proof_lem_avg_space_prefix_3} \\
& ~~\leq~~ \eps {\cal V} \ . \nonumber
\end{align}
Here, inequality~\eqref{eqn:proof_lem_avg_space_prefix_1} holds since ${\cal S}_{\mysuff} = \{ \ell_{\mymid} + 1, \ldots, \ell_M \}$, noting that this sequence of indices is increasing. Inequality~\eqref{eqn:proof_lem_avg_space_prefix_2} follows by recalling that $\ell_{\mymid}$ is the unique index up to which $\Delta$ of the classes ${\cal V}^{\eps}_{\ell_1}, \ldots, {\cal V}^{\eps}_{\ell_{\mymid}}$ are not empty, implying that $\ell_{\mymid} \geq \hat{\ell} + \Delta - 1$. Finally, inequality~\eqref{eqn:proof_lem_avg_space_prefix_3} is obtained by plugging in $\Delta = \lceil \log_{1 + \eps} ( \frac{ 125 }{ \eps^7 } ) \rceil$ and utilizing inequality~\eqref{eqn:LB_sum_gammaI_l1}.

\section{Additional Proofs from Section~\ref{sec:suffix_dense}}

\subsection{Proof of Lemma~\ref{lem:cost_U_dense}} \label{app:proof_lem_cost_U_dense}

We first observe that, since every commodity-vertex $i \in U$ has a unique adjacent edge $(i,\ell) \in {\cal E}^*$, in which case $C_{\myeoq,i}( \hat{T}_i ) = C_{\myeoq,i}( \hat{T}_{i \ell} ) = w_{i \ell}$, it follows that the total cost of the policies $\{ \hat{T}_i \}_{i \in U}$ can be written as $\sum_{i \in U}  C_{\myeoq,i}( \hat{T}_i ) = \sum_{(i,\ell) \in {\cal E}^*} w_{i \ell} = \opt({\cal I})$, where the second equality holds due to the optimality of ${\cal E}^*$. Therefore, to establish the desired claim, it suffices to argue that $\opt({\cal I}) \leq \sum_{i \in U} C_i( {\cal P}^{\eps}_i )$.

For this purpose, consider the set of edges ${\cal E}$, matching each commodity-vertex $i$ to the unique class-vertex $\ell$ for which $i \in {\cal V}_{\ell}^{\eps}$. This edge set is clearly a feasible solution to ${\cal I}$, implying that $\opt({\cal I}) \leq \sum_{(i,\ell) \in {\cal E}} w_{i \ell}$. We conclude the proof by showing that $w_{i \ell} \leq C_i( {\cal P}^{\eps}_i )$ for every edge $(i,\ell) \in {\cal E}$. Indeed, focusing on one such edge, since $i \in {\cal V}_{\ell}^{\eps}$, our definition of the volume class ${\cal V}_{\ell}^{\eps}$ in Section~\ref{subsec:alg_outline_2minus} guarantees that the  replenishment policy ${\cal P}^{\eps}_i$ satisfies
\[ \gamma_i \cdot \bar{I}( {\cal P}^{\eps}_i ) ~~\leq~~ \begin{cases}
\frac{1}{(1+\eps)^{\ell-1}} \cdot{\cal V}, \qquad & \text{when } \ell \in [L] \\
\frac{ \eps }{ n } \cdot {\cal V}, & \text{when } \ell = \infty
\end{cases} \]
The right-hand side of this inequality is identical to the right-hand side of the average space constraint~\eqref{eqn:occupied_space_constraint}. In contrast, the left-hand side here involves ${\cal P}^{\eps}_i$, which may not be a SOSI policy, whereas the left-hand side of constraint~\eqref{eqn:occupied_space_constraint} is only concerned with SOSI policies. However, by duplicating the arguments behind the proof of Lemma~\ref{lem:sosi_optimal}, it follows that there exists a SOSI policy $T_{i \ell}$ with $\bar{I}( T_{i \ell} ) \leq \bar{I}( {\cal P}^{\eps}_i )$ and $C_{\myeoq,i}( T_{i \ell} ) \leq C_i( {\cal P}^{\eps}_i )$. Consequently, $T_{i \ell}$ forms a feasible solution to the problem of minimizing $C_{\myeoq,i}( \cdot )$ subject to constraint~\eqref{eqn:occupied_space_constraint}, for which $\hat{T}_{i \ell}$ is an optimal solution. As a result, we indeed get $w_{ i \ell } = C_{\myeoq,i}( \hat{T}_{i \ell} ) \leq C_{\myeoq,i}( T_{i \ell} ) \leq C_i( {\cal P}^{\eps}_i )$.

\subsection{Proof of Lemma~\ref{lem:bound_space_suffix}} \label{app:proof_lem_bound_space_suffix}

To establish the desired claim, we observe that
\begin{align}
V_{\max}( {\cal P}_{\mysuff} ) & ~~=~~ \sum_{\ell \in {\cal S}_{\mysuff}} \sum_{i \in \tilde{\cal V}_{\ell}} \gamma_i \hat{T}_i \nonumber \\
& ~~\leq~~ 2 \cdot \sum_{\ell \in {\cal S}_{\mysuff}} \sum_{i \in \tilde{\cal V}_{\ell}} \left( \frac{1}{(1+\eps)^{\ell-1}} \cdot{\cal V} + \frac{ \eps }{ n } \cdot {\cal V} \right) \label{eqn:proof_space_suffix_1} \\
& ~~\leq~~ 2 \cdot \sum_{\ell \in {\cal S}_{\mysuff}} | {\cal V}_{\ell}^{\eps} | \cdot \frac{1}{(1+\eps)^{\ell-1}} \cdot{\cal V} + 2\eps {\cal V} \label{eqn:proof_space_suffix_2} \\
& ~~\leq~~ 4\eps {\cal V} \ . \label{eqn:proof_space_suffix_3}
\end{align}
Here, inequality~\eqref{eqn:proof_space_suffix_1} is an immediate consequence of property~\ref{prop:hatT_space}. Similarly, inequality~\eqref{eqn:proof_space_suffix_2} follows from property~\ref{prop:tildeV_size}, stating in particular that $|\tilde{\cal V}_{ \ell }| = |{\cal V}_{\ell}^{\eps} |$ for every $\ell \in {\cal S}_{\mysuff}$. Finally, inequality~\eqref{eqn:proof_space_suffix_3} is obtained by recalling that $\sum_{\ell \in {\cal S}_{\mysuff}} | {\cal V}_{\ell}^{\eps} | \cdot \frac{1}{(1+\eps)^{\ell-1}} \cdot{\cal V}$ is precisely the term appearing in~\eqref{eqn:proof_lem_avg_space_prefix_0}, which has already been upper-bounded by $\eps {\cal V}$ within the proof of Lemma~\ref{lem:avg_space_prefix}.

\subsection{Proof of Lemma~\ref{lem:main_result_Vell_light}} \label{app:proof_lem_main_result_Vell_light}

In terms of cost, due to setting $\tilde{\cal P}^{\ell}_i = \hat{T}_i$ for every commodity $i \in \tilde{\cal V}_{ \ell }$, we immediately get $C( \tilde{\cal P}^{\ell} ) = \sum_{i \in \tilde{\cal V}_{ \ell } } C_i ( \tilde{\cal P}^{\ell}_i ) = \sum_{i \in \tilde{\cal V}_{ \ell } } C_{\myeoq,i} ( \hat{T}_i )$. To derive an upper bound on the peak space requirement of $\tilde{\cal P}^{\ell}$, by recalling that $\{ \hat{T}_i \}_{ i \in \tilde{\cal V}_{ \ell } }$ are SOSI policies, we have
\begin{align}
V_{\max}( \tilde{\cal P}^{\ell} ) & ~~=~~ \sum_{i \in \tilde{\cal V}_{ \ell } } \gamma_i \hat{T}_i \nonumber \\
& ~~=~~ 2 \cdot \left( \sum_{i \in {\cal L}_{ \ell } } \gamma_i \cdot \bar{I}( \hat{T}_i ) + \sum_{i \in {\cal H}_{ \ell } } \gamma_i \cdot \bar{I}( \hat{T}_i ) \right) \nonumber \\
& ~~\leq~~ 2 \cdot \left( \frac{ 3 }{ 4 } \cdot |{\cal L}_{ \ell }| + |{\cal H}_{ \ell }| \right) \cdot \frac{1}{(1+\eps)^{\ell-1}} \cdot{\cal V} \label{eqn:proof_lem_main_result_Vell_light_eq2} \\
& ~~\leq~~ \frac{ 7 }{ 4 } \cdot | \tilde{\cal V}_{\ell} | \cdot \frac{1}{(1+\eps)^{\ell-1}} \cdot{\cal V} \ . \label{eqn:proof_lem_main_result_Vell_light_eq3}
\end{align}
Here, to obtain inequality~\eqref{eqn:proof_lem_main_result_Vell_light_eq2}, we note that $\gamma_i \cdot \bar{I}( \hat{T}_i ) \leq \frac{ 3 }{ 4 } \cdot \frac{1}{(1+\eps)^{\ell-1}} \cdot{\cal V}$ for every light commodity, whereas  $\gamma_i \cdot \bar{I}( \hat{T}_i ) \in (\frac{ 3 }{ 4 } \cdot \frac{1}{(1+\eps)^{\ell-1}} \cdot{\cal V}, \frac{1}{(1+\eps)^{\ell-1}} \cdot{\cal V}]$ for every heavy commodity. Inequality~\eqref{eqn:proof_lem_main_result_Vell_light_eq3} follows by recalling that $|{\cal L}_{ \ell }| + |{\cal H}_{ \ell }| = | \tilde{\cal V}_{ \ell } |$ and that $| {\cal L}_{ \ell } | \geq \frac{ | \tilde{\cal V}_{ \ell } | }{ 2 }$, by the hypothesis of Lemma~\ref{lem:main_result_Vell_light}.

\section{Additional Proofs from Section~\ref{sec:po2-sync}}

\subsection{Proof of Lemma~\ref{lem:construction_pair}} \label{app:proof_lem_construction_pair}

For ease of presentation, rather than directly considering the regime where $\frac{ \gamma_A T_A }{ \gamma_B T_B } \in 1 \pm \eps$, we will assume that  $\gamma_A T_A = \gamma_B T_B$. One can easily go through our proof with $\tilde{\gamma}_A = \frac{ \gamma_B T_B }{ T_A }$, which obviously satisfies $\tilde{\gamma}_A T_A = \gamma_B T_B$. Subsequently, since $\tilde{\gamma}_A \in (1 \pm \eps) \cdot \gamma_A$, by reverting to the original coefficient $\gamma_A$, we increase the joint occupied space by a factor of at most $1 + \eps$. With this assumption, we will construct dynamic replenishment policies ${\cal P}_A$ and ${\cal P}_B$ that satisfy
\[ V_{\max}( {\cal P}_A, {\cal P}_B ) ~~\leq~~ \frac{ 7 }{ 4 } \cdot ( \gamma_A \cdot \bar{I}( T_A ) + \gamma_B \cdot \bar{I}( T_B ) ) \quad \text{and} \quad \max \left\{ \frac{ C_A( {\cal P}_A ) }{ C_{\myeoq,A}( T_A )}, \frac{ C_B( {\cal P}_B ) }{ C_{\myeoq,B}( T_B)} \right\} ~~\leq~~ \frac{ 32 }{ 31} \ . \]
For convenience, we assume without loss of generality that $T_B \leq T_A = 1$. Given that $T_A / T_B$ is an integer power of $2$, our proof considers six different cases: $T_B = 1$, $T_B = \frac{ 1 }{ 2 }$, $T_B = \frac{ 1 }{ 4 }$, $T_B = \frac{ 1 }{ 8 }$, $T_B = \frac{ 1 }{ 16 }$, and $T_B = \frac{ 1 }{ 2^k }$ for some $k \geq 5$.

\paragraph{Case 1: \boldmath{${T_B = 1}$}.} We begin by considering the simplest possible scenario, since it is an instructive opportunity to explain our notation and terminology:
\begin{itemize}
    \item {\em The policy ${\cal P}_A$}: $A$-orders will be placed at the integer time points $0, 1, 2, \ldots$.

    \item {\em The policy ${\cal P}_B$}: $B$-orders will also have a gap of $1$ between successive orders. However, they will be offset by $0.5$, meaning that beyond a single order at time $0$, we place $B$-orders at $0.5, 1.5, 2.5, \ldots$.

    \item {\em Peak space requirement:} Since ${\cal P}_A$ and ${\cal P}_B$ have a joint cycle of length $1$, to detect where their peak space requirement is located, it suffices to examine a unit-length interval, say $[0.5,1.5)$. Moreover, since the inventory level of each commodity is decreasing between successive orders, the joint peak space requirement of ${\cal P}_A$ and ${\cal P}_B$ is attained at either $0.5$ or $1$. In these two points, we have:
    \begin{itemize}
        \item $V( {\cal P}_A, {\cal P}_B, 0.5 ) = \frac{ 1 }{ 2 } \cdot \gamma_A  T_A + \gamma_B  T_B  = \frac{ 3 }{ 2 } \cdot ( \gamma_A \cdot \bar{I}( T_A ) + \gamma_B \cdot \bar{I}( T_B ) )$.

        \item $V( {\cal P}_A, {\cal P}_B, 1 ) =  \gamma_A  T_A + \frac{ 1 }{ 2 } \cdot \gamma_B  T_B  = \frac{ 3 }{ 2 } \cdot ( \gamma_A \cdot \bar{I}( T_A ) + \gamma_B \cdot \bar{I}( T_B ) )$.
    \end{itemize}
    In both cases, we make use of $\gamma_A \cdot \bar{I}( T_A ) = \gamma_B \cdot \bar{I}( T_B )$, or equivalently $\gamma_A  T_A = \gamma_B   T_B$, which is an important relation to remember throughout this analysis.
    \item {\em Cost}: In terms of cost, since ${\cal P}_A$ and ${\cal P}_B$ are identical to $T_A$ and $T_B$ up to offsets, we have $C_A( {\cal P}_A ) = C_{\myeoq,A}( T_A )$ and $C_B( {\cal P}_B ) = C_{\myeoq,B}( T_B )$.

    \item {\em Summary}: $V_{\max}( {\cal P}_A, {\cal P}_B ) = \frac{ 3 }{ 2 } \cdot ( \gamma_A \cdot \bar{I}( T_A ) + \gamma_B \cdot \bar{I}( T_B ) )$ and $\max \{ \frac{ C_A( {\cal P}_A ) }{ C_{\myeoq,A}( T_A ) }, \frac{ C_B( {\cal P}_B ) }{ C_{\myeoq,B}( T_B ) } \} = 1$.
\end{itemize}
Figure~\ref{fig:case1} illustrates the resulting inventory dynamics, where the blue and red curves represent the individual inventory levels of commodities $A$ and $B$, respectively, and the violet curve depicts their total occupied space. For convenience, the $y$-axis is normalized by a factor of $\gamma_A T_A = \gamma_B T_B$, showing a reduced aggregate peak of $\frac{ 3 }{ 2}$.

\begin{figure}[htbp!]
    \centering

    \begin{minipage}[b]{0.48\textwidth}
        \centering
        \resizebox{\linewidth}{!}{%
        \begin{tikzpicture}[xscale=6, yscale=2]
            \def\H{1.5}

            \draw[->] (0.5,0) -- (1.6,0) node[right] {$t$};
            \draw[->] (0.5,-0.1) -- (0.5,2.5) node[above] {occupied space};

            \draw (0.5cm+1pt, 1.5*\H) -- (0.5cm-1pt, 1.5*\H) node[left] {$\frac{3}{2}$};
            \draw (0.5cm+1pt, \H) -- (0.5cm-1pt, \H) node[left] {$1$};
            \draw (0.5cm+1pt, 0.5*\H) -- (0.5cm-1pt, 0.5*\H) node[left] {$\frac{1}{2}$};

            \draw[dashed, gray!40] (0.5, 1.5*\H) -- (1.5, 1.5*\H);
            \draw[dotted, gray!60] (0.5, \H) -- (1.5, \H);

            \draw[thick, blue!60!cyan] (0.5, 0.5*\H) -- (1.0, 0) -- (1.0, \H) -- (1.5, 0.5*\H);

            \draw[thick, red!70!orange] (0.5, 0) -- (0.5, \H) -- (1.5, 0);

            \draw[very thick, violet]
                (0.5, 1.5*\H) -- (1.0, 0.5*\H) -- (1.0, 1.5*\H) -- (1.5, 0.5*\H);

            \foreach \x in {0.5, 1.0, 1.5}
                \draw (\x, 1pt) -- (\x, -1pt) node[below] {\x};

            \matrix [draw, fill=white, anchor=north east, font=\footnotesize, inner sep=3pt] at (1.6, 2.4) {
                \draw[very thick, violet] (0,0) -- (0.4,0) node[right, text=black] {Total space}; \\
                \draw[thick, blue!60!cyan] (0,0) -- (0.4,0) node[right, text=black] {Item A}; \\
                \draw[thick, red!70!orange] (0,0) -- (0.4,0) node[right, text=black] {Item B}; \\
            };
        \end{tikzpicture}%
        }
        \caption{Case 1}
        \label{fig:case1}
    \end{minipage}
    \hfill 
    \begin{minipage}[b]{0.48\textwidth}
        \centering
        \resizebox{\linewidth}{!}{%
        \begin{tikzpicture}[xscale=6, yscale=2]
            \def\H{1.5}

            \draw[->] (0.333,0) -- (1.45,0) node[right] {$t$};
            \draw[->] (0.333,-0.1) -- (0.333,3.0) node[above] {occupied space};

            \draw (0.333cm+1pt, 1.666*\H) -- (0.333cm-1pt, 1.666*\H) node[left] {$\frac{5}{3}$};
            \draw (0.333cm+1pt, \H) -- (0.333cm-1pt, \H) node[left] {$1$};
            \draw (0.333cm+1pt, 1.166*\H) -- (0.333cm-1pt, 1.166*\H) node[left] {$\frac{7}{6}$};

            \draw[dashed, gray!40] (0.333, 1.666*\H) -- (1.333, 1.666*\H);
            \draw[dotted, gray!60] (0.333, \H) -- (1.333, \H);

            \draw[thick, blue!60!cyan] (0.333, 1.0) -- (1.0, 0) -- (1.0, \H) -- (1.333, 1.0);

            \draw[thick, red!70!orange]
                (0.333, 0) -- (0.333, \H) -- (0.833, 0) -- (0.833, \H) -- (1.333, 0);

            \draw[very thick, violet]
                (0.333, 1.666*\H) -- (0.833, 0.166*\H) -- (0.833, 1.166*\H)
                -- (1.0, 0.666*\H) -- (1.0, 1.666*\H)
                -- (1.333, 0.666*\H);

            \draw (0.333, 1pt) -- (0.333, -1pt) node[below] {$\frac{1}{3}$};
            \draw (0.833, 1pt) -- (0.833, -1pt) node[below] {$\frac{5}{6}$};
            \draw (1.0, 1pt) -- (1.0, -1pt) node[below] {$1$};
            \draw (1.333, 1pt) -- (1.333, -1pt) node[below] {$1\frac{1}{3}$};

            \matrix [draw, fill=white, anchor=north east, font=\footnotesize, inner sep=3pt] at (1.45, 2.9) {
                \draw[very thick, violet] (0,0) -- (0.4,0) node[right, text=black] {Total space}; \\
                \draw[thick, blue!60!cyan] (0,0) -- (0.4,0) node[right, text=black] {Item A}; \\
                \draw[thick, red!70!orange] (0,0) -- (0.4,0) node[right, text=black] {Item B}; \\
            };
        \end{tikzpicture}%
        }
        \caption{Case 2}
        \label{fig:case2}
    \end{minipage}
\end{figure}

\paragraph{Case 2: \boldmath{${T_B = \frac{ 1 }{ 2 }}$}.} As schematically illustrated in Figure~\ref{fig:case2}, our construction proceeds as follows:
\begin{itemize}
    \item {\em The policy ${\cal P}_A$}: $A$-orders will be placed at $0, 1, 2, \ldots$.

    \item {\em The policy ${\cal P}_B$}: $B$-orders will be placed at $\frac{1}{3}, \frac{5}{6}, 1\frac{1}{3}, 1\frac{5}{6}, \ldots$. From this point on, we will not be mentioning the obvious singular order at time $0$.

    \item {\em Peak space requirement:} Once again,  ${\cal P}_A$ and ${\cal P}_B$ have a joint cycle of length $1$, and we examine the unit-length interval $[\frac{1}{3},1\frac{1}{3})$, where there are three ordering points to be tested, $\frac{1}{3}$, $\frac{5}{6}$, and $1$:
    \begin{itemize}
        \item $V( {\cal P}_A, {\cal P}_B, \frac{1}{3} ) = \frac{ 2 }{ 3 } \cdot \gamma_A  T_A + \gamma_B  T_B  = \frac{ 5 }{ 3 } \cdot ( \gamma_A \cdot \bar{I}( T_A ) + \gamma_B \cdot \bar{I}( T_B ) )$.

        \item $V( {\cal P}_A, {\cal P}_B, \frac{5}{6} ) = \frac{ 1 }{ 6 } \cdot \gamma_A  T_A + \gamma_B  T_B  = \frac{ 7 }{ 6 } \cdot ( \gamma_A \cdot \bar{I}( T_A ) + \gamma_B \cdot \bar{I}( T_B ) )$.

        \item $V( {\cal P}_A, {\cal P}_B, 1 ) =  \gamma_A  T_A + \frac{ 2 }{ 3 } \cdot \gamma_B  T_B  = \frac{ 5 }{ 3 } \cdot ( \gamma_A \cdot \bar{I}( T_A ) + \gamma_B \cdot \bar{I}( T_B ) )$.
    \end{itemize}

    \item {\em Cost}: Again,  ${\cal P}_A$ and ${\cal P}_B$ are identical to $T_A$ and $T_B$ up to offsets, and therefore $C_A( {\cal P}_A ) = C_{\myeoq,A}( T_A )$ and $C_B( {\cal P}_B ) = C_{\myeoq,B}( T_B )$.

    \item {\em Summary}: $V_{\max}( {\cal P}_A, {\cal P}_B ) = \frac{ 5 }{ 3 } \cdot ( \gamma_A \cdot \bar{I}( T_A ) + \gamma_B \cdot \bar{I}( T_B ) )$ and $\max \{ \frac{ C_A( {\cal P}_A ) }{ C_{\myeoq,A}( T_A ) }, \frac{ C_B( {\cal P}_B ) }{ C_{\myeoq,B}( T_B ) } \} = 1$.
\end{itemize}

\paragraph{Case 3: \boldmath{${T_B = \frac{ 1 }{ 4 }}$}.} As schematically illustrated in Figure~\ref{fig:case3}, our construction proceeds as follows:
\begin{itemize}
    \item {\em The policy ${\cal P}_A$}: For the first time, ${\cal P}_A$ will not be identical to $T_A$, but rather a $\frac{31}{32}$-scaling of this policy. As such, $A$-orders will be placed at $0, \frac{31}{32}, \frac{62}{32}, \frac{93}{32} \ldots$.

    \item {\em The policy ${\cal P}_B$}: Here, for the first time, we will be employing a non-SOSI policy. Specifically, the cycle length of ${\cal P}_B$ will be $\frac{ 31 }{ 32}$, which is similar to that of ${\cal P}_A$, and its offset will be $\frac{ 5 }{ 32 }$. Each such cycle will be filled by four $B$-orders, the first of length $\frac{7}{8} \cdot T_B = \frac{ 7 }{ 32 }$, and the next three of length $T_B = \frac{ 1 }{ 4 }$, noting that these quantities indeed sum up to $\frac{ 31 }{ 32}$.

    \item {\em Peak space requirement:} ${\cal P}_A$ and ${\cal P}_B$ have a joint cycle of length $\frac{ 31 }{ 32}$, and for convenience, we examine the interval $[\frac{ 5 }{ 32 }, \frac{ 36 }{ 32 })$, where there are five ordering points to be tested, $\frac{ 5 }{ 32 }$, $\frac{ 12 }{ 32 }$, $\frac{ 20 }{ 32 }$, $\frac{ 28 }{ 32 }$, and $\frac{ 31 }{ 32 }$:
    \begin{itemize}
        \item $V( {\cal P}_A, {\cal P}_B, \frac{ 5 }{ 32 } ) = \frac{ 26 }{ 31 } \cdot \frac{ 31 }{ 32 } \cdot \gamma_A  T_A + \frac{ 7 }{ 8 } \cdot \gamma_B  T_B  = \frac{ 27 }{ 16 } \cdot ( \gamma_A \cdot \bar{I}( T_A ) + \gamma_B \cdot \bar{I}( T_B ) )$.

        \item $V( {\cal P}_A, {\cal P}_B, \frac{ 12 }{ 32 } ) = \frac{ 19 }{ 31 } \cdot \frac{ 31 }{ 32 } \cdot \gamma_A  T_A + \gamma_B  T_B  = \frac{ 51 }{ 32 } \cdot ( \gamma_A \cdot \bar{I}( T_A ) + \gamma_B \cdot \bar{I}( T_B ) )$.

        \item $V( {\cal P}_A, {\cal P}_B, \frac{ 20 }{ 32 } ) = \frac{ 11 }{ 31 } \cdot \frac{ 31 }{ 32 } \cdot \gamma_A  T_A + \gamma_B  T_B  = \frac{ 43 }{ 32 } \cdot ( \gamma_A \cdot \bar{I}( T_A ) + \gamma_B \cdot \bar{I}( T_B ) )$.

        \item $V( {\cal P}_A, {\cal P}_B, \frac{28}{32} ) = \frac{ 3 }{ 31 } \cdot \frac{ 31 }{ 32 } \cdot \gamma_A  T_A + \gamma_B  T_B  = \frac{ 34 }{ 32 } \cdot ( \gamma_A \cdot \bar{I}( T_A ) + \gamma_B \cdot \bar{I}( T_B ) )$.

        \item $V( {\cal P}_A, {\cal P}_B, \frac{ 31 }{ 32 } ) = \frac{ 31 }{ 32 } \cdot \gamma_A  T_A + \frac{ 5 }{ 8 } \cdot \gamma_B  T_B  = \frac{ 51 }{ 32 } \cdot ( \gamma_A \cdot \bar{I}( T_A ) + \gamma_B \cdot \bar{I}( T_B ) )$.
    \end{itemize}

    \item {\em Cost}: Since ${\cal P}_A$ is a $\frac{ 31 }{ 32 }$-scaling of $T_A$, we have $C_A( {\cal P}_A ) \leq \frac{ 32 }{ 31 } \cdot C_{\myeoq,A}( T_A )$. In addition, given the construction of ${\cal P}_B$, this policy is a $\frac{7}{8}$-scaling of $T_B$ on a $\frac{7}{31}$-fraction of its cycle, and it coincides with $T_B$ on the remaining fraction, meaning that
    \[ C_B( {\cal P}_B ) ~~\leq~~ \left( \frac{7}{31} \cdot \frac{ 8 }{ 7 } + \frac{24}{31}  \right) \cdot C_{\myeoq,B}( T_B ) ~~=~~ \frac{ 32 }{ 31 } \cdot C_{\myeoq,B}( T_B )  \ .  \]

    \item {\em Summary}: $V_{\max}( {\cal P}_A, {\cal P}_B ) = \frac{ 27 }{ 16 } \cdot ( \gamma_A \cdot \bar{I}( T_A ) + \gamma_B \cdot \bar{I}( T_B ) )$ and $\max \{ \frac{ C_A( {\cal P}_A ) }{ C_{\myeoq,A}( T_A ) }, \frac{ C_B( {\cal P}_B ) }{ C_{\myeoq,B}( T_B ) } \} \leq \frac{ 32 }{ 31 }$.
\end{itemize}

\begin{figure}[htbp!]
    \centering

    \begin{minipage}[b]{0.48\textwidth}
        \centering
        \resizebox{\linewidth}{!}{%
        \begin{tikzpicture}[xscale=8, yscale=2.5]

            \def\H{1.5}

            \draw[->] (0.15,0) -- (1.4,0) node[right] {$t$};
            \draw[->] (0.156,-0.1) -- (0.156,3.0) node[above] {occupied space};

            \draw (0.156cm+1pt, 1.6875*\H) -- (0.156cm-1pt, 1.6875*\H) node[left] {$27/16$};
            \draw (0.156cm+1pt, 1.59375*\H) -- (0.156cm-1pt, 1.59375*\H) node[left] {$51/32$};
            \draw (0.156cm+1pt, \H) -- (0.156cm-1pt, \H) node[left] {$1$};

            \draw[dashed, gray!40] (0.156, 1.6875*\H) -- (1.125, 1.6875*\H);
            \draw[dotted, gray!60] (0.156, \H) -- (1.125, \H);

            \draw[thick, blue!60!cyan]
                (0.15625, 0.8125*\H) -- (0.96875, 0) -- (0.96875, 0.96875*\H) -- (1.125, 0.8125*\H);

            \draw[thick, red!70!orange]
                (0.15625, 0.875*\H) -- (0.375, 0) -- (0.375, \H)
                -- (0.625, 0) -- (0.625, \H)
                -- (0.875, 0) -- (0.875, \H)
                -- (1.125, 0);

            \draw[very thick, violet]
                (0.15625, 1.6875*\H)
                -- (0.375, 0.59375*\H) -- (0.375, 1.59375*\H)
                -- (0.625, 0.34375*\H) -- (0.625, 1.34375*\H)
                -- (0.875, 0.09375*\H) -- (0.875, 1.09375*\H)
                -- (0.96875, 0.625*\H) -- (0.96875, 1.59375*\H)
                -- (1.125, 0.8125*\H);

            \draw (0.15625, 1pt) -- (0.15625, -1pt) node[below] {$\frac{5}{32}$};
            \draw (0.375, 1pt) -- (0.375, -1pt) node[below] {$\frac{12}{32}$};
            \draw (0.625, 1pt) -- (0.625, -1pt) node[below] {$\frac{20}{32}$};
            \draw (0.875, 1pt) -- (0.875, -1pt) node[below] {$\frac{28}{32}$};
            \draw (0.96875, 1pt) -- (0.96875, -1pt) node[below] {$\frac{31}{32}$};
            \draw (1.125, 1pt) -- (1.125, -1pt) node[below] {$\frac{36}{32}$};

            \matrix [draw, fill=white, anchor=north east, font=\footnotesize, inner sep=3pt] at (1.35, 2.9) {
                \draw[very thick, violet] (0,0) -- (0.4,0) node[right, text=black] {Total space}; \\
                \draw[thick, blue!60!cyan] (0,0) -- (0.4,0) node[right, text=black] {Item A}; \\
                \draw[thick, red!70!orange] (0,0) -- (0.4,0) node[right, text=black] {Item B}; \\
            };
        \end{tikzpicture}%
        }
        \caption{Case 3}
        \label{fig:case3}
    \end{minipage}
    \hfill 
    \begin{minipage}[b]{0.48\textwidth}
        \centering
        \resizebox{\linewidth}{!}{%
        \begin{tikzpicture}[xscale=8, yscale=2.5]

            \def\H{1.5}

            \draw[->] (0.09,0) -- (1.4,0) node[right] {$t$};
            \draw[->] (0.09375,-0.1) -- (0.09375,3.0) node[above] {occupied space};

            \draw (0.09375cm+1pt, 1.7195*\H) -- (0.09375cm-1pt, 1.7195*\H) node[left] {$2201/1280$};
            \draw (0.09375cm+1pt, \H) -- (0.09375cm-1pt, \H) node[left] {$1$};

            \draw[dashed, gray!40] (0.09375, 1.7195*\H) -- (1.0625, 1.7195*\H);
            \draw[dotted, gray!60] (0.09375, \H) -- (1.0625, \H);

            \draw[thick, blue!60!cyan]
                (0.09375, 0.875*\H) -- (0.96875, 0) -- (0.96875, 0.96875*\H) -- (1.0625, 0.875*\H);

            \draw[thick, red!70!orange]
                (0.09375, 0.84375*\H) -- (0.1992, 0) -- (0.1992, 0.95*\H)
                -- (0.3179, 0) -- (0.3179, \H)
                -- (0.4429, 0) -- (0.4429, \H)
                -- (0.5679, 0) -- (0.5679, \H)
                -- (0.6929, 0) -- (0.6929, \H)
                -- (0.8179, 0) -- (0.8179, 0.95625*\H)
                -- (0.9375, 0) -- (0.9375, \H)
                -- (1.0625, 0);

            \draw[very thick, violet]
                (0.09375, 1.71875*\H)
                -- (0.1992, 0.77*\H) -- (0.1992, 1.7195*\H) 
                -- (0.3179, 0.77*\H) -- (0.3179, 1.65*\H)
                -- (0.4429, 0.52*\H) -- (0.4429, 1.52*\H)
                -- (0.5679, 0.39*\H) -- (0.5679, 1.39*\H)
                -- (0.6929, 0.27*\H) -- (0.6929, 1.27*\H)
                -- (0.8179, 0.14*\H) -- (0.8179, 1.1*\H)
                -- (0.9375, 0.03*\H) -- (0.9375, 1.03*\H)
                -- (0.96875, 0.75*\H) -- (0.96875, 1.71875*\H)
                -- (1.0625, 0.875*\H);

            \draw (0.09375, 1pt) -- (0.09375, -1pt) node[below] {$\frac{3}{32}$};
            \draw (0.1992, 1pt) -- (0.1992, -1pt) node[below] {$\frac{51}{256}$};
            \draw (0.96875, 1pt) -- (0.96875, -1pt) node[below] {$\frac{31}{32}$};
            \draw (1.0625, 1pt) -- (1.0625, -1pt) node[below] {$\frac{34}{32}$};

            \matrix [draw, fill=white, anchor=north east, font=\footnotesize, inner sep=3pt] at (1.35, 2.9) {
                \draw[very thick, violet] (0,0) -- (0.4,0) node[right, text=black] {Total space}; \\
                \draw[thick, blue!60!cyan] (0,0) -- (0.4,0) node[right, text=black] {Item A}; \\
                \draw[thick, red!70!orange] (0,0) -- (0.4,0) node[right, text=black] {Item B}; \\
            };
        \end{tikzpicture}%
        }
        \caption{Case 4}
        \label{fig:case4}
    \end{minipage}
\end{figure}

\paragraph{Case 4: \boldmath{${T_B = \frac{ 1 }{ 8 }}$}.} As schematically illustrated in Figure~\ref{fig:case4}, our construction proceeds as follows:
\begin{itemize}
    \item {\em The policy ${\cal P}_A$}: Similarly to case~3, the policy ${\cal P}_A$ will be a $\frac{31}{32}$-scaling of $T_A$, with $A$-orders placed at $0, \frac{31}{32}, \frac{62}{32}, \frac{93}{32} \ldots$.

    \item {\em The policy ${\cal P}_B$}: Similarly to case~3, the cycle length of ${\cal P}_B$ will be $\frac{ 31 }{ 32}$ as well, with an offset of $\frac{ 3 }{ 32 }$. Each such cycle will be filled from left to right by eight $B$-orders of length: $\frac{ 27 }{ 32 } \cdot T_B$, $\frac{ 19 }{ 20 } \cdot T_B$, $T_B$, $T_B$, $T_B$, $T_B$, $\frac{ 153 }{ 160 } \cdot T_B$, and $T_B$. Since $T_B = \frac{ 1 }{ 8 }$, these quantities indeed sum up to $\frac{ 31 }{ 32}$.

    \item {\em Peak space requirement:} ${\cal P}_A$ and ${\cal P}_B$ have a joint cycle of length $\frac{ 31 }{ 32}$, and for convenience, we examine the interval $[\frac{ 3 }{ 32 }, \frac{ 34 }{ 32 })$, where there are nine ordering points to be tested, $\frac{ 3 }{ 32 }$, $\frac{ 51 }{ 256 }$, $\frac{ 407 }{ 1280 }$, $\frac{ 567 }{ 1280 }$, $\frac{ 727 }{ 1280 }$, $\frac{ 887 }{ 1280 }$, $\frac{ 1047 }{ 1280 }$, $\frac{ 15 }{ 16 }$, and $\frac{ 31 }{ 32 }$:
    \begin{itemize}
        \item $V( {\cal P}_A, {\cal P}_B, \frac{3 }{ 32 } ) = \frac{ 28 }{ 31 } \cdot \frac{ 31 }{ 32 } \cdot \gamma_A  T_A + \frac{ 27 }{ 32 } \cdot \gamma_B  T_B  = \frac{ 55 }{ 32 } \cdot ( \gamma_A \cdot \bar{I}( T_A ) + \gamma_B \cdot \bar{I}( T_B ) )$.

        \item $V( {\cal P}_A, {\cal P}_B, \frac{ 51 }{ 256 } ) = (1 - \frac{ 51/256 }{31/32}) \cdot \frac{ 31 }{32}\cdot \gamma_A  T_A + \frac{ 19 }{ 20 } \cdot \gamma_B  T_B  = \frac{ 2201 }{ 1280 } \cdot ( \gamma_A \cdot \bar{I}( T_A ) + \gamma_B \cdot \bar{I}( T_B ) )$.

        \item $V( {\cal P}_A, {\cal P}_B, \frac{ 407 }{ 1280 } ) = (1 - \frac{ 407/1280 }{31/32}) \cdot \frac{ 31 }{32}\cdot \gamma_A  T_A + \gamma_B  T_B  = \frac{ 2113 }{ 1280 } \cdot ( \gamma_A \cdot \bar{I}( T_A ) + \gamma_B \cdot \bar{I}( T_B ) )$.

        \item $V( {\cal P}_A, {\cal P}_B, \frac{ 567 }{ 1280 } ) = (1 - \frac{ 567/1280 }{31/32}) \cdot \frac{ 31 }{32}\cdot \gamma_A  T_A + \gamma_B  T_B  = \frac{ 1953 }{ 1280 } \cdot ( \gamma_A \cdot \bar{I}( T_A ) + \gamma_B \cdot \bar{I}( T_B ) )$.

        \item $V( {\cal P}_A, {\cal P}_B, \frac{ 727 }{ 1280 } ) = (1 - \frac{ 727/1280 }{31/32}) \cdot \frac{ 31 }{32}\cdot \gamma_A  T_A + \gamma_B  T_B  = \frac{ 1793 }{ 1280 } \cdot ( \gamma_A \cdot \bar{I}( T_A ) + \gamma_B \cdot \bar{I}( T_B ) )$.

        \item $V( {\cal P}_A, {\cal P}_B, \frac{ 887 }{ 1280 } ) = (1 - \frac{ 887/1280 }{31/32}) \cdot \frac{ 31 }{32}\cdot \gamma_A  T_A + \gamma_B  T_B  = \frac{ 1633 }{ 1280 } \cdot ( \gamma_A \cdot \bar{I}( T_A ) + \gamma_B \cdot \bar{I}( T_B ) )$.

        \item $V( {\cal P}_A, {\cal P}_B, \frac{ 1047 }{ 1280 } ) = (1 - \frac{ 1047/1280 }{31/32}) \cdot \frac{ 31 }{32}\cdot \gamma_A  T_A + \frac{ 153 }{ 160 } \cdot \gamma_B  T_B  = \frac{ 1417 }{ 1280 } \cdot ( \gamma_A \cdot \bar{I}( T_A ) + \gamma_B \cdot \bar{I}( T_B ) )$.

        \item $V( {\cal P}_A, {\cal P}_B, \frac{ 15 }{ 16 } ) = (1 - \frac{ 15/16 }{31/32}) \cdot \frac{ 31 }{32}\cdot \gamma_A  T_A +  \gamma_B  T_B  = \frac{ 33 }{ 32 } \cdot ( \gamma_A \cdot \bar{I}( T_A ) + \gamma_B \cdot \bar{I}( T_B ) )$.

        \item $V( {\cal P}_A, {\cal P}_B, \frac{ 31 }{ 32 } ) = \frac{ 31 }{ 32 } \cdot \gamma_A  T_A + \frac{ 3 }{ 4 } \cdot \gamma_B  T_B  = \frac{ 55 }{ 32 } \cdot ( \gamma_A \cdot \bar{I}( T_A ) + \gamma_B \cdot \bar{I}( T_B ) )$.
    \end{itemize}

    \item {\em Cost}: Since ${\cal P}_A$ is a $\frac{ 31 }{ 32 }$-scaling of $T_A$, we have $C_A( {\cal P}_A ) \leq \frac{ 32 }{ 31 } \cdot C_{\myeoq,A}( T_A )$.  In addition, given the construction of ${\cal P}_B$, this policy is a $\frac{27}{32}$-scaling of $T_B$ on a $\frac{ 27 / 256 }{ 31/32 }$-fraction of its cycle. Along the same lines, ${\cal P}_B$ is a $\frac{19}{20}$-scaling of $T_B$ on a $\frac{ 19 / 160}{ 31/32 }$-fraction, so on and so forth. Therefore,
    \begin{align*}
    C_B( {\cal P}_B ) & ~~\leq~~ \left( \frac{ 27 / 256 }{ 31/32 } \cdot \frac{ 32 }{ 27 } + \frac{ 19 / 160}{ 31/32 } \cdot \frac{ 20 }{ 19 } + 5 \cdot \frac{ 1 / 8}{ 31/32 } + \frac{ 153 / 1280 }{ 31/32 } \cdot \frac{ 160 }{ 153 } \right) \cdot C_{\myeoq,B}( T_B ) \\
    & ~~=~~   \frac{ 32 }{ 31 } \cdot C_{\myeoq,B}( T_B ) \ .
    \end{align*}

    \item {\em Summary}: $V_{\max}( {\cal P}_A, {\cal P}_B ) = \frac{ 2201 }{ 1280 } \cdot ( \gamma_A \cdot \bar{I}( T_A ) + \gamma_B \cdot \bar{I}( T_B ) )$ and $\max \{ \frac{ C_A( {\cal P}_A ) }{ C_{\myeoq,A}( T_A ) }, \frac{ C_B( {\cal P}_B ) }{ C_{\myeoq,B}( T_B ) } \} \leq \frac{ 32 }{ 31 }$.
\end{itemize}

\paragraph{Case 5: \boldmath{${T_B = \frac{ 1 }{ 16 }}$}.} As schematically illustrated in Figure~\ref{fig:case5}, our construction proceeds as follows:
\begin{itemize}
    \item {\em The policy ${\cal P}_A$}: Once again, the policy ${\cal P}_A$ will be a $\frac{31}{32}$-scaling of $T_A$, with $A$-orders placed at $0, \frac{31}{32}, \frac{62}{32}, \frac{93}{32} \ldots$.

    \item {\em The policy ${\cal P}_B$}: The cycle length of ${\cal P}_B$ will be $\frac{ 31 }{ 32}$ as well. Each such cycle, say $[0, \frac{ 31 }{ 32 })$, is partitioned into three segments: $[0, \frac{ 9 }{ 32 })$, $[\frac{ 9 }{ 32 }, \frac{ 23 }{ 32 })$, and $[\frac{ 23 }{ 32 },\frac{ 31 }{32})$, filled by $B$-orders as follows:
    \begin{itemize}
        \item In the left segment $[0, \frac{ 9 }{ 32 })$, we place six $B$-orders, each of length $\frac{3}{4} \cdot T_B$.

        \item In the middle segment $[\frac{ 9 }{ 32 }, \frac{ 23 }{ 32 })$, we place seven $B$-orders, each of length $T_B$.

        \item In the right segment $[\frac{ 23 }{ 32 },\frac{ 31 }{32})$, we place three $B$-orders, each of length $\frac{4}{3} \cdot T_B$.
    \end{itemize}

    \item {\em Peak space requirement:} ${\cal P}_A$ and ${\cal P}_B$ have a joint cycle of length $\frac{ 31 }{ 32}$, and for convenience, we examine $[0, \frac{ 31 }{ 32 })$. While this interval has quite a few ordering points, we observe that since the inventory level of commodity $A$ is decreasing across $[0, \frac{ 31 }{ 32 })$, and since we repeat the same policy for commodity $B$ in each of the  segments $[0, \frac{ 9 }{ 32 })$, $[\frac{ 9 }{ 32 }, \frac{ 23 }{ 32 })$, and $[\frac{ 23 }{ 32 },\frac{ 31 }{32})$, the peak space requirement of ${\cal P}_A$ and ${\cal P}_B$ will be attained at one of the points $0$, $\frac{ 9 }{ 32 }$, and $\frac{ 23 }{ 32 }$.
    \begin{itemize}
        \item $V( {\cal P}_A, {\cal P}_B, 0 ) = \gamma_A  T_A + \frac{ 3 }{ 4 } \cdot \gamma_B  T_B  = \frac{ 7 }{ 4 } \cdot ( \gamma_A \cdot \bar{I}( T_A ) + \gamma_B \cdot \bar{I}( T_B ) )$.

        \item $V( {\cal P}_A, {\cal P}_B, \frac{9}{32} ) = \frac{22}{31} \cdot \frac{31}{32} \cdot \gamma_A  T_A + \gamma_B  T_B  = \frac{ 27 }{ 16 } \cdot ( \gamma_A \cdot \bar{I}( T_A ) + \gamma_B \cdot \bar{I}( T_B ) )$.

        \item $V( {\cal P}_A, {\cal P}_B, \frac{23}{32} ) = \frac{8}{31} \cdot \frac{31}{32} \cdot \gamma_A  T_A + \frac{4}{3} \cdot \gamma_B  T_B  = \frac{ 19 }{ 12 } \cdot ( \gamma_A \cdot \bar{I}( T_A ) + \gamma_B \cdot \bar{I}( T_B ) )$.
    \end{itemize}

    \item {\em Cost}: Since ${\cal P}_A$ is a $\frac{ 31 }{ 32 }$-scaling of $T_A$, we have $C_A( {\cal P}_A ) \leq \frac{ 32 }{ 31 } \cdot C_{\myeoq,A}( T_A )$.  In addition, given the construction of ${\cal P}_B$, this policy is a $\frac{3}{4}$-scaling of $T_B$ on a $\frac{ 9 }{ 31 }$-fraction of its cycle, $\frac{4}{3}$-scaling of $T_B$ on a $\frac{ 8 }{ 31 }$-fraction, and coincides with $T_B$ on the remaining $\frac{ 14 }{ 31 }$-fraction. Therefore,
    \begin{align*}
    C_B( {\cal P}_B ) & ~~=~~ \frac{ 9 }{ 31 } \cdot C_{\myeoq,B} \left( \frac{ 3 }{ 4 } \cdot T_B \right) + \frac{ 8 }{ 31 } \cdot C_{\myeoq,B} \left( \frac{ 4 }{ 3 } \cdot T_B \right) + \frac{ 14 }{ 31 } \cdot C_{\myeoq,B} (  T_B )  \\
    & ~~=~~ \frac{ 8 }{ 31 } \cdot \underbrace{ \left( C_{\myeoq,B} \left( \frac{ 3 }{ 4 } \cdot T_B \right) + C_{\myeoq,B} \left( \frac{ 4 }{ 3 } \cdot T_B \right) \right) }_{ = (\frac{4}{3} + \frac{3}{4}) \cdot C_{\myeoq,B}(T_B) } \\
    & \qquad \qquad \mbox{} + \frac{ 1 }{ 31 } \cdot C_{\myeoq,B} \left( \frac{ 3 }{ 4 } \cdot T_B \right) + \frac{ 14 }{ 31 } \cdot C_{\myeoq,B} (  T_B ) \\
    & ~~\leq~~ \left( \frac{ 50 }{ 93 } + \frac{ 1 }{ 31 } \cdot \frac{4}{3} + \frac{ 14 }{ 31 } \right) \cdot C_{\myeoq,B} (  T_B ) \\
    & ~~=~~ \frac{ 32 }{ 31 } \cdot C_{\myeoq,B} (  T_B ) \ .
    \end{align*}

    \item {\em Summary}: $V_{\max}( {\cal P}_A, {\cal P}_B ) = \frac{ 7 }{ 4 } \cdot ( \gamma_A \cdot \bar{I}( T_A ) + \gamma_B \cdot \bar{I}( T_B ) )$ and $\max \{ \frac{ C_A( {\cal P}_A ) }{ C_{\myeoq,A}( T_A ) }, \frac{ C_B( {\cal P}_B ) }{ C_{\myeoq,B}( T_B ) } \} \leq \frac{ 32 }{ 31 }$.
\end{itemize}

\begin{figure}[htbp!]
    \centering

    \begin{minipage}[b]{0.48\textwidth}
        \centering
        \resizebox{\linewidth}{!}{%
        \begin{tikzpicture}[xscale=8, yscale=2.5]

            \def\H{1.5}
            \def\Tmax{0.96875} 

            \draw[->] (0,0) -- (1.1,0) node[right] {$t$};
            \draw[->] (0,-0.1) -- (0,3.0) node[above] {occupied space};

            \draw (1pt, 1.75*\H) -- (-1pt, 1.75*\H) node[left] {$7/4$};
            \draw (1pt, \H) -- (-1pt, \H) node[left] {$1$};

            \draw[dashed, gray!40] (0, 1.75*\H) -- (0.96875, 1.75*\H);
            \draw[dotted, gray!60] (0, \H) -- (0.96875, \H);

            \draw[thick, blue!60!cyan] (0, \H) -- (0.96875, 0);

            \draw[thick, red!70!orange] (0, 0.75*\H)
                -- (0.046875, 0) -- (0.046875, 0.75*\H)
                -- (0.09375, 0) -- (0.09375, 0.75*\H)
                -- (0.140625, 0) -- (0.140625, 0.75*\H)
                -- (0.1875, 0) -- (0.1875, 0.75*\H)
                -- (0.234375, 0) -- (0.234375, 0.75*\H)
                -- (0.28125, 0) -- (0.28125, \H) 
                -- (0.34375, 0) -- (0.34375, \H)
                -- (0.40625, 0) -- (0.40625, \H)
                -- (0.46875, 0) -- (0.46875, \H)
                -- (0.53125, 0) -- (0.53125, \H)
                -- (0.59375, 0) -- (0.59375, \H)
                -- (0.65625, 0) -- (0.65625, \H)
                -- (0.71875, 0) -- (0.71875, 1.3333*\H) 
                -- (0.80208, 0) -- (0.80208, 1.3333*\H)
                -- (0.88541, 0) -- (0.88541, 1.3333*\H)
                -- (0.96875, 0);

            \draw[very thick, violet]
                (0, 1.75*\H)
                -- (0.046875, { (1 - 1.0322*0.046875)*\H }) -- (0.046875, { (1 - 1.0322*0.046875)*\H + 0.75*\H })
                -- (0.09375,  { (1 - 1.0322*0.09375)*\H  }) -- (0.09375,  { (1 - 1.0322*0.09375)*\H  + 0.75*\H })
                -- (0.140625, { (1 - 1.0322*0.140625)*\H }) -- (0.140625, { (1 - 1.0322*0.140625)*\H + 0.75*\H })
                -- (0.1875,   { (1 - 1.0322*0.1875)*\H   }) -- (0.1875,   { (1 - 1.0322*0.1875)*\H   + 0.75*\H })
                -- (0.234375, { (1 - 1.0322*0.234375)*\H }) -- (0.234375, { (1 - 1.0322*0.234375)*\H + 0.75*\H })
                -- (0.28125,  { (1 - 1.0322*0.28125)*\H  }) -- (0.28125,  { (1 - 1.0322*0.28125)*\H  + \H })
                -- (0.34375,  { (1 - 1.0322*0.34375)*\H  }) -- (0.34375,  { (1 - 1.0322*0.34375)*\H  + \H })
                -- (0.40625,  { (1 - 1.0322*0.40625)*\H  }) -- (0.40625,  { (1 - 1.0322*0.40625)*\H  + \H })
                -- (0.46875,  { (1 - 1.0322*0.46875)*\H  }) -- (0.46875,  { (1 - 1.0322*0.46875)*\H  + \H })
                -- (0.53125,  { (1 - 1.0322*0.53125)*\H  }) -- (0.53125,  { (1 - 1.0322*0.53125)*\H  + \H })
                -- (0.59375,  { (1 - 1.0322*0.59375)*\H  }) -- (0.59375,  { (1 - 1.0322*0.59375)*\H  + \H })
                -- (0.65625,  { (1 - 1.0322*0.65625)*\H  }) -- (0.65625,  { (1 - 1.0322*0.65625)*\H  + \H })
                -- (0.71875,  { (1 - 1.0322*0.71875)*\H  }) -- (0.71875,  { (1 - 1.0322*0.71875)*\H  + 1.3333*\H })
                -- (0.80208,  { (1 - 1.0322*0.80208)*\H  }) -- (0.80208,  { (1 - 1.0322*0.80208)*\H  + 1.3333*\H })
                -- (0.88541,  { (1 - 1.0322*0.88541)*\H  }) -- (0.88541,  { (1 - 1.0322*0.88541)*\H  + 1.3333*\H })
                -- (0.96875, 0);

            \draw (0.28125, 1pt) -- (0.28125, -1pt) node[below] {$\frac{9}{32}$};
            \draw (0.71875, 1pt) -- (0.71875, -1pt) node[below] {$\frac{23}{32}$};
            \draw (0.96875, 1pt) -- (0.96875, -1pt) node[below] {$\frac{31}{32}$};

            \matrix [draw, fill=white, anchor=north east, font=\footnotesize, inner sep=3pt] at (1.1, 2.9) {
                \draw[very thick, violet] (0,0) -- (0.4,0) node[right, text=black] {Total space}; \\
                \draw[thick, blue!60!cyan] (0,0) -- (0.4,0) node[right, text=black] {Item A}; \\
                \draw[thick, red!70!orange] (0,0) -- (0.4,0) node[right, text=black] {Item B}; \\
            };
        \end{tikzpicture}%
        }
        \caption{Case 5}
        \label{fig:case5}
    \end{minipage}
    \hfill 
    \begin{minipage}[b]{0.48\textwidth}
        \centering
        \resizebox{\linewidth}{!}{%
        \begin{tikzpicture}[xscale=8, yscale=2.5]

            \def\H{1.5}
            \def\w{0.03125} 

            \draw[->] (0,0) -- (1.1,0) node[right] {$t$};
            \draw[->] (0,-0.1) -- (0,3.0) node[above] {occupied space};

            \draw (1pt, 1.75*\H) -- (-1pt, 1.75*\H) node[left] {$7/4$};
            \draw (1pt, \H) -- (-1pt, \H) node[left] {$1$};

            \draw[dashed, gray!40] (0, 1.75*\H) -- (1.0, 1.75*\H);
            \draw[dotted, gray!60] (0, \H) -- (1.0, \H);

            \draw[very thick, violet] (0, 1.75*\H)
            \foreach \i in {1,...,12} {
                -- ({\i*\w}, {(1-\i*\w)*\H})
                -- ({\i*\w}, {(1-\i*\w)*\H + 0.75*\H})
            }
            -- ({12*\w}, {(1-(12*\w))*\H})
            -- ({12*\w}, {(1-(12*\w))*\H + \H})
            \foreach \i in {1,...,8} {
                -- ({12*\w+\i*\w}, {(1-(12*\w+\i*\w))*\H})
                -- ({12*\w+\i*\w}, {(1-(12*\w+\i*\w))*\H + \H})
            }
            -- ({20*\w}, {(1-(20*\w))*\H})
            -- ({20*\w}, {(1-(20*\w))*\H + 1.3333*\H})
            \foreach \i in {1,...,12} {
                -- ({20*\w+\i*\w}, {(1-(20*\w+\i*\w))*\H})
                -- ({20*\w+\i*\w}, {(1-(20*\w+\i*\w))*\H + 1.3333*\H})
            }
            -- (1.0, 0);

            \draw[thick, blue!60!cyan] (0, \H) -- (1, 0);

            \draw[thick, red!70!orange] (0, 0.75*\H)
            \foreach \i in {1,...,12} { -- ({\i*\w}, 0) -- ({\i*\w}, 0.75*\H) };

            \draw[thick, red!70!orange] (0.375, 0.75*\H) -- (0.375, 0) -- (0.375, \H)
            \foreach \i in {1,...,8} { -- ({0.375 + \i*\w}, 0) -- ({0.375 + \i*\w}, \H) };

            \draw[thick, red!70!orange] (0.625, \H) -- (0.625, 0) -- (0.625, 1.3333*\H)
            \foreach \i in {1,...,12} { -- ({0.625 + \i*\w}, 0) -- ({0.625 + \i*\w}, 1.3333*\H) };

            \draw (0.375, 1pt) -- (0.375, -1pt) node[below] {$\frac{3}{8}$};
            \draw (0.625, 1pt) -- (0.625, -1pt) node[below] {$\frac{5}{8}$};
            \draw (1.0, 1pt) -- (1.0, -1pt) node[below] {$1$};

            \matrix [draw, fill=white, anchor=north east, font=\footnotesize, inner sep=3pt] at (1.1, 2.9) {
                \draw[very thick, violet] (0,0) -- (0.4,0) node[right, text=black] {Total space}; \\
                \draw[thick, blue!60!cyan] (0,0) -- (0.4,0) node[right, text=black] {Item A}; \\
                \draw[thick, red!70!orange] (0,0) -- (0.4,0) node[right, text=black] {Item B}; \\
            };
        \end{tikzpicture}%
        }
        \caption{Case 6}
        \label{fig:case6}
    \end{minipage}
\end{figure}

\paragraph{Case 6: \boldmath{${T_B = \frac{ 1 }{ 2^k }}$} for \boldmath{${k \geq 5}$}.} As schematically illustrated in Figure~\ref{fig:case6}, our construction proceeds as follows:
\begin{itemize}
    \item {\em The policy ${\cal P}_A$}: $A$-orders will be placed at the integer time points $0, 1, 2, \ldots$.

    \item {\em The policy ${\cal P}_B$}: The cycle length of ${\cal P}_B$ will be $1$ as well. Each such cycle, say $[0, 1)$, is partitioned into three segments: $[0, \frac{ 3 }{ 8 })$, $[\frac{ 3 }{ 8 }, \frac{ 5 }{ 8 })$, and $[\frac{ 5 }{ 8 },1)$, filled by $B$-orders as follows:
    \begin{itemize}
        \item In the left segment $[0, \frac{ 3 }{ 8 })$, we place $\frac{ 1 }{ 2T_B }$ orders, each of length $\frac{3}{4} \cdot T_B$.

        \item In the middle segment $[\frac{ 3 }{ 8 }, \frac{ 5 }{ 8 })$, we place $\frac{ 1 }{ 4T_B }$ orders, each of length $T_B$.

        \item In the right segment $[\frac{ 5 }{ 8 },1)$, we place $\frac{ 9 }{ 32T_B }$ orders, each of length $\frac{4}{3} \cdot T_B$.
    \end{itemize}
    It is important to note that, since $T_B = \frac{ 1 }{ 2^k }$ for $k \geq 5$, we ensure that each of the terms  $\frac{ 1 }{ 2T_B }$,  $\frac{ 1 }{ 4T_B }$, and $\frac{ 9 }{ 32T_B }$ is indeed an integer.

    \item {\em Peak space requirement:} As in case~5,  since the inventory level of commodity $A$ is decreasing across $[0, 1)$, and since we repeat the same $B$-policy in each of the segments $[0, \frac{ 3 }{ 8 })$, $[\frac{ 3 }{ 8 }, \frac{ 5 }{ 8 })$, and $[\frac{ 5 }{ 8 },1)$, the peak space requirement of ${\cal P}_A$ and ${\cal P}_B$ will be attained at one of the points $0$, $\frac{ 3 }{ 8 }$, and $\frac{ 5 }{ 8 }$.
    \begin{itemize}
        \item $V( {\cal P}_A, {\cal P}_B, 0 ) = \gamma_A  T_A + \frac{ 3 }{ 4 } \cdot \gamma_B  T_B  = \frac{ 7 }{ 4 } \cdot ( \gamma_A \cdot \bar{I}( T_A ) + \gamma_B \cdot \bar{I}( T_B ) )$.

        \item $V( {\cal P}_A, {\cal P}_B, \frac{3}{8} ) = \frac{5}{8} \cdot \gamma_A  T_A + \gamma_B  T_B  = \frac{ 13 }{ 8 } \cdot ( \gamma_A \cdot \bar{I}( T_A ) + \gamma_B \cdot \bar{I}( T_B ) )$.

        \item $V( {\cal P}_A, {\cal P}_B, \frac{5}{8} ) = \frac{3}{8} \cdot \gamma_A  T_A + \frac{4}{3} \cdot \gamma_B  T_B  = \frac{ 41 }{ 24 } \cdot ( \gamma_A \cdot \bar{I}( T_A ) + \gamma_B \cdot \bar{I}( T_B ) )$.
    \end{itemize}

    \item {\em Cost}: Since ${\cal P}_A$ coincides with $T_A$, we have $C_A( {\cal P}_A ) = C_{\myeoq,A}( T_A )$.  In addition, given the construction of ${\cal P}_B$, this policy is a $\frac{3}{4}$-scaling of $T_B$ on a $\frac{ 3 }{ 8 }$-fraction of its cycle, $\frac{4}{3}$-scaling of $T_B$ on a $\frac{ 3 }{ 8 }$-fraction, and coincides with $T_B$ on the remaining $\frac{ 1 }{ 4 }$-fraction. Therefore,
        \begin{align*}
        C_B( {\cal P}_B ) & ~~=~~ \frac{ 3 }{ 8 } \cdot \underbrace{ \left( C_{\myeoq,B} \left( \frac{ 4 }{ 3 } \cdot T_B \right) + C_{\myeoq,B} \left( \frac{ 3 }{ 4 } \cdot T_B \right) \right) }_{ = (\frac{4}{3} + \frac{3}{4}) \cdot C_{\myeoq,B}(T_B) } + \frac{ 1 }{ 4 } \cdot C_{\myeoq,B}(T_B) \\
        & ~~=~~  \frac{ 33 }{ 32 } \cdot C_{\myeoq,B}( T_B ) \ .
        \end{align*}

    \item {\em Summary}: $V_{\max}( {\cal P}_A, {\cal P}_B ) = \frac{ 7 }{ 4 } \cdot ( \gamma_A \cdot \bar{I}( T_A ) + \gamma_B \cdot \bar{I}( T_B ) )$ and $\max \{ \frac{ C_A( {\cal P}_A ) }{ C_{\myeoq,A}( T_A ) }, \frac{ C_B( {\cal P}_B ) }{ C_{\myeoq,B}( T_B ) } \} \leq \frac{ 33 }{ 32 }$.
\end{itemize}

\subsection{Proof of Claim~\ref{clm:bound_on_far_pairs}} \label{app:proof_lem_bound_on_far_pairs}

By Claim~\ref{clm:properties_PO2}(2), we know that $T_i^{ \Theta_{\ell, q} } \in [ \frac{ \hat{T}_i }{ \sqrt{2} } , \sqrt{2} \hat{T}_i ]$, for every commodity $i \in {\cal H}_{ \ell ,q}$. In addition, since any such commodity is heavy, $\gamma_i \hat{T}_i = 2\gamma_i \cdot \bar{I}( \hat{T}_i ) \in 2 \cdot [\frac{3}{4},1] \cdot \frac{1}{(1+\eps)^{\ell-1}} \cdot{\cal V}$. It follows that all elements of the sequence~\eqref{eqn:sequence_gamma_T} are bounded within the interval $[\frac{3 }{2\sqrt{2}}, 2\sqrt{2}] \cdot \frac{1}{(1+\eps)^{\ell-1}} \cdot{\cal V}$, whose endpoints differ by a multiplicative factor of $\frac{ 8 }{ 3 }$. Given this bound, we conclude that $|{\cal F}_{ \ell ,q}^{ \Pi_{ \ell, q } }| \leq  \log_{ 1+\eps } (\frac{ 8 }{ 3 }) \leq \frac{ 11 }{ 10\eps }$.

\subsection{Proof of Lemma~\ref{lem:success_Aell}} \label{app:proof_lem_success_Aell}

Let us consider the collection of random variables $X_1, \ldots, X_Q$, where $X_q = \sum_{i \in {\cal H}_{\ell,q} } \gamma_i  T_i^{ \Theta_{\ell, q} }$. Recalling that ${\cal A}_{\ell}$ stands for the event
\[ \sum_{i \in {\cal H}_{\ell} } \gamma_i   T_i^{ \Theta_{\ell} } ~~\leq~~ (1 + \eps) \cdot \frac{ 1 }{ \sqrt{2} \ln 2 } \cdot \sum_{i \in {\cal H}_{\ell} } \gamma_i   \hat{T}_i \ , \]
we observe that the left-hand side of this inequality is exactly $\sum_{q \in [Q]} X_q$. In addition, $X_1, \ldots, X_Q$ are mutually independent, since our power-of-$2$ rounding procedure is employed for each subset ${\cal H}_{\ell,q}$ independently of any other subset. For these random variables, we establish the next auxiliary claim, whose proof is given in Appendix~\ref{app:proof_clm_auxiliary_success_Aell}. For readability, we make use of the shorthand notation $\eta_q = \sum_{i \in {\cal H}_{\ell,q} } \gamma_i \hat{T}_i$, with $\eta_{\max}$ and $\eta_{\min}$ being the maximum and minimum of these quantities over $q \in [Q]$.

\begin{claim} \label{clm:auxiliary_success_Aell}
\begin{enumerate}
    \item $\exsubpar{ \Theta_\ell }{ X_q } = \frac{ \eta_q }{ \sqrt{2} \ln 2 }$.

    \item $X_q \leq \sqrt{2} \eta_q$.

    \item $\frac{ \eta_{\min}}{ \eta_{\max} } \geq \frac{ 1 }{ 2 }$.
\end{enumerate}
\end{claim}

Given this claim, we argue that $\prsubpar{ \Theta_\ell }{ \bar{\cal A}_{ \ell }} \leq \frac{ \eps }{ 10 }$, by noting that
\begin{align}
\prsub{ \Theta_\ell }{ \bar{\cal A}_{ \ell }} & ~~=~~ \prsub{ \Theta_\ell }{  \sum_{i \in {\cal H}_{\ell} } \gamma_i   T_i^{ \Theta_{\ell} } >  (1 + \eps) \cdot \frac{ 1 }{ \sqrt{2} \ln 2 } \cdot \sum_{i \in {\cal H}_{\ell} } \gamma_i   \hat{T}_i } \nonumber \\
& ~~=~~ \prsub{ \Theta_\ell }{ \sum_{q \in [Q]} \frac{ X_q }{ \sqrt{2}\eta_{\max} }  >  (1 + \eps) \cdot \sum_{q \in [Q]} \frac{ \exsubpar{ \Theta_\ell }{ X_q } }{ \sqrt{2}\eta_{\max} } } \label{eqn:proof_lem_success_Aell_1}  \\
& ~~\leq~~ \exp \left( -\frac{ \eps^2 }{3} \cdot \sum_{q \in [Q]} \frac{ \exsubpar{ \Theta_\ell }{ X_q } }{ \sqrt{2} \eta_{\max} } \right) \label{eqn:proof_lem_success_Aell_2}   \\
& ~~\leq~~ \exp \left( -\frac{ \eps^2 Q }{6\ln 2} \cdot \frac{  \eta_{\min} }{ \eta_{\max} } \right) \label{eqn:proof_lem_success_Aell_3}   \\
& ~~\leq~~ \exp \left( - 2 \ln \left( \frac{1}{ \eps } \right) \right) \label{eqn:proof_lem_success_Aell_4} \\
& ~~\leq~~ \frac{ \eps }{ 10 }  \ . \nonumber
\end{align}
Here, equality~\eqref{eqn:proof_lem_success_Aell_1} is obtained by  combining the definitions of $X_q$ and $\eta_q$ along with Claim~\ref{clm:auxiliary_success_Aell}(1). To arrive at inequality~\eqref{eqn:proof_lem_success_Aell_2}, we make use of a dimension-free Chernoff-Hoeffding inequality stated in \citet[Thm.~1.1 and Ex.~1.1]{DubhashiP09}. Specifically, letting $Y_1, \ldots, Y_n$ be independent $[0,1]$-bounded random variables, for every $\eps \in (0,1)$, we have
\[ \pr{ \sum_{i \in [n]} Y_i > (1 + \eps) \cdot \ex{ \sum_{i \in [n]} Y_i } } ~~\leq~~ \exp \left( - \frac{ \eps^2 }{ 3 } \cdot \ex{ \sum_{i \in [n]} Y_i } \right) \ . \]
In this context, it is worth mentioning that $\{ \frac{ X_q }{\sqrt{2}\eta_{\max} } \}_{q \in [Q]}$ are mutually independent. In addition, these random variables are $[0,1]$-bounded, by Claim~\ref{clm:auxiliary_success_Aell}(2). Inequality~\eqref{eqn:proof_lem_success_Aell_3} holds since $\exsubpar{ \Theta_\ell }{ X_q } = \frac{ \eta_q }{ \sqrt{2} \ln 2 } \geq \frac{ \eta_{\min} }{ \sqrt{2} \ln 2 }$, by Claim~\ref{clm:auxiliary_success_Aell}(1). Finally, inequality~\eqref{eqn:proof_lem_success_Aell_4} follows by noting that $Q = \frac{ 20 \ln (1/\eps) }{ \eps^2 }$ and that $\frac{ \eta_{\min}}{ \eta_{\max} } \geq \frac{ 1 }{ 2 }$, by Claim~\ref{clm:auxiliary_success_Aell}(3).

\subsection{Proof of Claim~\ref{clm:auxiliary_success_Aell}} \label{app:proof_clm_auxiliary_success_Aell}

By definition of $X_q$, we have
\[ \exsub{ \Theta_\ell }{ X_q } ~~=~~ \exsub{ \Theta_{ \ell, q } }{ \sum_{i \in {\cal H}_{\ell,q} } \gamma_i   T_i^{ \Theta_{\ell, q} } } ~~=~~ \frac{ 1 }{ \sqrt{2} \ln 2 } \cdot \sum_{i \in {\cal H}_{\ell,q} } \gamma_i  \hat{T}_i ~~=~~ \frac{ \eta_q }{ \sqrt{2} \ln 2 } \ , \] where the second equality holds since $\exsubpar{ \Theta_{\ell, q} } { T_i^{ \Theta_{\ell, q} } } = \frac{ 1 }{ \sqrt{2} \ln 2 } \cdot \hat{T}_i$ , by Claim~\ref{clm:properties_PO2}(1). Similarly,
\[ X_q ~~=~~ \sum_{i \in {\cal H}_{\ell,q} } \gamma_i   T_i^{ \Theta_{\ell, q} } ~~\leq~~ \sqrt{2} \cdot \sum_{i \in {\cal H}_{\ell,q} } \gamma_i   \hat{T}_i ~~=~~ \sqrt{2} \eta_q \ , \]
where the inequality above holds since $T_i^{ \Theta_{\ell, q} } \leq \sqrt{2} \hat{T}_i$,  by  Claim~\ref{clm:properties_PO2}(2).

Now, in regard to the relation between $\eta_{\max}$ and $\eta_{\min}$, on the one hand,
\[ \eta_{\max} ~~=~~ \max_{q \in [Q]} \left\{ \sum_{i \in {\cal H}_{\ell,q} } \gamma_i   \hat{T}_i \right\} ~~\leq~~ \left( \left\lfloor \frac{ |{\cal H}_{ \ell }| }{ Q } \right\rfloor+1 \right) \cdot \frac{2}{(1+\eps)^{\ell-1}} \cdot{\cal V} ~~\leq~~  \frac{ |{\cal H}_{ \ell }| }{ Q } \cdot \frac{2(1+\eps)}{(1+\eps)^{\ell-1}} \cdot{\cal V}  \ . \]
Here, the first inequality is obtained by noting that each subset ${\cal H}_{\ell,q}$ is of size either $\lfloor \frac{ |{\cal H}_{ \ell }| }{ Q } \rfloor$ or $\lceil \frac{ |{\cal H}_{ \ell }| }{ Q } \rceil$, as explained in Section~\ref{subsec:pow-rounding}. In addition, $\gamma_i \cdot \bar{I}( \hat{T}_i ) \leq \frac{1}{(1+\eps)^{\ell-1}} \cdot{\cal V}$ for every commodity $i \in {\cal H}_{\ell,q} \subseteq \tilde{\cal V}_{\ell}$, according to property~\ref{prop:hatT_space}. On the other hand,
\[ \eta_{\min} ~~=~~ \min_{q \in [Q]} \left\{ \sum_{i \in {\cal H}_{\ell,q} } \gamma_i  \hat{T}_i \right\} ~~\geq~~ \left\lfloor \frac{ |{\cal H}_{ \ell }| }{ Q } \right\rfloor \cdot \frac{3}{2(1+\eps)^{\ell-1}} \cdot{\cal V} ~~\geq~~  \frac{ |{\cal H}_{ \ell }| }{ Q } \cdot \frac{3(1-\eps)}{2(1+\eps)^{\ell-1}} \cdot{\cal V} \ , \]
where the first inequality holds since $\gamma_i \cdot \bar{I}( \hat{T}_i ) \geq \frac{ 3 }{ 4 } \cdot \frac{1}{(1+\eps)^{\ell-1}} \cdot{\cal V}$, by definition of
${\cal H}_{\ell}$. Combining these bounds on $\eta_{\max}$ and $\eta_{\min}$, it follows that $\frac{ \eta_{\min}}{ \eta_{\max} } \geq \frac{ 3(1-\eps) }{ 4(1 + \eps) } \geq \frac{ 1 }{ 2 }$.

\subsection{Proof of Lemma~\ref{lem:PO2_bound_Vmax}} \label{app:proof_lem_PO2_bound_Vmax}

\paragraph{Bounding \boldmath{${V_{\max}( {\cal P}^{\ell-} )}$}.} Starting with the simple claim, we derive an upper bound on the peak space requirement of ${\cal P}^{\ell-}$ by recalling that ${\cal P}_i^{\ell-} = \alpha \hat{T}_i =  \frac{ 7/8 }{ \sqrt{2} \ln 2 } \cdot \hat{T}_i$ for every commodity $i \in \tilde{\cal V}_{\ell}$. Therefore,
\[ V_{\max}( {\cal P}^{\ell-} )  ~~=~~ \frac{ 7/8 }{ \sqrt{2} \ln 2 } \cdot \sum_{i \in \tilde{\cal V}_{\ell}} \gamma_i  \hat{T}_i ~~\leq~~ \frac{ 7/4 }{ \sqrt{2} \ln 2 } \cdot | \tilde{\cal V}_{\ell} | \cdot \frac{1}{(1+\eps)^{\ell-1}} \cdot{\cal V} \ , \]
where the inequality above holds since $\gamma_i \cdot \bar{I}( \hat{T}_i ) \leq \frac{1}{(1+\eps)^{\ell-1}} \cdot{\cal V}$, by property~\ref{prop:hatT_space}.

\paragraph{Bounding \boldmath{${V_{\max}( {\cal P}^{\ell+} )}$}.} To upper-bound the random peak space requirement associated with $\tilde{\cal P}^{\ell+}$, note that
\begin{align}
V_{\max}( \tilde{\cal P}^{\ell+} ) & ~~\leq~~ \underbrace{ \left[ \left. \sum_{q \in [Q]} \sum_{ \{ \Pi_{ \ell, q }(2i-1), \Pi_{ \ell, q }(2i) \} \in {\cal N}_{ \ell ,q}^{ \Pi_{ \ell, q } } } V_{\max}( {\cal P}_{ \Pi_{ \ell, q }(2i-1) }^{\ell,q}, {\cal P}_{ \Pi_{ \ell, q }(2i)}^{\ell,q} ) \right| {\cal A}_{\ell} \right] }_{ \text{type 1} }  \nonumber \\
& \phantom{~~\leq~~} \mbox{} + \underbrace{ \left[ \left. \sum_{ i \in {\cal T}_2^{ \Theta_\ell } } \gamma_i  T_i^{ \Theta_{\ell} } \right| {\cal A}_{\ell} \right] }_{ \text{type 2} } + \underbrace{ \left[ \left. \sum_{ i \in {\cal T}_3^{ \Theta_\ell } } \gamma_i  T_i^{ \Theta_{\ell} } \right| {\cal A}_{\ell} \right] }_{ \text{type 3} } + \underbrace{ \vphantom{\left[ \left. \sum_{ i \in {\cal T}_3^{ \Theta_\ell } } \gamma_i  T_i^{ \Theta_{\ell} } \right| {\cal A}_{\ell} \right]} \sum_{ i \in {\cal T}_4 } \gamma_i  \hat{T}_i }_{ \text{type 4} } \ . \label{eqn:proof_lem_PO2_bound_Vmax_eq6}
\end{align}
We proceed by separately bounding each of these terms.
\begin{enumerate}
    \item {\em Bounding type-1}: By Corollary~\ref{cor:construct_sub1_near}(1), we have
    \begin{align}
    [\text{type 1}] ~~& \leq~~
     (1 + \eps) \cdot \frac{ 7 }{ 8 } \cdot \left[ \sum_{q \in [Q]} \sum_{ \{ \Pi_{ \ell, q }(2i-1), \Pi_{ \ell, q }(2i) \} \in {\cal N}_{ \ell ,q}^{ \Pi_{ \ell, q } } } \left(  \gamma_{ \Pi_{ \ell, q }(2i-1) } \cdot T_{ \Pi_{ \ell, q }(2i-1)}^{ \Theta_{\ell, q} } \right. \right. \nonumber \\
    ~~& \qquad \qquad \qquad \qquad \qquad \qquad \qquad \qquad \qquad \left. \left. \vphantom{ \sum_{ \{ \Pi_{ \ell, q }(2i-1), \Pi_{ \ell, q }(2i) \} \in {\cal N}_{ \ell ,q}^{ \Pi_{ \ell, q } } } } \left. \mbox{} + \gamma_{ \Pi_{ \ell, q }(2i) } \cdot T_{ \Pi_{ \ell, q }(2i)}^{ \Theta_{\ell, q} } \right) \right| {\cal A}_{\ell} \right] \nonumber \\
    ~~& \leq~~ (1 + \eps) \cdot \frac{ 7 }{ 8 } \cdot \left[ \left. \sum_{i \in {\cal H}_{\ell}} \gamma_i T_i^{ \Theta_{\ell} } \right| {\cal A}_{\ell} \right] \label{eqn:proof_lem_PO2_bound_Vmax_eq1} \\
    ~~& \leq~~ (1 + 3\eps) \cdot \frac{ 7/8 }{ \sqrt{2} \ln 2 } \cdot \sum_{i \in {\cal H}_{\ell}} \gamma_i  \hat{T}_i \label{eqn:proof_lem_PO2_bound_Vmax_eq2} \\
    ~~& \leq~~ (1 + 3\eps) \cdot \frac{ 7/4 }{ \sqrt{2} \ln 2 } \cdot | {\cal H}_{\ell} | \cdot \frac{1}{(1+\eps)^{\ell-1}} \cdot{\cal V} \ . \label{eqn:proof_lem_PO2_bound_Vmax_eq3}
    \end{align}
    Here, inequality~\eqref{eqn:proof_lem_PO2_bound_Vmax_eq1} utilizes the basic observation that, for every $q \in [Q]$, the set of commodities residing in the random collection of near pairs ${\cal N}_{ \ell ,q}^{ \Pi_{ \ell, q } }$ is a subset of ${\cal H}_{\ell,q}$. Inequality~\eqref{eqn:proof_lem_PO2_bound_Vmax_eq2} follows by recalling that ${\cal A}_{\ell}$ stands for the event where $\sum_{i \in {\cal H}_{\ell} } \gamma_i T_i^{ \Theta_{\ell} } \leq (1 + \eps) \cdot \frac{ 1 }{ \sqrt{2} \ln 2 } \cdot \sum_{i \in {\cal H}_{\ell} } \gamma_i   \hat{T}_i$. Finally, inequality~\eqref{eqn:proof_lem_PO2_bound_Vmax_eq3} holds since $\gamma_i \cdot \bar{I}( \hat{T}_i ) \leq \frac{1}{(1+\eps)^{\ell-1}} \cdot{\cal V}$, by property~\ref{prop:hatT_space}.

    \item {\em Bounding type-2}: To bound this term, we recall that ${\cal T}_2^{ \Theta_\ell }$ stands for the collection of commodities that belong to far pairs across ${\cal H}_{ \ell , 1}, \ldots, {\cal H}_{\ell,Q}$. Therefore, by Claim~\ref{clm:bound_on_far_pairs}, we have $| {\cal T}_2^{ \Theta_\ell } | = 2 \cdot \sum_{q \in [Q]} | {\cal F}_{ \ell ,q}^{ \Pi_{ \ell, q } } | \leq \frac{ 11Q }{ 5\eps }$. Consequently,
    \begin{align}
    [\text{type 2}] & ~~\leq~~ \sqrt{2} \cdot \sum_{ i \in {\cal T}_2^{ \Theta_\ell } } \gamma_i  \hat{T}_i \label{eqn:proof_lem_PO2_bound_Vmax_eq3.5} \\
    & ~~\leq~~ \frac{ 22 \sqrt{2} Q }{ 5\eps } \cdot \frac{1}{(1+\eps)^{\ell-1}} \cdot{\cal V} \label{eqn:proof_lem_PO2_bound_Vmax_eq4} \\
    & ~~\leq~~ 4\eps \cdot |{\cal H}_{\ell}| \cdot \frac{1}{(1+\eps)^{\ell-1}} \cdot{\cal V} \label{eqn:proof_lem_PO2_bound_Vmax_eq5} \\
    & ~~\leq~~ 4\eps \cdot |\tilde{\cal V}_{\ell}| \cdot \frac{1}{(1+\eps)^{\ell-1}} \cdot{\cal V} \ . \nonumber
    \end{align}
    Here, inequality~\eqref{eqn:proof_lem_PO2_bound_Vmax_eq3.5} holds since $T_i^{ \Theta_{\ell} } \leq \sqrt{2} \hat{T}_i$ almost surely, by Claim~\ref{clm:properties_PO2}(2). Inequality~\eqref{eqn:proof_lem_PO2_bound_Vmax_eq4} is obtained by recalling that $\gamma_i \cdot \bar{I}( \hat{T}_i ) \leq \frac{1}{(1+\eps)^{\ell-1}} \cdot{\cal V}$, by property~\ref{prop:hatT_space}. Finally, inequality~\eqref{eqn:proof_lem_PO2_bound_Vmax_eq5} follows due to having $\lfloor \frac{ |{\cal H}_{ \ell }| }{ Q } \rfloor \geq \frac{ 2 }{ \eps^2 }$, as shown in~\eqref{eqn:ratio_HQ_eps}.

    \item {\em Bounding type-3}: To bound this term, we recall that ${\cal T}_3^{ \Theta_\ell }$ contains at most one commodity from each of the subsets ${\cal H}_{ \ell , 1}, \ldots, {\cal H}_{\ell,Q}$, implying that $|{\cal T}_3^{ \Theta_\ell }| \leq Q$. As a result, based on arguments similar to those of the previous item,
    \[ [\text{type 3}] ~~\leq~~ 2 \sqrt{2} Q \cdot \frac{1}{(1+\eps)^{\ell-1}} \cdot{\cal V} ~~\leq~~ \sqrt{2}\eps^2 \cdot |\tilde{\cal V}_{\ell}| \cdot \frac{1}{(1+\eps)^{\ell-1}} \cdot{\cal V} \ . \]

    \item {\em Bounding type-4}: By definition, we know that  $\gamma_i \cdot \bar{I}( \hat{T}_i ) \leq \frac{ 3 }{ 4 } \cdot \frac{1}{(1+\eps)^{\ell-1}} \cdot{\cal V}$ for every commodity $i \in {\cal L}_{\ell}$. Hence,
    \[ [\text{type 4}] ~~\leq~~ \frac{ 3 }{ 2 } \cdot | {\cal L}_{\ell} | \cdot  \frac{1}{(1+\eps)^{\ell-1}} \cdot{\cal V} \ .  \]
\end{enumerate}
In summary, by plugging the above-mentioned bounds into inequality~\eqref{eqn:proof_lem_PO2_bound_Vmax_eq6}, we conclude that
\begin{align*}
V_{\max}( {\cal P}^{\ell+} ) & ~~\leq~~ \left( (1 + 3\eps) \cdot \frac{ 7/4 }{ \sqrt{2} \ln 2 } \cdot | {\cal H}_{\ell} | + 4\eps \cdot |\tilde{\cal V}_{\ell}| + \sqrt{2}\eps^2 \cdot |\tilde{\cal V}_{\ell}| + \frac{ 3 }{ 2 } \cdot | {\cal L}_{\ell} | \right) \cdot  \frac{1}{(1+\eps)^{\ell-1}} \cdot{\cal V} \\
& ~~\leq~~  (1 + 6\eps) \cdot  \frac{ 7/4 }{ \sqrt{2} \ln 2 } \cdot | \tilde{\cal V}_{\ell} | \cdot \frac{1}{(1+\eps)^{\ell-1}} \cdot{\cal V} \ .
\end{align*}

\subsection{Proof of Lemma~\ref{lem:PO2_bound_cost}} \label{app:proof_lem_PO2_bound_cost}

\paragraph{The cost of \boldmath{${{\cal P}^{\ell-}}$}.} Since ${\cal P}^{\ell-}$ is a deterministic policy, with ${\cal P}_i^{\ell-} = \alpha \hat{T}_i$ for every commodity $i \in \tilde{\cal V}_{\ell}$, we have
\begin{equation} \label{eqn:proof_lem_PO2_bound_cost_eq4}
C( {\cal P}^{\ell-} ) ~~=~~ \sum_{ i \in \tilde{\cal V}_{\ell} } C_{\myeoq,i}( \alpha \hat{T}_i ) ~~\leq~~ \max \left\{ \alpha, \frac{ 1 }{ \alpha } \right\} \cdot \sum_{ i \in \tilde{\cal V}_{\ell} } C_{\myeoq,i}( \hat{T}_i ) ~~\leq~~ 2 \cdot \sum_{ i \in \tilde{\cal V}_{\ell} } C_{\myeoq,i}( \hat{T}_i ) \ ,
\end{equation}
where the first inequality follows from Claim~\ref{clm:EOQ_properties}(3), and the second inequality is obtained by recalling that $\alpha =  \frac{ 7/8 }{ \sqrt{2} \ln 2 } \approx 0.892$.

\paragraph{The expected cost of \boldmath{${{\cal P}^{\ell+}}$}.} We obtain an upper bound on the expected long-run average cost of the policy ${\cal P}^{\ell+}$ by initially ignoring the  conditioning on ${\cal A}_{\ell}$. With respect to the latter, we have
\begin{align}
& \exsub{ \Theta_{\ell} }{ C( {\cal P}^{\ell} ) } \nonumber \\
& \qquad \leq~~ \underbrace{ \exsub{ \Theta_{\ell} }{ \sum_{q \in [Q]} \sum_{ \{ \Pi_{ \ell, q }(2i-1), \Pi_{ \ell, q }(2i) \} \in {\cal N}_{ \ell ,q}^{ \Pi_{ \ell, q } } } \left( C_{ \Pi_{ \ell, q }(2i-1) }( {\cal P}^{\ell,q}_{ \Pi_{ \ell, q }(2i-1) } ) + C_{ \Pi_{ \ell, q }(2i) }( {\cal P}^{\ell,q}_{ \Pi_{ \ell, q }(2i) } ) \right) } }_{ \text{type 1} } \nonumber \\
& \qquad \qquad  \mbox{} + \underbrace{ \exsub{ \Theta_{\ell} }{ \sum_{ i \in {\cal T}_2^{ \Theta_\ell } \cup {\cal T}_3^{ \Theta_\ell }  } C_{\myeoq,i}( T_i^{ \Theta_{\ell }} ) } }_{ \text{types 2 and 3} } + \underbrace{ \vphantom{\exsub{ \Theta_{\ell} }{ \sum_{ i \in {\cal T}_2^{ \Theta_\ell } \cup {\cal T}_3^{ \Theta_\ell }  } C_{\myeoq,i}( T_i^{ \Theta_{\ell }} ) }} \sum_{ i \in {\cal T}_4 } C_{\myeoq,i}( \hat{T}_i ) }_{ \text{type 4} }\nonumber \\
& \qquad \leq~~ \frac{ 32 }{ 31 } \cdot \exsubnop{ \Theta_{\ell} } \left[ \sum_{q \in [Q]} \sum_{ \{ \Pi_{ \ell, q }(2i-1), \Pi_{ \ell, q }(2i) \} \in {\cal N}_{ \ell ,q}^{ \Pi_{ \ell, q } } }  \Big(  C_{ \myeoq, \Pi_{ \ell, q }(2i-1) }( T_{ \Pi_{ \ell, q }(2i-1)}^{ \Theta_{\ell, q} } )  \right. \nonumber \\
& \qquad \qquad \left. \vphantom{\sum_{q \in [Q]} \sum_{ \{ \Pi_{ \ell, q }(2i-1), \Pi_{ \ell, q }(2i) \} \in {\cal N}_{ \ell ,q}^{ \Pi_{ \ell, q } } } }
\mbox{} + C_{ \myeoq, \Pi_{ \ell, q }(2i) }( T_{ \Pi_{ \ell, q }(2i)}^{ \Theta_{\ell, q} } ) \Big) \right] + \exsub{ \Theta_{\ell} }{ \sum_{ i \in {\cal T}_2^{ \Theta_\ell } \cup {\cal T}_3^{ \Theta_\ell }  } C_{\myeoq,i}( T_i^{ \Theta_{\ell }} ) }  + \sum_{ i \in {\cal T}_4 } C_{\myeoq,i}( \hat{T}_i ) \label{eqn:proof_lem_PO2_bound_cost_eq1} \\
& \qquad \leq~~ \frac{ 32 }{ 31 } \cdot \exsub{ \Theta_{\ell} }{ \sum_{ i \in {\cal H}_{\ell} } C_{\myeoq,i}( T_i^{ \Theta_{\ell}} ) } + \sum_{ i \in {\cal L}_{\ell} } C_{\myeoq,i}( \hat{T}_i )\nonumber \\
& \qquad =~~ \frac{ 32/31 }{ \sqrt{2} \ln 2 } \cdot \sum_{ i \in {\cal H}_{\ell} } C_{\myeoq,i}( \hat{T}_i ) + \sum_{ i \in {\cal L}_{\ell} } C_{\myeoq,i}( \hat{T}_i ) \label{eqn:proof_lem_PO2_bound_cost_eq2} \\
& \qquad \leq~~ \frac{ 32/31 }{ \sqrt{2} \ln 2} \cdot \sum_{i \in \tilde{\cal V}_{ \ell } } C_{\myeoq,i}( \hat{T}_i ) \ . \nonumber
\end{align}
Here, inequalities~\eqref{eqn:proof_lem_PO2_bound_cost_eq1} and~\eqref{eqn:proof_lem_PO2_bound_cost_eq2} respectively follow from Corollary~\ref{cor:construct_sub1_near}(2) and Claim~\ref{clm:properties_PO2}(1).

Now, to derive an upper bound on the expected cost of the conditional policy ${\cal P}^{\ell+} = [ {\cal P}^{\ell} | {\cal A}_{\ell} ]$, we observe that
\[ \exsub{ \Theta_{\ell} }{ C( {\cal P}^{\ell} ) } ~~\geq~~ \prsub{ \Theta_\ell }{ {\cal A}_{\ell} } \cdot \exsub{ \Theta_{\ell} }{ \left. C( {\cal P}^{\ell} ) \right| {\cal A}_{\ell} }  ~~\geq~~ \left( 1 - \frac{ \eps }{ 10 } \right) \cdot \exsub{ \Theta_{\ell} }{ C( {\cal P}^{\ell+} ) } \ ,  \]
where the last inequality follows from Lemma~\ref{lem:success_Aell}. Therefore,
\begin{equation} \label{eqn:ex_PLplus_UB}
\exsub{ \Theta_{\ell} }{ C( {\cal P}^{\ell+} ) } ~~\leq~~ \left( 1 + \frac{ \eps }{ 5 } \right) \cdot \frac{ 32/31 }{ \sqrt{2} \ln 2} \cdot \sum_{i \in \tilde{\cal V}_{ \ell } } C_{\myeoq,i}( \hat{T}_i ) \ .
\end{equation}

\paragraph{Putting everything together.} We remind the reader that the policy $\tilde{\cal P}^{ \ell }$ we return is precisely ${\cal P}^{\ell+}$ when the event ${\cal A}_{\ell}$ occurs; otherwise, it coincides with the deterministic policy ${\cal P}^{\ell-}$. Therefore,
\begin{align}
\exsub{ \Theta_{\ell} }{ C( \tilde{\cal P}^{\ell} ) } & ~~=~~ \prsub{ \Theta_\ell }{ {\cal A}_{\ell} } \cdot \exsub{ \Theta_{\ell} }{ C( {\cal P}^{\ell+} ) } + \prsub{ \Theta_\ell }{ \bar{\cal A}_{\ell} } \cdot  C( {\cal P}^{\ell-} ) \nonumber \\
& ~~\leq~~ \exsub{ \Theta_{\ell} }{ C( {\cal P}^{\ell+} ) } + \prsub{ \Theta_\ell }{ \bar{\cal A}_{\ell} } \cdot  C( {\cal P}^{\ell-} ) \nonumber \\
& ~~\leq~~ \left( 1 + \frac{ \eps }{ 5 } \right) \cdot \frac{ 32/31 }{ \sqrt{2} \ln 2} \cdot \sum_{i \in \tilde{\cal V}_{ \ell } } C_{\myeoq,i}( \hat{T}_i ) + \frac{ \eps }{ 5 } \cdot \sum_{ i \in \tilde{\cal V}_{\ell} } C_{\myeoq,i}( \hat{T}_i ) \label{eqn:proof_lem_PO2_bound_cost_eq3} \\
& ~~\leq~~ \left( 1 + \frac{ 2\eps }{ 5 } \right) \cdot \frac{ 32/31 }{ \sqrt{2} \ln 2} \cdot \sum_{i \in \tilde{\cal V}_{ \ell } } C_{\myeoq,i}( \hat{T}_i ) \ , \nonumber
\end{align}
where inequality~\eqref{eqn:proof_lem_PO2_bound_cost_eq3} is obtained by recalling that $\prsubpar{ \Theta_\ell }{ \bar{\cal A}_{\ell} } \leq \frac{ \eps }{10}$, according to Lemma~\ref{lem:success_Aell}, and by plugging-in~\eqref{eqn:proof_lem_PO2_bound_cost_eq4} and~\eqref{eqn:ex_PLplus_UB}.

\end{document}